\documentclass[twocolumn,trackchanges]{aastex701}

\usepackage{comment}
\usepackage{float}
\usepackage{amsmath}
\usepackage{booktabs}
\usepackage{multirow}
\usepackage{array}
\usepackage[utf8]{inputenc}
\usepackage[percent]{overpic}

\begin{document}

\title{Verification of the Polarimetric Capability of the East Asia VLBI Network}

\author[orcid=0009-0005-7254-0174]{Yunjeong Lee}
\affiliation{School of Space Research, Kyung Hee University, 1732, Deogyeong-daero, Giheung-gu, Yongin-si, Gyeonggi-do 17104, Republic of Korea}
\affiliation{G-LAMP NEXUS Institute, Kyung Hee University, Yongin, 17104, Republic of Korea}
\email{apfhddldbswj@khu.ac.kr}

\author[0000-0001-6558-9053]{Jongho Park}
\affiliation{School of Space Research, Kyung Hee University, 1732, Deogyeong-daero, Giheung-gu, Yongin-si, Gyeonggi-do 17104, Republic of Korea}
\affiliation{G-LAMP NEXUS Institute, Kyung Hee University, Yongin, 17104, Republic of Korea}
\affiliation{Institute of Astronomy and Astrophysics, Academia Sinica, P.O. Box 23-141, Taipei 10617, Taiwan, R. O. C.}
\email{jparkastro@khu.ac.kr}

\author[0000-0003-1157-4109]{Do-Young Byun}
\affiliation{Korea Astronomy and Space Science Institute, Daedeok-daero 776, Yuseong-gu, Daejeon 34055, Republic of Korea}
\affiliation{University of Science and Technology, Gajeong-ro 217, Yuseong-gu, Daejeon 34113, Republic of Korea}
\email{bdy@kasi.re.kr}

\author[0000-0001-9799-765X]{Minchul Kam}
\affiliation{Institute of Astronomy and Astrophysics, Academia Sinica, P.O. Box 23-141, Taipei 10617, Taiwan, R. O. C.}
\email{mkam@asiaa.sinica.edu.tw}

\author[0000-0001-6906-772X]{Kazuhiro Hada}
\affiliation{Graduate School of Science, Nagoya City University, Yamanohata 1, Mizuho-cho, Mizuho-ku, Nagoya, 467-8501, Aichi, Japan}
\affiliation{Mizusawa VLBI Observatory, National Astronomical Observatory of Japan, 2-12 Hoshigaoka-cho, Mizusawa, Oshu, 023-0861, Iwate, Japan}
\email{hada@nsc.nagoya-cu.ac.jp}

\author[0000-0001-6993-1696]{Juan Carlos Algaba}
\affiliation{Department of Physics, Faculty of Science, Universiti Malaya, 50603 Kuala Lumpur, Malaysia}
\email{algaba@um.edu.my}

\author[0000-0001-7556-8504]{Sanghyun Kim}
\affiliation{Korea Astronomy and Space Science Institute, 776 Daedeok-daero, Yuseong-gu, Daejeon 34055, Republic of Korea}
\email{sanghkim@kasi.re.kr}

\author[0000-0003-3540-8746]{Zhiqiang Shen}
\affiliation{Shanghai Astronomical Observatory, Chinese Academy of Sciences, Shanghai 200030, China}
\affiliation{School of Astronomy and Space Science, University of Chinese Academy of Sciences, Beijing 100049, China}
\email{zshen@shao.ac.cn}

\author[0000-0002-4991-9638]{Junghwan Oh}
\affiliation{Joint Institute for VLBI ERIC (JIVE), Oude Hoogeveensedijk 4, 7991 PD Dwingeloo, The Netherlands}
\email{oh@jive.eu}

\author[0000-0002-0112-4836]{Sincheol Kang}
\affiliation{Korea Astronomy and Space Science Institute, 776 Daedeok-daero, Yuseong-gu, Daejeon 34055, Republic of Korea}
\email{kang87@kasi.re.kr}

\author[0009-0005-7629-8450]{Hyeon-Woo Jeong}
\affiliation{Astronomy and Space Science, University of Science and Technology, 217 Gajeong-ro, Yuseong-gu, Daejeon 34113,
Republic of Korea}
\affiliation{Korea Astronomy and Space Science Institute, 776 Daedeok-daero, Yuseong-gu, Daejeon 34055, Republic of Korea}
\email{hwjeong@kasi.re.kr}

\author[0009-0002-1871-5824]{Whee Yeon Cheong}
\affiliation{Astronomy and Space Science, University of Science and Technology, 217 Gajeong-ro, Yuseong-gu, Daejeon 34113,
Republic of Korea}
\affiliation{Korea Astronomy and Space Science Institute, 776 Daedeok-daero, Yuseong-gu, Daejeon 34055, Republic of Korea}
\email{wheeyeon@kasi.re.kr}

\author[0000-0002-6269-594X]{Sang-Sung Lee}
\affiliation{Astronomy and Space Science, University of Science and Technology, 217 Gajeong-ro, Yuseong-gu, Daejeon 34113,
Republic of Korea}
\affiliation{Korea Astronomy and Space Science Institute, 776 Daedeok-daero, Yuseong-gu, Daejeon 34055, Republic of Korea}
\email{sslee@kasi.re.kr}

\correspondingauthor{Jongho Park}
\email{jparkastro@khu.ac.kr}

\begin{abstract}

The East Asia VLBI Network (EAVN) has recently enabled dual-polarization observations at $22$ and $43\,\mathrm{GHz}$. We present the first systematic verification of its polarimetric performance using EAVN observations of M87, 3C~279, 3C~273, and OJ~287, calibrated with the GPCAL pipeline and evaluated against near-contemporaneous VLBA images at comparable frequencies. Most stations show stable polarimetric leakages with amplitudes of $5$--$10\%$ over monthly timescales. While several VERA stations exhibit D-term phase variations between epochs, we attribute these to field-rotator (FR) offsets and demonstrate that phase stability is restored after applying the analytically derived FR corrections. The resulting linear-polarization morphologies and EVPAs broadly agree with the VLBA results within uncertainties; fractional polarization measured by the EAVN tends to be slightly higher near polarization peaks. Although exact one-to-one comparisons are limited by moderate frequency and epoch differences, the combined evidence indicates robust EAVN polarimetric calibration and imaging capabilities at $22$ and $43\,\mathrm{GHz}$. These results support the scientific capability of EAVN polarimetry and lay the groundwork for expanded, higher-fidelity polarimetric studies in East Asia.


\end{abstract}

\section{Introduction}
\label{sec:intro}

Active Galactic Nuclei (AGNs), which emit intense radiation across the whole electromagnetic spectrum, are believed to be powered by the accretion of matter onto supermassive black holes at their centers \citep{rees1984}. A fraction of AGNs host relativistic jets, powerful outflows launched from the vicinity of the black holes, which remain one of the most intriguing topics and are being actively investigated regarding their formation, collimation, and acceleration mechanisms \citep{Blandford2019}.

Theoretically, the formation of relativistic jets is known to be closely associated with magnetic fields \citep{blandford_1977, blandford_1982}. To probe the magnetic field structures of AGN jets, high-resolution polarimetric observations are essential. Various studies utilizing Very Long Baseline Interferometric (VLBI) observations have indeed provided important information regarding the jet magnetic fields \citep[e.g.,][]{Asada2002, lister2005, Jorstad2007, Clausen-Brown2011, hovatta2012, Cawthorne2013, Park2019, Park2021c, Park2026, eht2021}.

The East Asia VLBI Network (EAVN) is a VLBI array that has undergone significant development in recent years. The concept of the EAVN was first proposed in 2003 \citep{an2018}, and the EAVN has been realized as a structured and coordinated network since 2016 through the integration of the Korean VLBI Network (KVN), the VLBI Exploration of Radio Astrometry (VERA), and the Chinese VLBI Network (CVN), enabling systematic and intensive observations. The EAVN has demonstrated strong performance for total intensity imaging, achieving good angular resolution and sensitivity, at both 22 and 43 GHz (corresponding to K and Q bands, respectively; \citealt{cui2021}). With the installation of right-hand circular polarization (RCP) receivers, dual-polarization mode observations at K and Q bands became available for open use starting in the 2023A semester (\citealt{Akiyama2022}; see also EAVN Status Report 2023A\footnote{\url{https://eavn.kasi.re.kr/status_report/files/Status_Report_EAVN_2023A.pdf}}). However, the performance of EAVN polarimetry has not yet been verified. Therefore, in this study, we aim to assess the quality of EAVN polarization data and present the initial polarization imaging results based on recent EAVN observations.

One of the most critical factors in deriving good, science-ready VLBI polarization data is the accurate estimation of instrumental polarization, also known as antenna polarimetric leakages or "D-terms" \citep{Thompson2017}. This effect occurs because a single polarization feed also responds to signals from the other polarization and can produce spurious signals at the same order of magnitude as the intrinsic polarization of the source. In this work, we utilize the recently developed instrumental polarization calibration pipeline, the Generalized CALibration Pipeline (GPCAL; \citealt{park2021, Park2023a, Park2023b}), which was designed to overcome the limitations of conventional VLBI polarization calibration methods.

In this paper, we present the initial results of EAVN polarimetric observations of several AGNs conducted in 2023 and 2024. We compare the resulting linear polarization images with those obtained from the VLBA, specifically from the Monitoring Of Jets in Active galactic nuclei with VLBA Experiments (MOJAVE; \citealt{lister2018}) and the Blazars Entering the Astrophysical Multi-Messenger Era (BEAM-ME; \citealt{Jorstad2017, Weaver2022}) databases. In Section~\ref{sec:2}, we provide details of the EAVN polarimetric observations and data reduction. Section~\ref{sec:3} presents the results and discussion. Section~\ref{sec:4} summarizes our study and outlines future prospects for EAVN polarimetry.

\begin{deluxetable*}{lcccc}[t!]
\tablecaption{Summary of Observation Sessions. \label{tab:1}}
\tablehead{
\colhead{Obs.Code} & \colhead{Frequency} & \colhead{Date} & \colhead{UT Time} & \colhead{Stations} \\
\colhead{} & \colhead{(GHz)} & \colhead{} & \colhead{(d:hh:mm)} & \colhead{}
}
\startdata
a2323a (Session A) & 22 & 24 Nov 2023 & 0:19:05--1:04:05 & KaVA+NSRT \\
a2323b (Session B) & 43 & 25 Nov 2023 & 0:19:05--1:04:05 & KaVA \\
a2323c (Session C) & 22 & 13 Jan 2024 & 0:15:50--1:00:50 & KaVA+NSRT \\
a2323d (Session D) & 43 & 14 Jan 2024 & 0:15:45--1:00:45 & KaVA \\
\enddata
\end{deluxetable*}

\begin{figure*}[t!]
\centering
\includegraphics[width=0.24\textwidth]{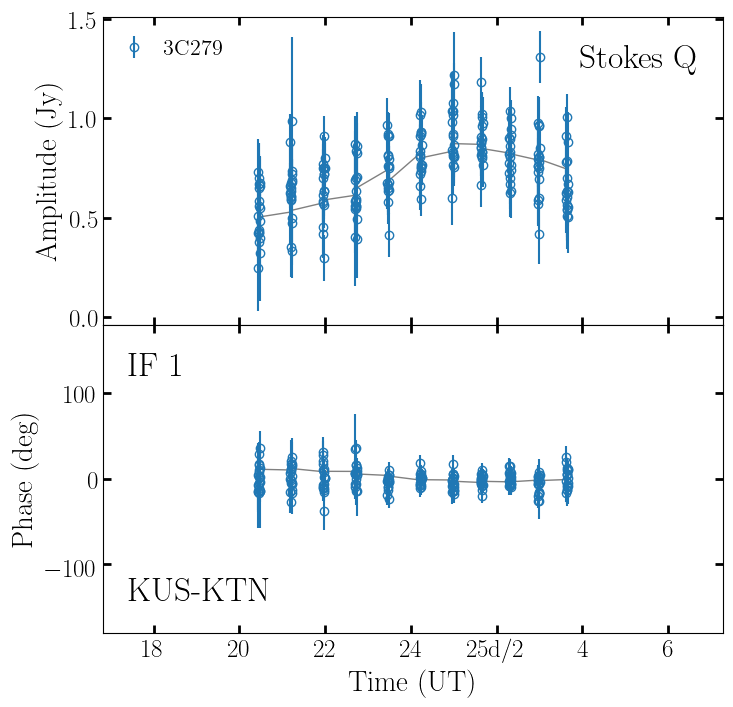}\hfill
\includegraphics[width=0.24\textwidth]{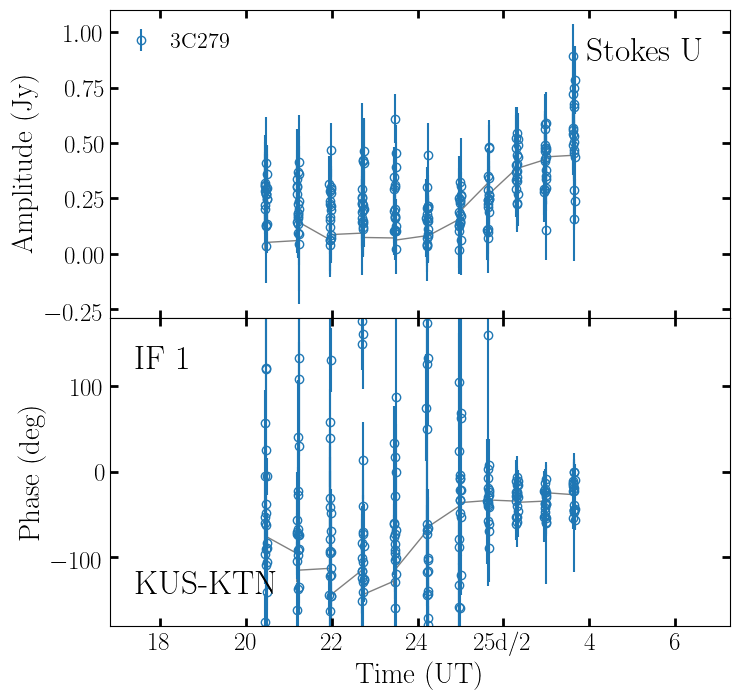}\hfill
\includegraphics[width=0.24\textwidth]{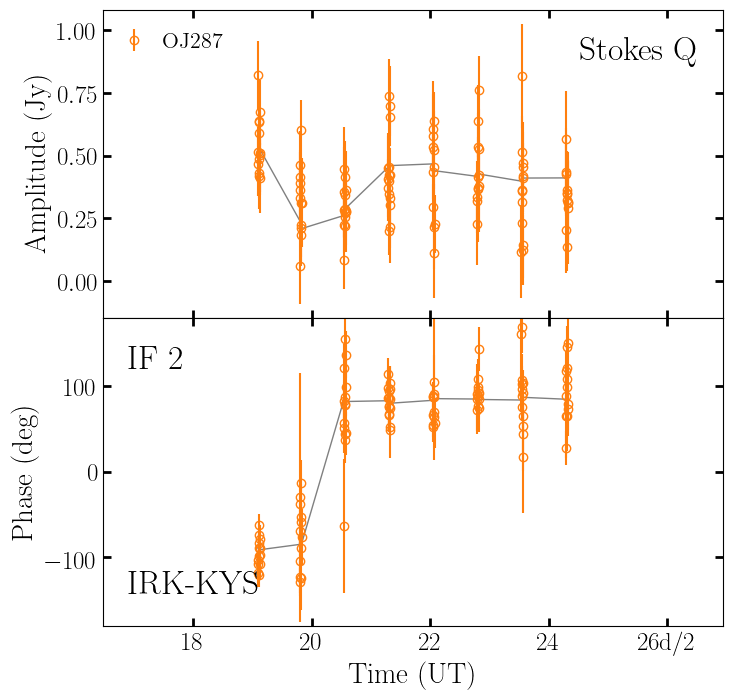}\hfill
\includegraphics[width=0.24\textwidth]{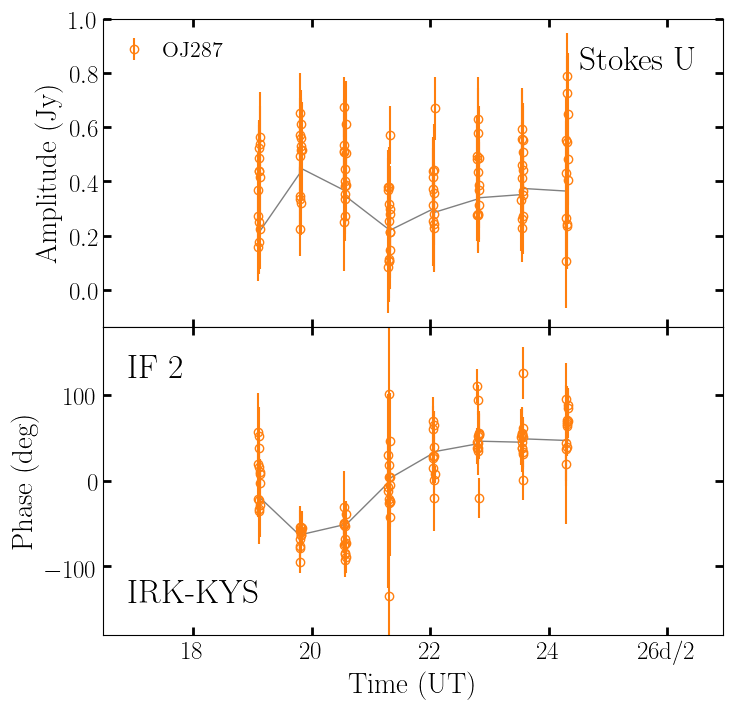}

\vspace{0.5em}

\includegraphics[width=0.24\textwidth]{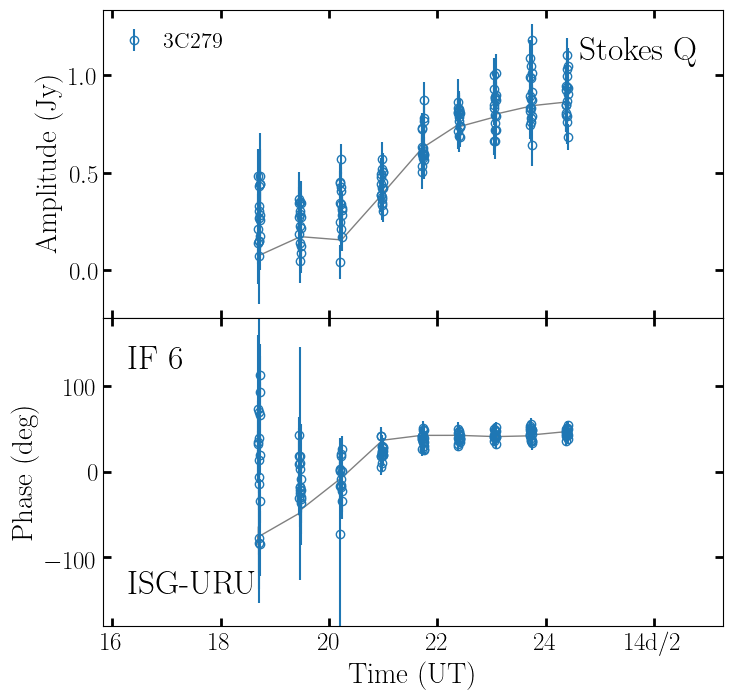}\hfill
\includegraphics[width=0.24\textwidth]{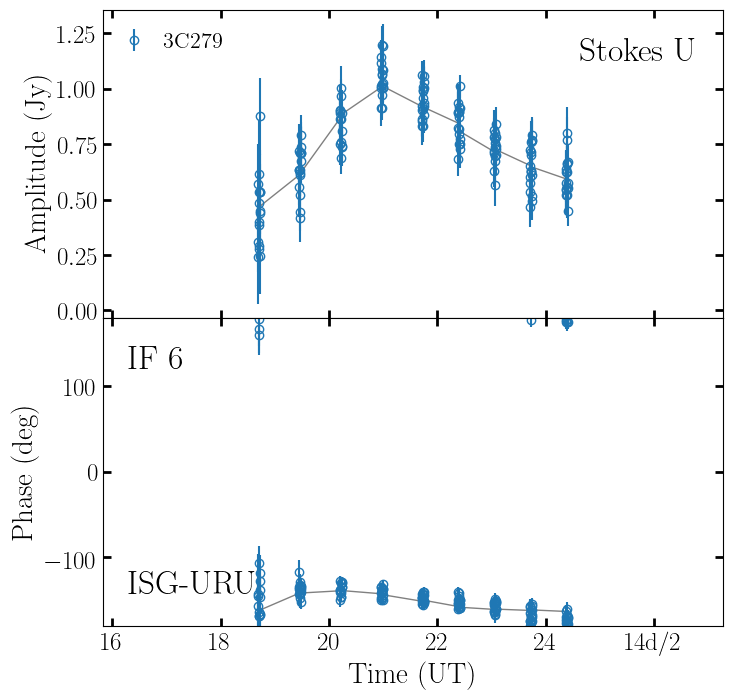}\hfill
\includegraphics[width=0.24\textwidth]{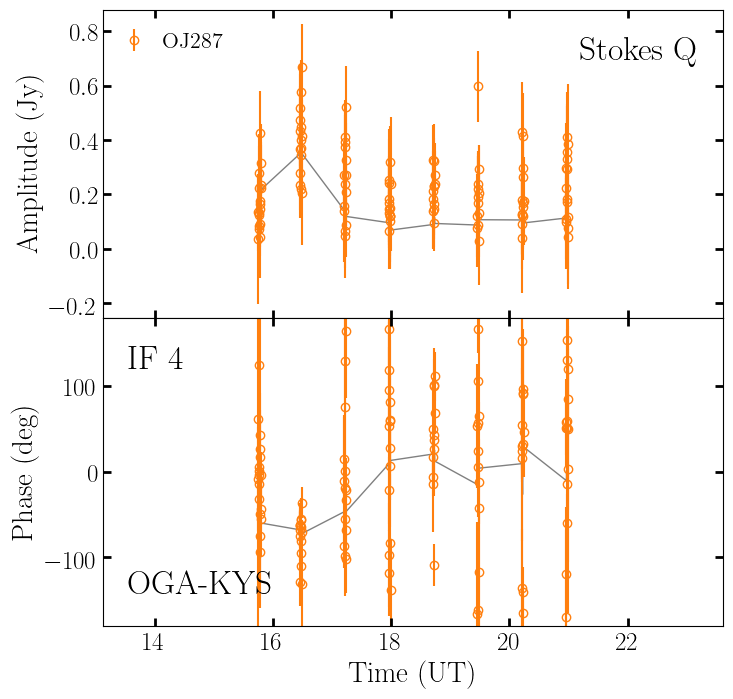}\hfill
\includegraphics[width=0.24\textwidth]{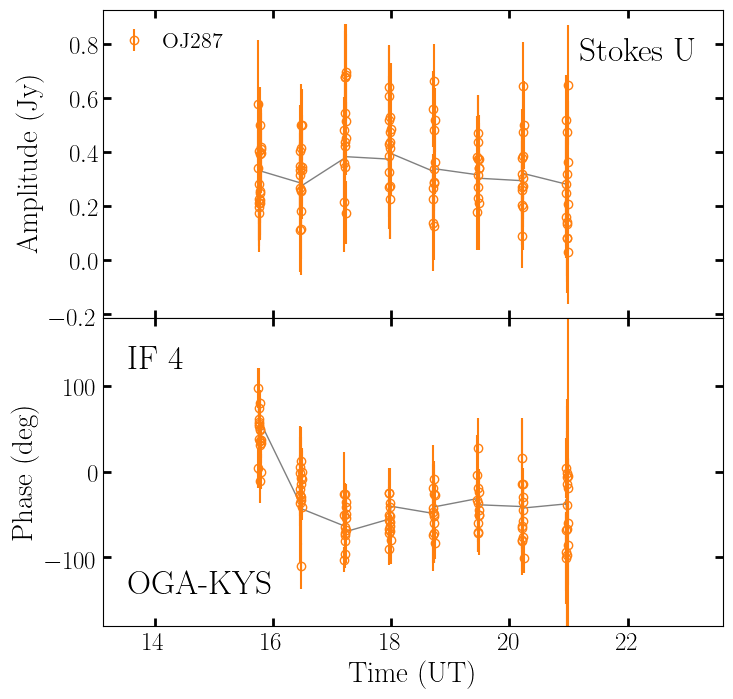}

\caption{Visibility amplitudes (top) and phases (bottom) of the Stokes $Q$ and $U$ for selected baselines and sources. The gray solid lines indicate the best-fit instrumental polarization model obtained by GPCAL.}
\label{fig:1}
\end{figure*}

\begin{figure*}[p]
\centering

\begin{minipage}{0.48\textwidth}
    \centering
    \begin{overpic}[width=\linewidth]{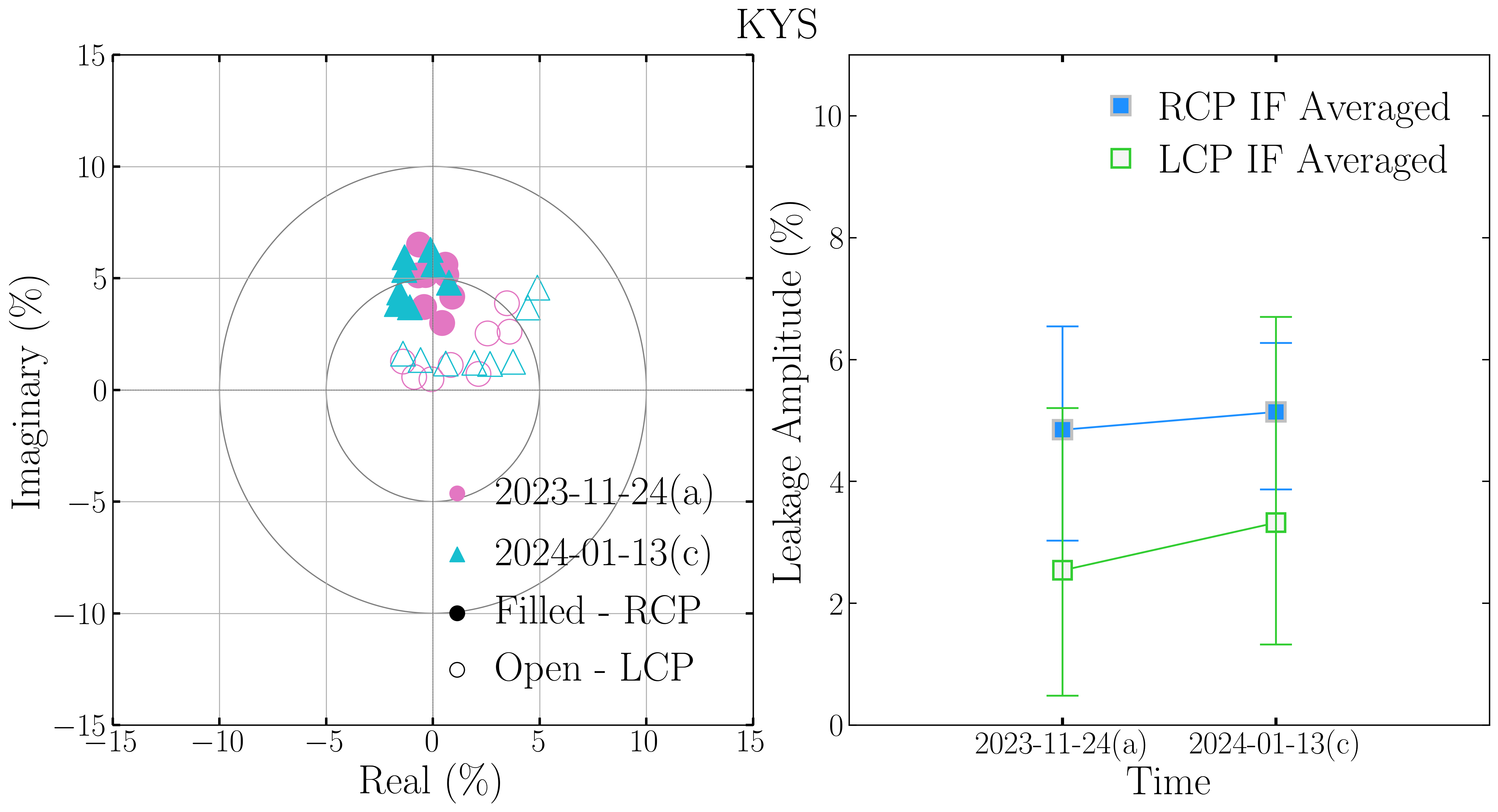}
        \put(3,55){\small (a)}
    \end{overpic}
\end{minipage}\hfill
\begin{minipage}{0.48\textwidth}
    \centering
    \begin{overpic}[width=\linewidth]{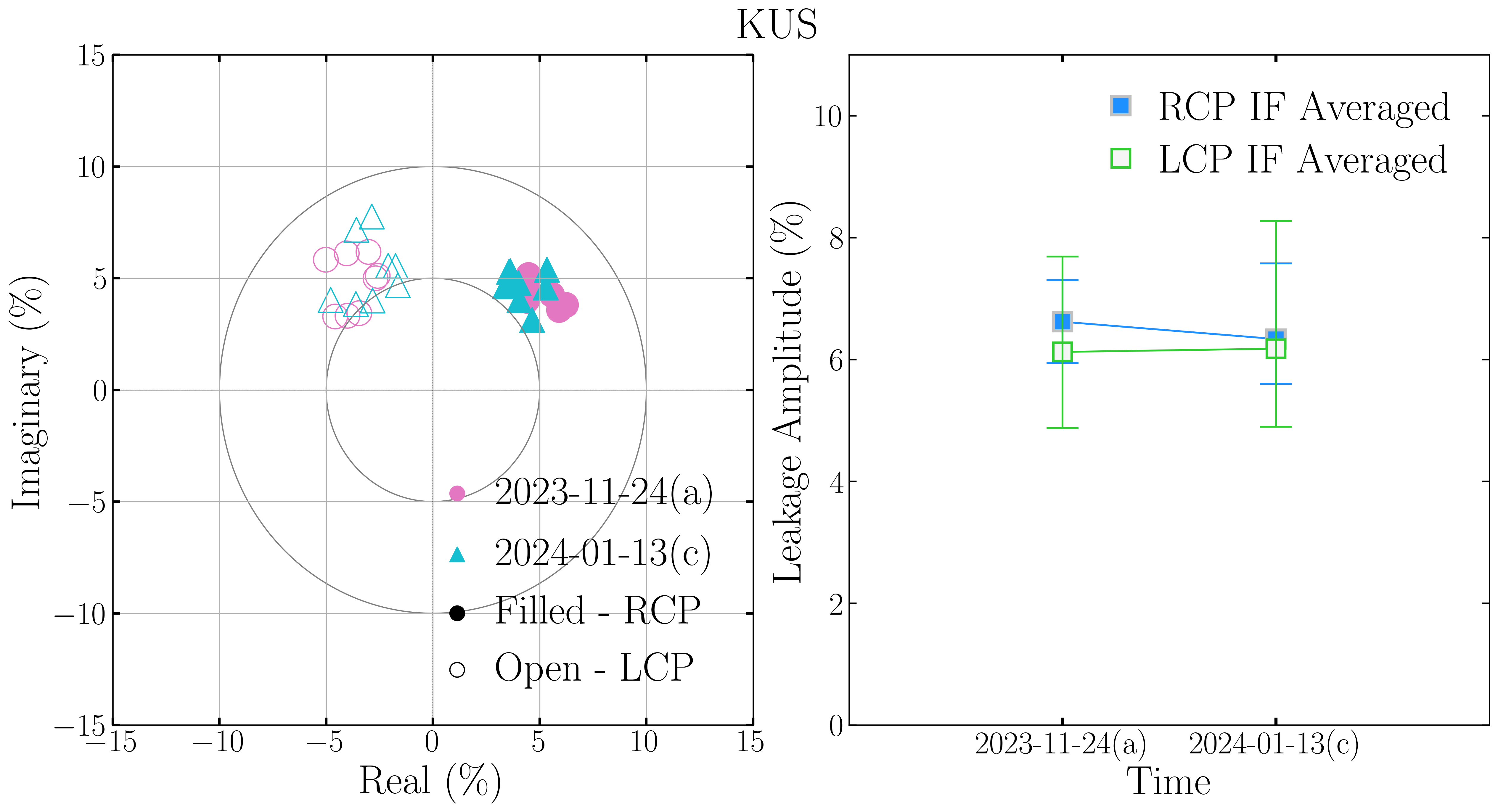}
        \put(3,55){\small (b)}
    \end{overpic}
\end{minipage}

\vspace{1em}

\begin{minipage}{0.48\textwidth}
    \centering
    \begin{overpic}[width=\linewidth]{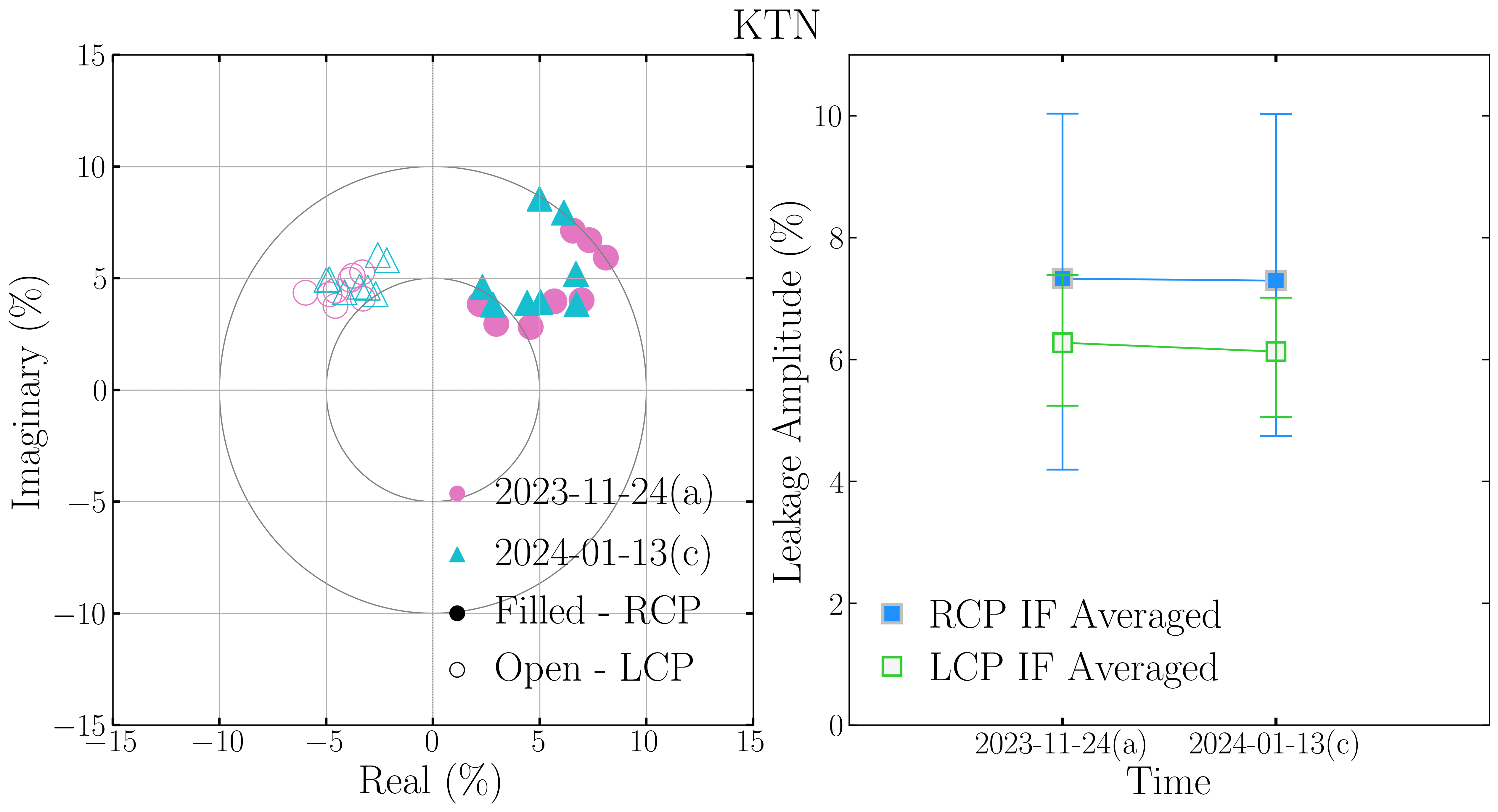}
        \put(3,55){\small (c)}
    \end{overpic}
\end{minipage}\hfill
\begin{minipage}{0.48\textwidth}
    \centering
    \begin{overpic}[width=\linewidth]{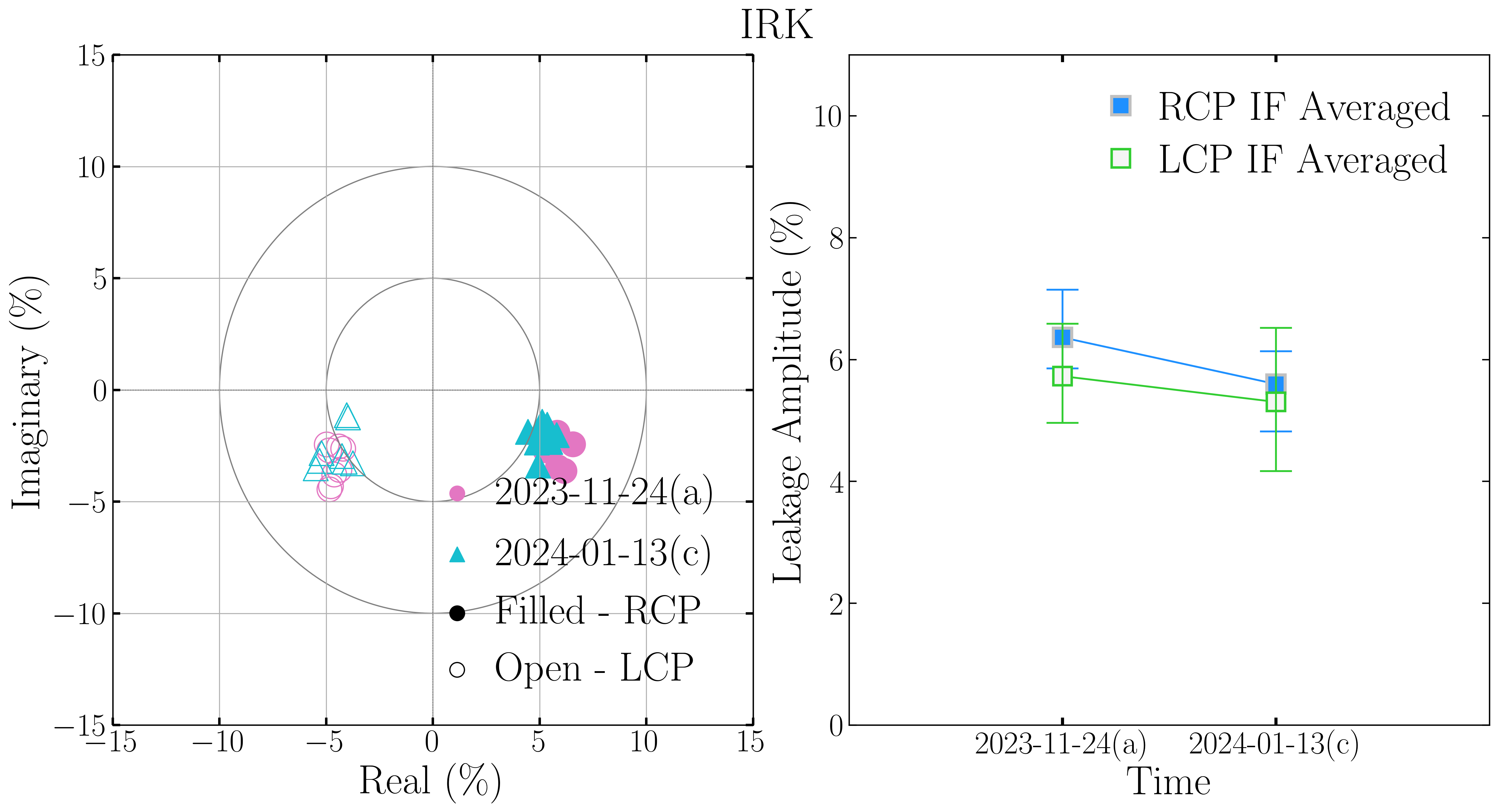}
        \put(3,55){\small (d)}
    \end{overpic}
\end{minipage}

\vspace{1em}

\begin{minipage}{0.48\textwidth}
    \centering
    \begin{overpic}[width=\linewidth]{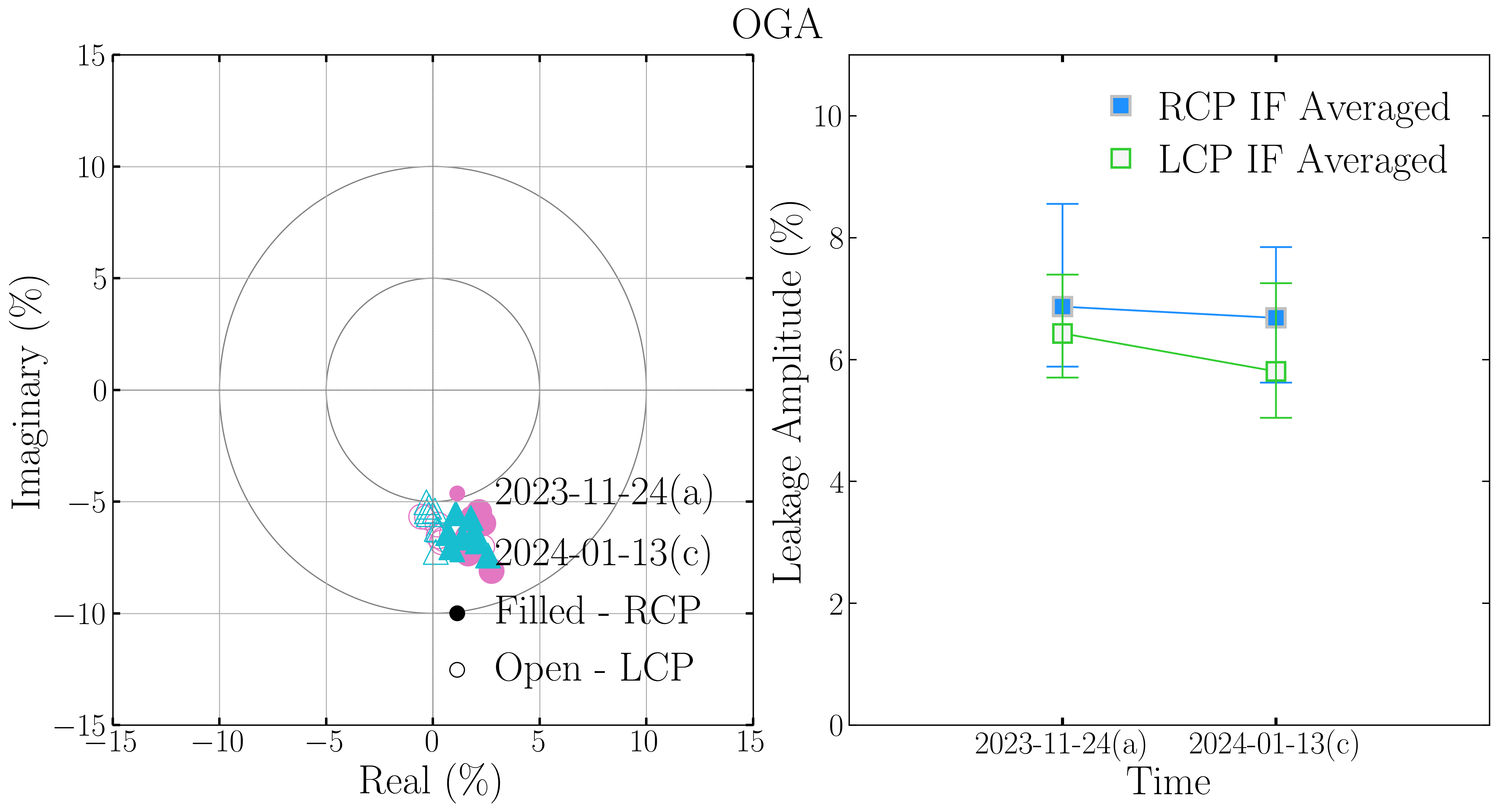}
        \put(3,55){\small (e)}
    \end{overpic}
\end{minipage}\hfill
\begin{minipage}{0.48\textwidth}
    \centering
    \begin{overpic}[width=\linewidth]{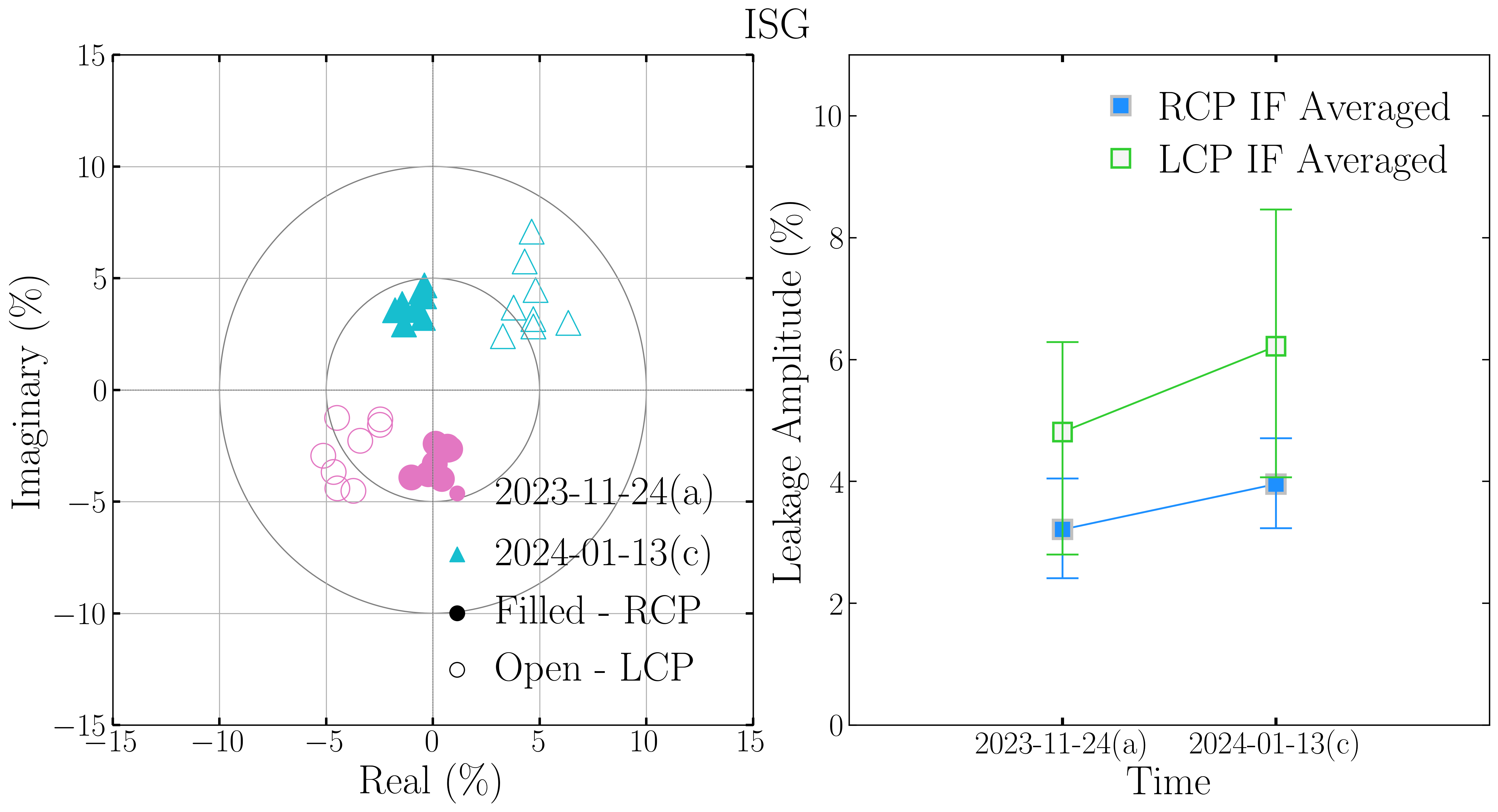}
        \put(3,55){\small (f)}
    \end{overpic}
\end{minipage}

\vspace{1em}

\begin{minipage}{0.48\textwidth}
    \centering
    \begin{overpic}[width=\linewidth]{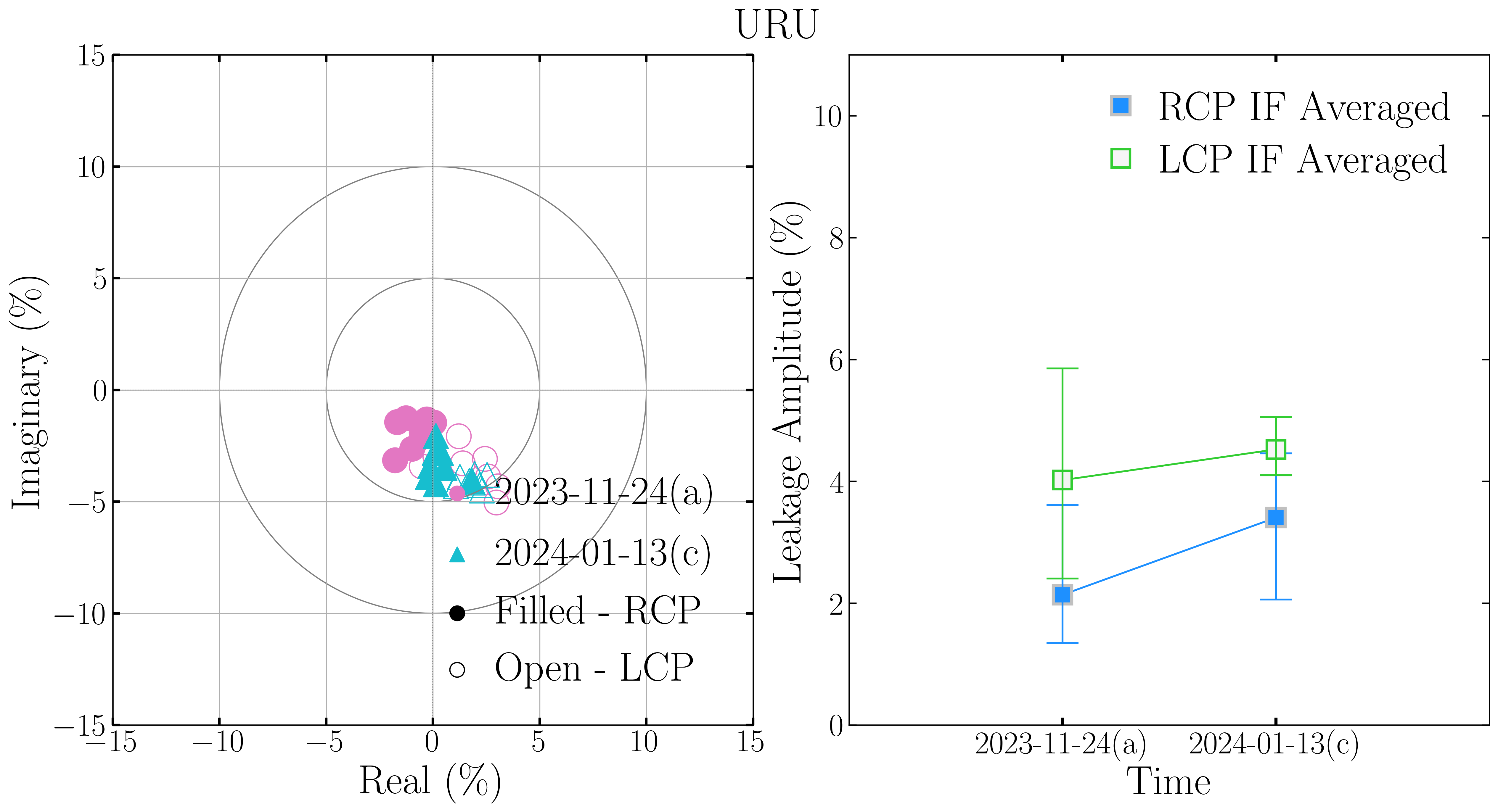}
        \put(3,55){\small (g)}
    \end{overpic}
\end{minipage}

\caption{Time variability of the D-terms at 22 GHz from session A to C for all antennas: (a) KYS, (b) KUS, (c) KTN, (d) IRK, (e) OGA, (f) ISG, (g) URU. Left panels show the change in the D-term phase on the complex plane, where the central gray circles indicate 5\% and 10\% leakage amplitudes. Right panels show the corresponding change in D-term amplitude over time.}

\label{fig:3}
\end{figure*}

\begin{figure*}[t!]
\centering

\begin{minipage}{0.48\textwidth}
    \centering
    \begin{overpic}[width=\linewidth]{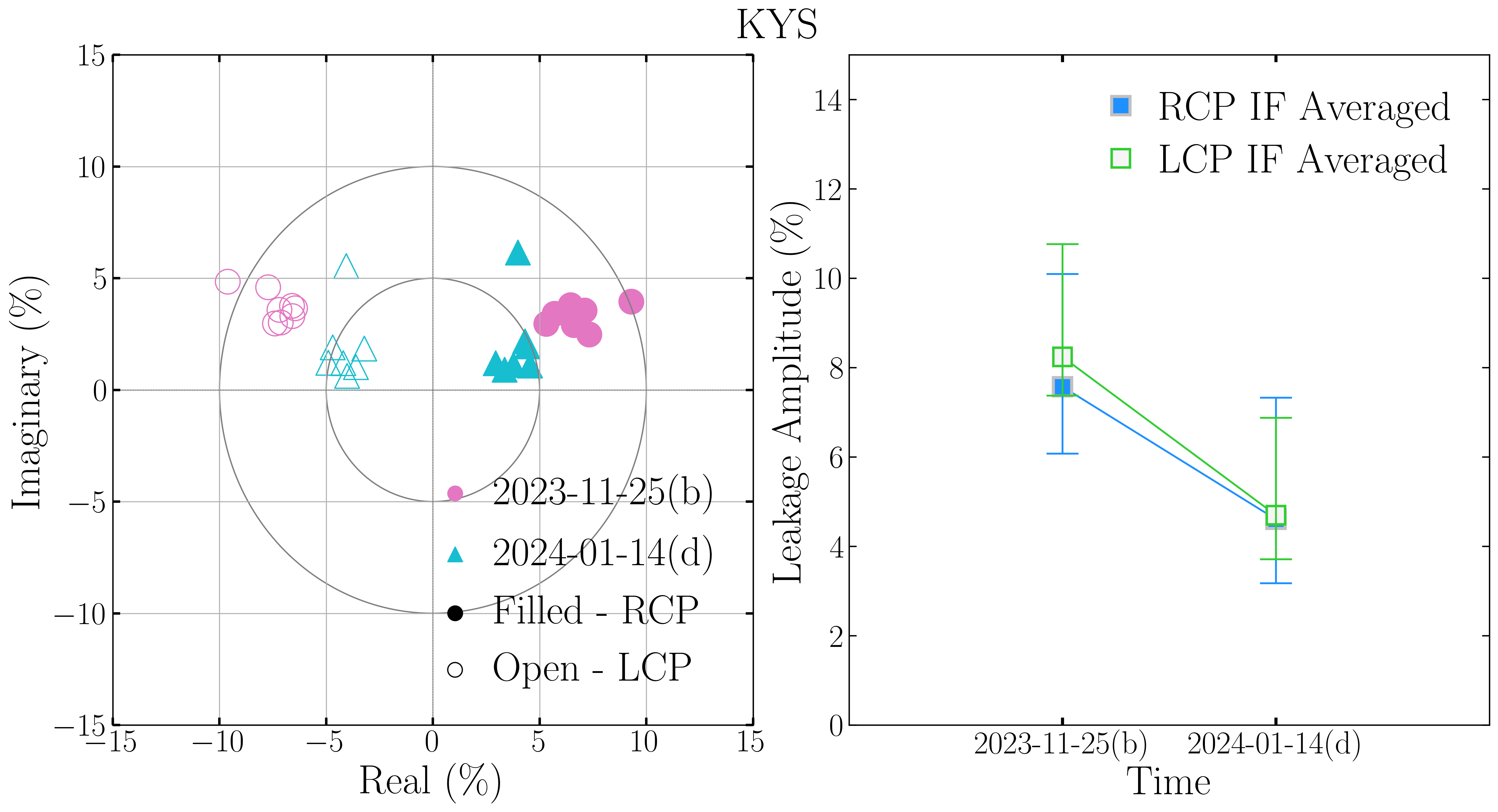}
        \put(3,55){\small (a)}
    \end{overpic}
\end{minipage}\hfill
\begin{minipage}{0.48\textwidth}
    \centering
    \begin{overpic}[width=\linewidth]{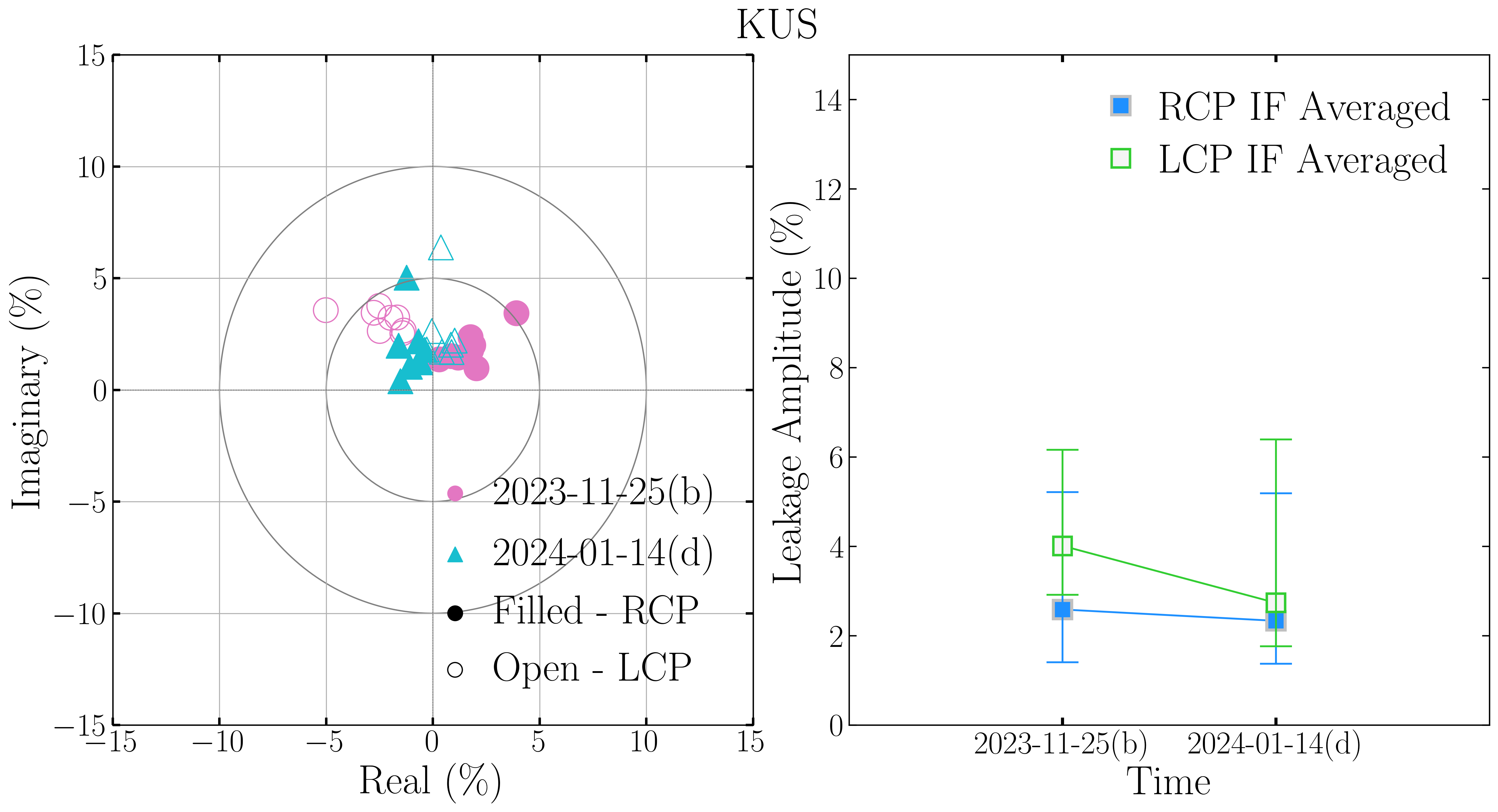}
        \put(3,55){\small (b)}
    \end{overpic}
\end{minipage}

\vspace{1em}

\begin{minipage}{0.48\textwidth}
    \centering
    \begin{overpic}[width=\linewidth]{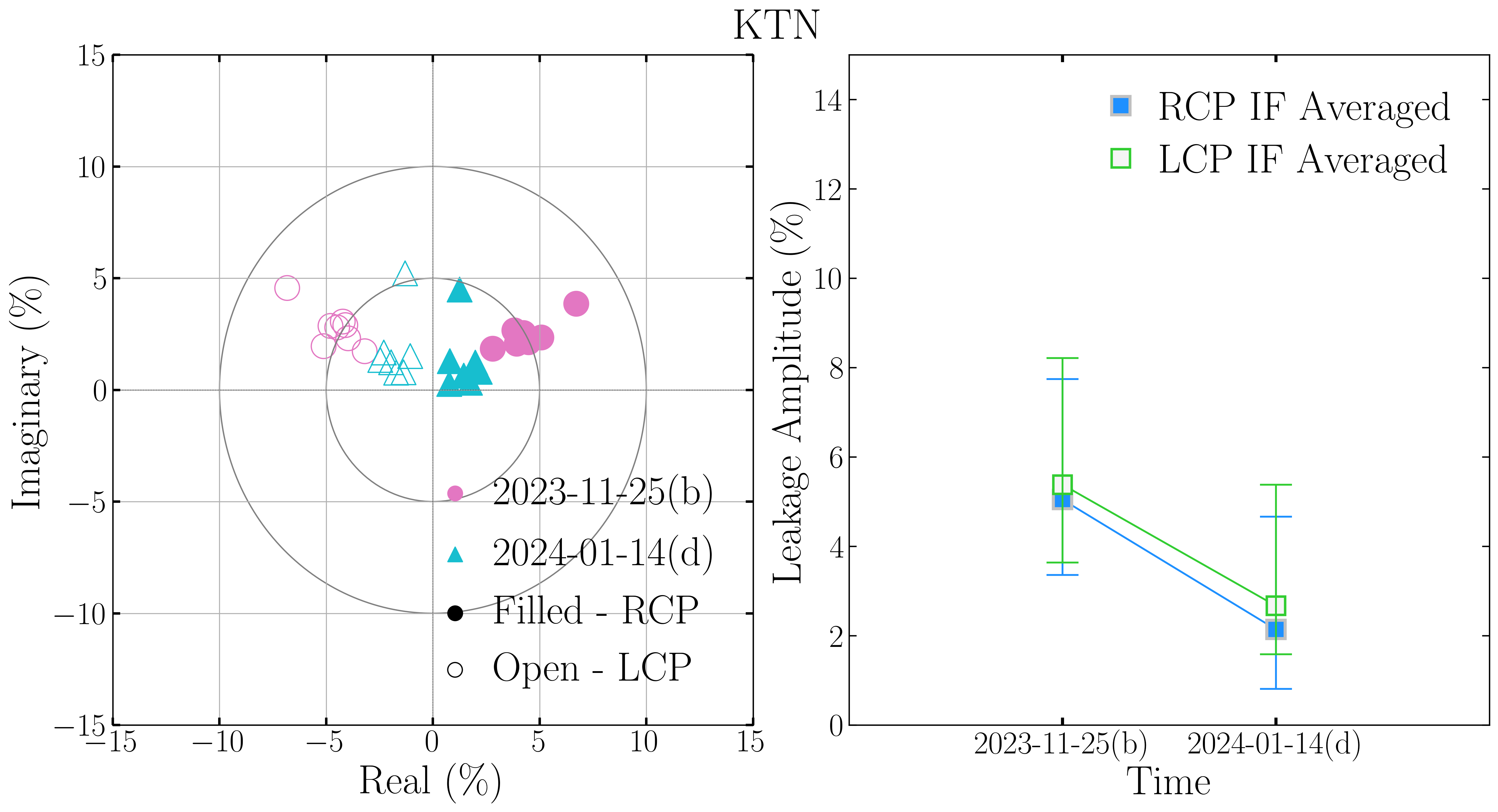}
        \put(3,55){\small (c)}
    \end{overpic}
\end{minipage}\hfill
\begin{minipage}{0.48\textwidth}
    \centering
    \begin{overpic}[width=\linewidth]{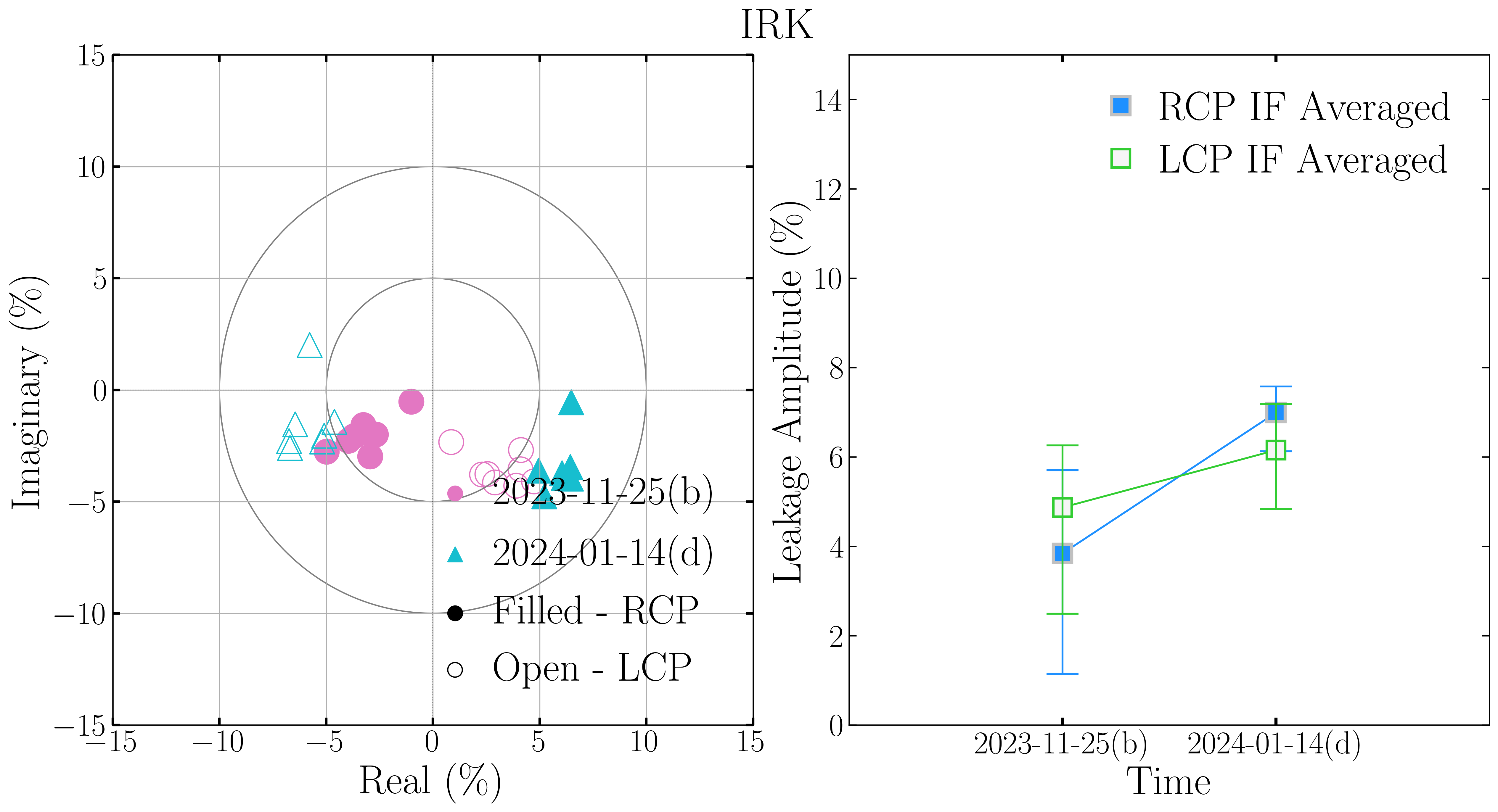}
        \put(3,55){\small (d)}
    \end{overpic}
\end{minipage}

\vspace{1em}

\begin{minipage}{0.48\textwidth}
    \centering
    \begin{overpic}[width=\linewidth]{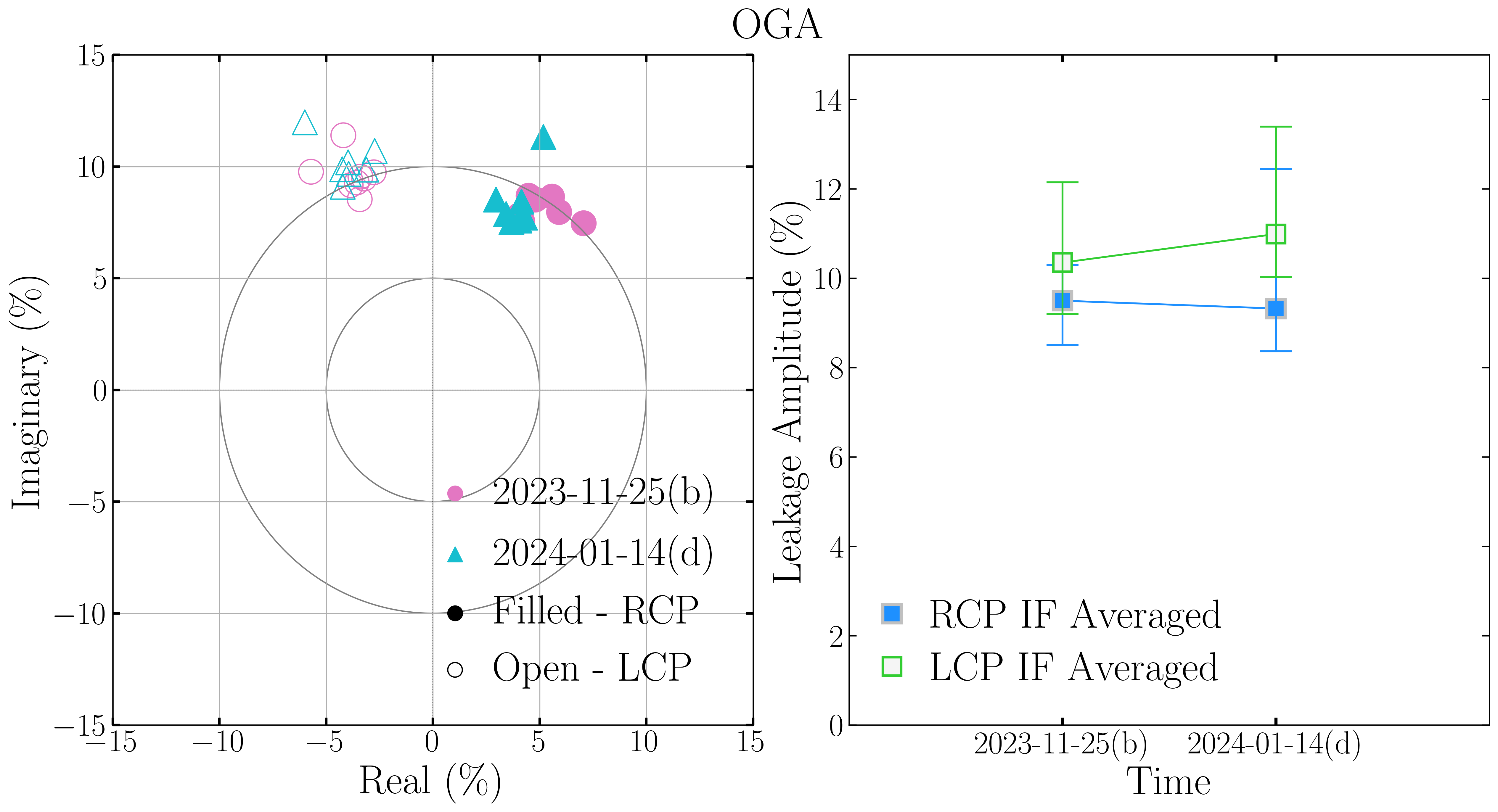}
        \put(3,55){\small (e)}
    \end{overpic}
\end{minipage}\hfill
\begin{minipage}{0.48\textwidth}
    \centering
    \begin{overpic}[width=\linewidth]{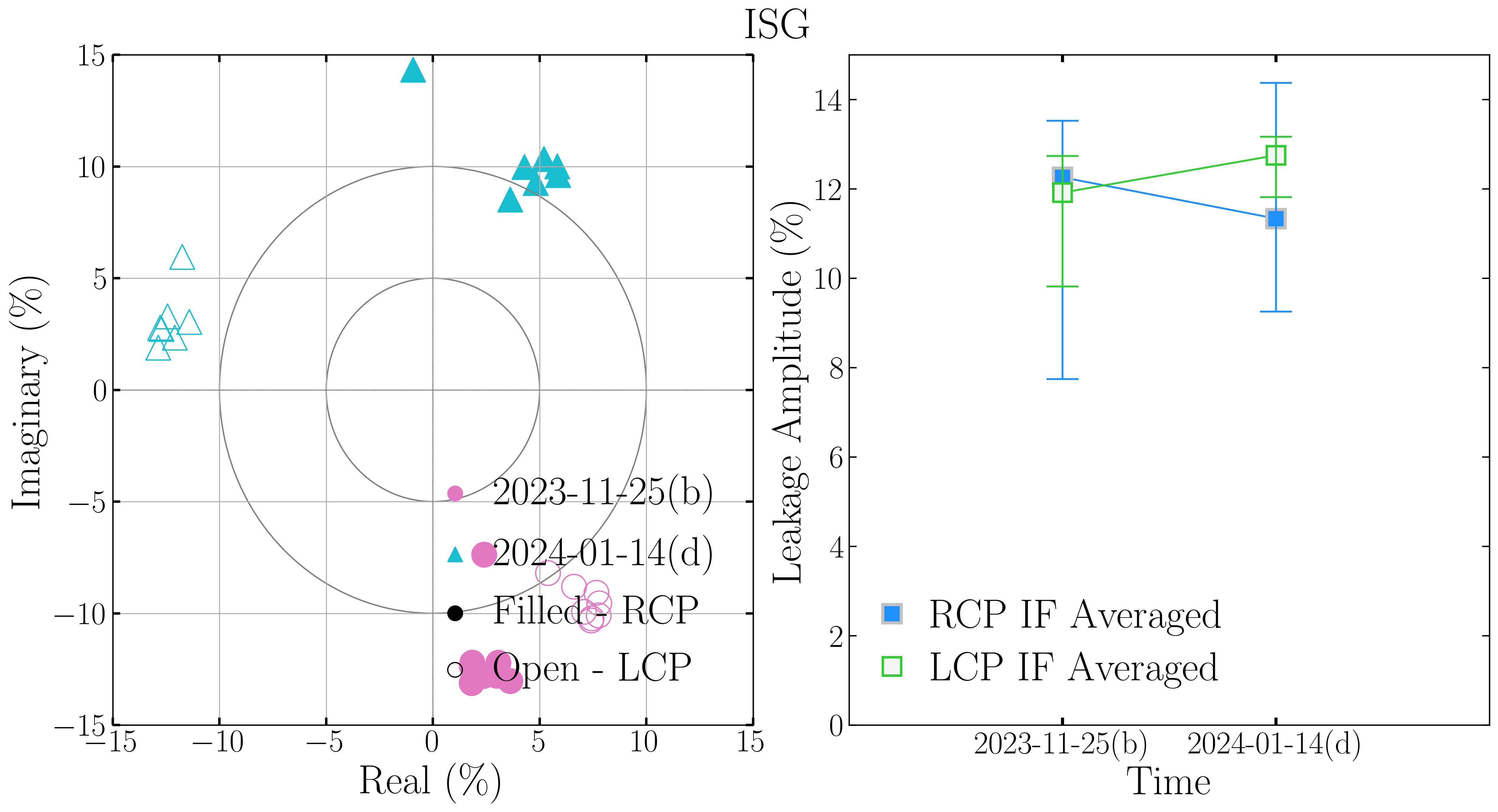}
        \put(3,55){\small (f)}
    \end{overpic}
\end{minipage}

\caption{Same as Figure~\ref{fig:3} but at 43 GHz from session B to D. The results for all antennas are shown: (a) KYS, (b) KUS, (c) KTN, (d) IRK, (e) OGA, (f) ISG.}
\label{fig:4}
\end{figure*}

\section{Observations and Data Reduction}\label{sec:2}

\subsection{EAVN Observations}\label{sec:2.1}

We conducted observations in two epochs at each of the K and Q bands, as summarized in \autoref{tab:1}. The primary target was M87, observed along with three bright calibrators: 3C 273, 3C 279, and OJ 287. The K-band observations were carried out on November 24, 2023, and January 13, 2024 (observation codes: a2323a, a2323c; hereafter, sessions A and C), while the Q-band observations were conducted on November 25, 2023, and January 14, 2024 (observation codes: a2323b, a2323d; hereafter, sessions B and D). For all sessions, the Mizusawa (MIZ) station did not participate due to technical issues with the motor system. Consequently, the K-band observations were performed with seven antennas—comprising the KVN and VERA Array (KaVA; \citealt{Niinuma2014}), excluding MIZ, together with the Nanshan 26m Radio Telescope (NSRT)—while the Q-band observations involved six antennas from KaVA (also excluding MIZ). Tianma was not included in these observations as M87 provides sufficient brightness at these frequencies, and the NSRT provided the necessary resolution.

Each of the two K-band sessions lasted 9 hours and consisted of 13 cycles. Scan durations were 2.5 minutes for 3C 273 and OJ 287, 3 minutes for 3C 279, and 30 minutes for M87, with the final cycle comprising a single 20 minute M87 scan. Similarly, the 9-hour Q-band sessions consisted of 13 cycles with 3-minute scans for all three calibrators and 30-minute scans for M87. Across all sessions, 3C 279 was excluded from the first two cycles, while OJ 287 was excluded from the ninth cycle onward. The data were recorded at a rate of 1 Gbps. The total bandwidth of 128 MHz for each polarization was divided into eight 16 MHz baseband channels (referred to as intermediate frequencies; IFs). All correlation processing was carried out at the Korea-Japan Correlation Center (KJCC) \citep{lee2015}.

We performed standard data reduction using the NRAO Astronomical Image Processing System (AIPS; \citealt{greisen2003}), following previous VLBI studies \citep[e.g.,][]{Park2021b, Park2024}. The KVN Yonsei station served as the reference antenna for phase calibration for sessions A, C, and D, while the KVN Ulsan station was used for session B. Iterative imaging and self-calibration were conducted using the CLEAN algorithm implemented in Difmap \citep{shepherd1994}. We note that the antenna system temperature for the NSRT could not be measured due to a failure in the data acquisition system. To address this, we scaled the visibility amplitudes of the NSRT baselines via self-calibration, utilizing a CLEAN model constructed from the remaining baselines, and performed iterative CLEAN and self-calibration until the model accurately fitted the NSRT baselines.

\subsection{Polarimetric Leakage and EVPA Calibration}\label{sec:2.2}

We performed instrumental polarization calibration using GPCAL. Initial D-term estimation was conducted using M87, assuming it to be unpolarized\footnote{For details on the frequency- and time-dependent D-term calibration, see \citealt{Park2023a, Park2023b}.} \citep[e.g.,][]{Park2021c}. This was followed by 10 iterations of instrumental polarization self-calibration using the bright, highly polarized calibrators 3C 273, 3C 279, and OJ 287. Instrumental polarization self-calibration, the core methodology underlying GPCAL, separates the intrinsic source polarization from antenna leakages through an iterative procedure. Specifically, it utilizes tentative Stokes $Q$ and $U$ CLEAN models to predict the source's contribution to the observed cross-hand visibilities. By minimizing the difference between these model predictions and the actual cross-hand data, the pipeline isolates and updates the antenna-dependent D-terms. This cycle of polarimetric imaging and D-term solving is repeated until the solutions converge.
In Figure~\ref{fig:1}, we present the Stokes $Q$ and $U$ visibilities of 3C 279 and OJ 287 for selected baselines, along with the corresponding best-fit instrumental polarization model obtained by GPCAL. The fitting results are satisfactory for both short baselines, such as KUS-KTN, and long baselines, such as ISG-URU; the reduced chi-square values of the fitting are generally close to unity, with a maximum value of 1.34. This result indicates that the polarimetric data obtained by the EAVN follow the standard Radio Interferometric Measurement Equation (RIME; \citealt{smirnov2011}) on which GPCAL is based. 

We note that no significant linear polarization was detected in M87. This suggests that our EAVN data either lack the requisite sensitivity, or are limited by remaining systematic uncertainties such as residual polarimetric leakage, to detect the faint linear polarization of the jet \citep[e.g.,][]{ZT2002, Park2019, Park2021c, Park2026, Nikonov2023}. To constrain this non-detection, we estimate conservative upper limits on the polarized intensity of 4.51 and 3.29 mJy beam$^{-1}$ for the K-band sessions (A and C), and 6.50 and 4.04 mJy beam$^{-1}$ for the Q-band sessions (B and D). Defined as seven times the rms noise in the polarized intensity ($7\sigma_P$)\footnote{In well-calibrated, moderately to highly polarized sources (e.g., our calibrators), true polarization signals appear as distinct peaks in the Stokes $Q$ and $U$ dirty images that spatially coincide with the Stokes $I$ emission. In contrast, the M87 dirty images exhibit scattered spurious patterns, a characteristic signature of residual polarimetric leakage. Because these systematic errors typically follow non-Gaussian distributions with heavy tails, we adopted a highly conservative detection threshold of $7\sigma_P$ rather than a standard statistical limit (e.g., $3\sigma$ or $5\sigma$) to robustly bound the upper limit.}, these correspond to fractional polarization upper limits of 0.51\% and 0.35\% at 22 GHz, and 0.76\% and 0.49\% at 43 GHz at the core of the M87 jet. Given that \citealt{park2021} reported an intrinsic fractional polarization of 0.24\%--0.58\% for the core at 43 GHz, our non-detection is physically consistent with the source properties, as the expected polarization falls near or below our estimated upper limits.

We performed electric vector position angle (EVPA) calibration to correct for the EVPA offset arising from the phase difference between the two polarizations at the reference antenna. This was achieved by utilizing KVN single-dish data obtained as part of the Plasma-physics of Active Galactic Nuclei (PAGaN) blazar monitoring project \citep{Park2018}. Specifically, we used (i) the EVPAs of 3C 279 and OJ 287 measured on November 20, 2023, for sessions A and B, and (ii) those measured on January 17, 2024, for sessions C and D. In both cases, the single-dish data were obtained within 5 days of the corresponding VLBI sessions. The EVPAs were measured simultaneously at 22 and 43~GHz using the KVN Tamna station in dual-polarization mode. The single-dish data were reduced following the procedures described in \citealt{Kam2023}. For EVPA calibration, we compared the integrated EVPA values from the EAVN images of OJ 287 and 3C 279 with those from the KVN single-dish observations for each session.

Since the EVPA calibration method assumes that the calibrators are sufficiently compact such that both the interferometer and the single-dish observation detect the same amount of polarized flux, we compared the total CLEANed polarized flux density recovered from the EAVN images with the single-dish measurements across all sessions.
In most sessions, the EAVN polarized flux recovery exceeded 80\% of the single-dish values, with both measurements consistent within an upper limit of 1.4$\sigma$. We note that for OJ 287 in session D, the EAVN polarized flux density was 3.32 times larger than the single-dish value, showing a rather significant discrepancy of about 2.7$\sigma$. We attribute this discrepancy to a relatively large uncertainty in the polarized flux measurement for this particular source. Nevertheless, the EVPA correction factors derived independently from 3C 279 and OJ 287 are consistent within 2°-3°, ensuring that our EVPA calibration remains robust.

In addition to the EAVN stations presented in this paper, we utilized one additional EAVN observation conducted on November 24, 2024, at Q-band, hereafter referred to as session E (observation code: a2414b). Further details of these data will be presented elsewhere (Kam et al. in preparation). The purpose of including this additional observation is to verify the long-term stability of the D-terms of the EAVN stations. Althogh such stability has been established for the VLBA, where polarimetric leakage arises from instrumental properties that typically do not vary significantly on monthly timescales \citep{Gomez2002}, this has not yet been confirmed for the EAVN.

\subsection{VLBA Data}\label{sec:2.3}

For comparison with the resulting EAVN polarization images, we selected archival datasets obtained with the VLBA: specifically, 15 GHz data from the MOJAVE \citep{lister2018} program and 43 GHz data from the BEAM-ME \citep{Jorstad2017, Weaver2022} program, both of which are well known for their robust linear polarization imaging capabilities. For each EAVN observation, we chose the datasets closest in time. For the K-band observations, both sessions were compared with MOJAVE data from December 15, 2023 (VLBA code: BL286BB); however, OJ 287 from session A was compared with MOJAVE data from November 14, 2023 (VLBA code: BL286BA), as it was the closest in date. For the Q-band observations, session B was compared with BEAM-ME data from November 25, 2023, and session D with BEAM-ME data from December 15, 2023.

Since the MOJAVE visibility data were calibrated and reduced but not imaged, we performed imaging using the CLEAN algorithm on these data. For the BEAM-ME data, we adopted the CLEAN models for Stokes $I$ available in the database and performed CLEAN on the Stokes $Q$ and $U$ data to generate the reference polarization images.

The VLBA data have a smaller synthesized beam compared to the EAVN data. To compare the images, we convolved the CLEAN models of the VLBA images with the synthesized beam of the corresponding EAVN data.

\section{Results}\label{sec:3}

\subsection{Instrumental Polarization}\label{sec:3.1}
In \autoref{tab:2}, we present the averaged D-term amplitudes and median phases over eight IFs for all antennas\footnote{We note that the D-terms exhibit a moderate frequency dependence, especially in the K-band (see \autoref{appendix:amplitude}). We present the IF-averaged D-terms as representative values in this table.}.
Most antennas exhibit relatively stable D-term amplitudes in the range of 5--10\%, consistent with previous results from KVN-only and VERA-only observations \citep{kim2015, Park2018, hagiwara2022, Takamura2023}.

We assessed the stability of the EAVN antenna polarimetric leakages over monthly timescales using our multi-epoch observations (Figure~\ref{fig:3} and~\ref{fig:4}). Since the D-terms were determined prior to EVPA calibration, their phases include an arbitrary offset relative to the intrinsic leakages, corresponding to the R-L phase difference at the reference antenna \citep{Gomez2002}. Therefore, we aligned the D-term phases to a common reference to allow for a direct comparison across epochs. We found that both the D-term amplitudes and phases are generally stable over time. However, we observed that the D-term phases vary significantly between epochs for some VERA stations (ISG at 22~GHz and IRK/ISG at 43~GHz). We attribute this phase shift to the field rotator setup of those VERA stations , which will be addressed in more detail in Section~\ref{sec:4.1}.

\begin{deluxetable*}{ccccccccc}
\tablecaption{Averaged amplitude (\%) and median phase ($^\circ$) of D-term. \label{tab:2}}
\tablehead{
\colhead{} & \colhead{Session} & \colhead{IRK} & \colhead{OGA} & \colhead{ISG} & \colhead{KYS} & \colhead{KUS} & \colhead{KTN} & \colhead{URU}
}
\startdata
\multirow{4}{*}{RCP} & A & 6.36~(-25.7) & 6.87~(-71.8) & 3.21~(-85.6) & 4.85~(89.0) & 6.62~(43.7) & 7.33~(39.4) & 2.14~(-108) \\
 & C & 5.60~(-21.2) & 6.68~(-75.5) & 3.95~(101) & 5.14~(103) & 6.33~(47.7) & 7.29~(46.9) & 3.40~(-87.2) \\
 & B & 3.84~(-151) & 9.50~(60.9) & 12.26~(-78.2) & 7.57~(25.6) & 2.59~(48.9) & 5.04~(30.4) & \nodata \\
 & D & 6.99~(-31.6) & 9.32~(63.9) & 11.33~(62.9) & 4.60~(19.4) & 2.33~(111) & 2.14~(22.2) & \nodata \\
\hline
\multirow{4}{*}{LCP} & A & 5.72~(-144) & 6.43~(-87.1) & 4.81~(-147) & 2.54~(50.7) & 6.12~(127) & 6.27~(132) & 4.02~(-59.6) \\
 & C & 5.30~(-148) & 5.80~(-90.5) & 6.22~(39.6) & 3.32~(41.4) & 6.18~(114) & 6.13~(125) & 4.52~(-64.4) \\
 & B & 4.87~(-51.3) & 10.35~(111) & 11.92~(-53.5) & 8.24~(153) & 4.01~(123) & 5.38~(149) & \nodata \\
 & D & 6.15~(-158) & 10.99~(112) & 12.75~(166) & 4.70~(161) & 2.73~(71.8) & 2.67~(146) & \nodata \\
\enddata
\end{deluxetable*}

\begin{figure}[t]
\centering
\includegraphics[width=\linewidth]{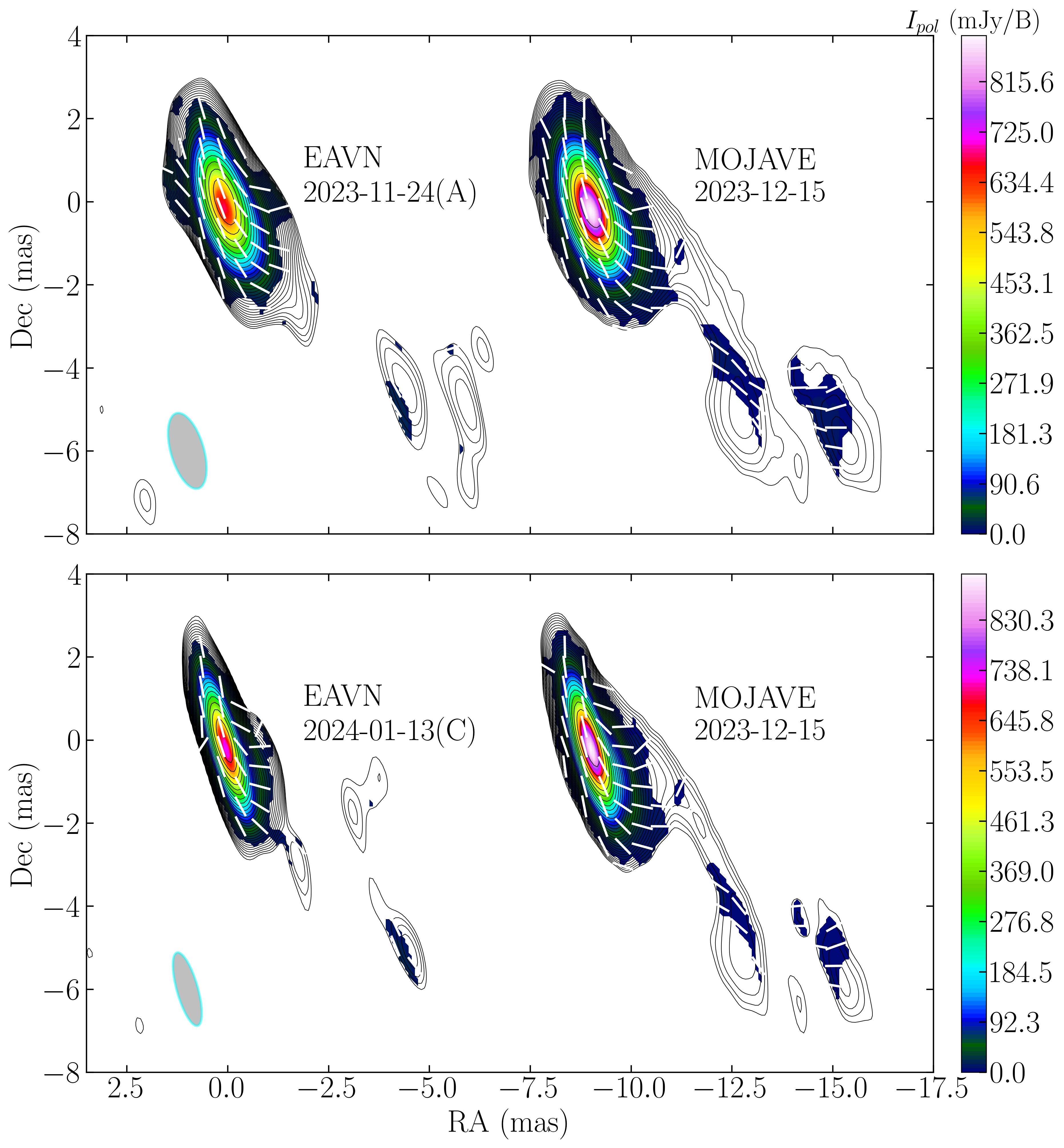}
\caption{Linear polarization intensities and polarization vector orientations (white bars) of 3C 279 at 22~GHz. The upper and lower panels show results from sessions A and C, respectively. The images on the right for comparison show the results at 15~GHz from the VLBA. The gray ellipses outlined in cyan in the lower left corners represent the restoring beam sizes for the two images: 1.89~mas $\times$ 0.81~mas with a position angle of 17$^\circ$ for the upper panel, and 1.83~mas $\times$ 0.52~mas with 16$^\circ$ for the lower panel.}
\label{fig:5}
\end{figure}

\begin{figure}[t]
\centering
\includegraphics[width=\linewidth]{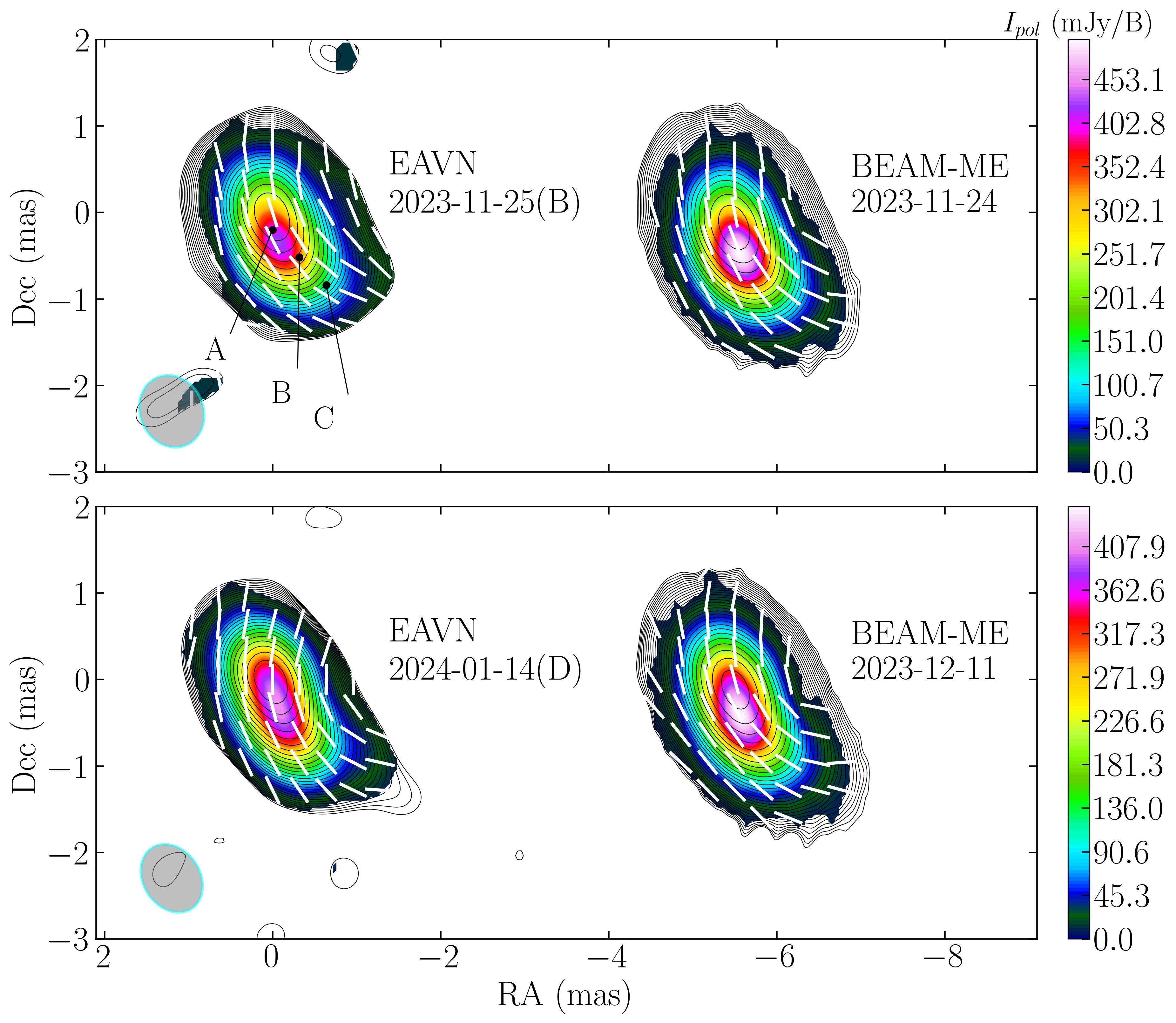}
\caption{Linear polarization intensities and polarization vector orientations (white bars) of 3C 279 at 43~GHz. The upper and lower panels show results from sessions B and D, respectively. The images on the right for comparison show the results at 43~GHz from the VLBA. The restoring beam sizes are 0.85~mas $\times$ 0.75~mas with a position angle of 29$^\circ$ for the upper panel, and 0.85~mas $\times$ 0.67~mas with 37$^\circ$ for the lower panel. Points A--C in the upper panel show the locations where polarization properties are compared in Figure~\ref{fig:12}.}
\label{fig:6}
\end{figure}

\subsection{Linear Polarization Images}\label{sec:3.2}

\subsubsection{3C 279}\label{sec:3.2.1}
3C 279 is a flat-spectrum radio quasar (FSRQ) at a redshift of $z=0.536$ \citep{strauss1992}, well known for its extreme variability and strong polarization across the entire electromagnetic spectrum from radio to $\gamma$-rays \citep{chatterjee2008,hayashida2012,kiehlmann2016,larionov2020}.

The EAVN linear polarization images of 3C 279 at 22~GHz are shown in \autoref{fig:5}, revealing a core-dominated structure with a jet extending toward the southwest. Comparison with the contemporaneous 15~GHz VLBA data from the MOJAVE database shows excellent agreement in total intensity, linear polarization intensity, and EVPA structure.


Similarly, the 43~GHz linear polarization images are presented in \autoref{fig:6}. These images also exhibit EVPAs aligned with the jet axis, and the polarization intensity and EVPA distributions are in good agreement with the corresponding VLBA polarization images. In \autoref{appendix:residual}, fractional residual maps of Stokes $Q$ and $U$ are provided for three sources, including 3C 279, to facilitate a quantitative comparison between the VLBA and EAVN polarization images. These maps are based on only Session B observations, which were conducted at comparable frequencies and just one day apart. As shown in these maps, the fractional residuals tend to be close to zero where the polarized intensity peaks (i.e., where the fractional polarization uncertainty is smallest) and show increasing residuals with alternating signs around the peak, implying that the two polarization images are in good agreement with each other without exhibiting systematic trends in the residuals. These results demonstrate that the EAVN has a good linear polarization imaging capability at both K and Q bands.


\subsubsection{3C 273}\label{sec:3.2.2}
\begin{figure*}[!t]
\centering
\includegraphics[width=0.9\linewidth]{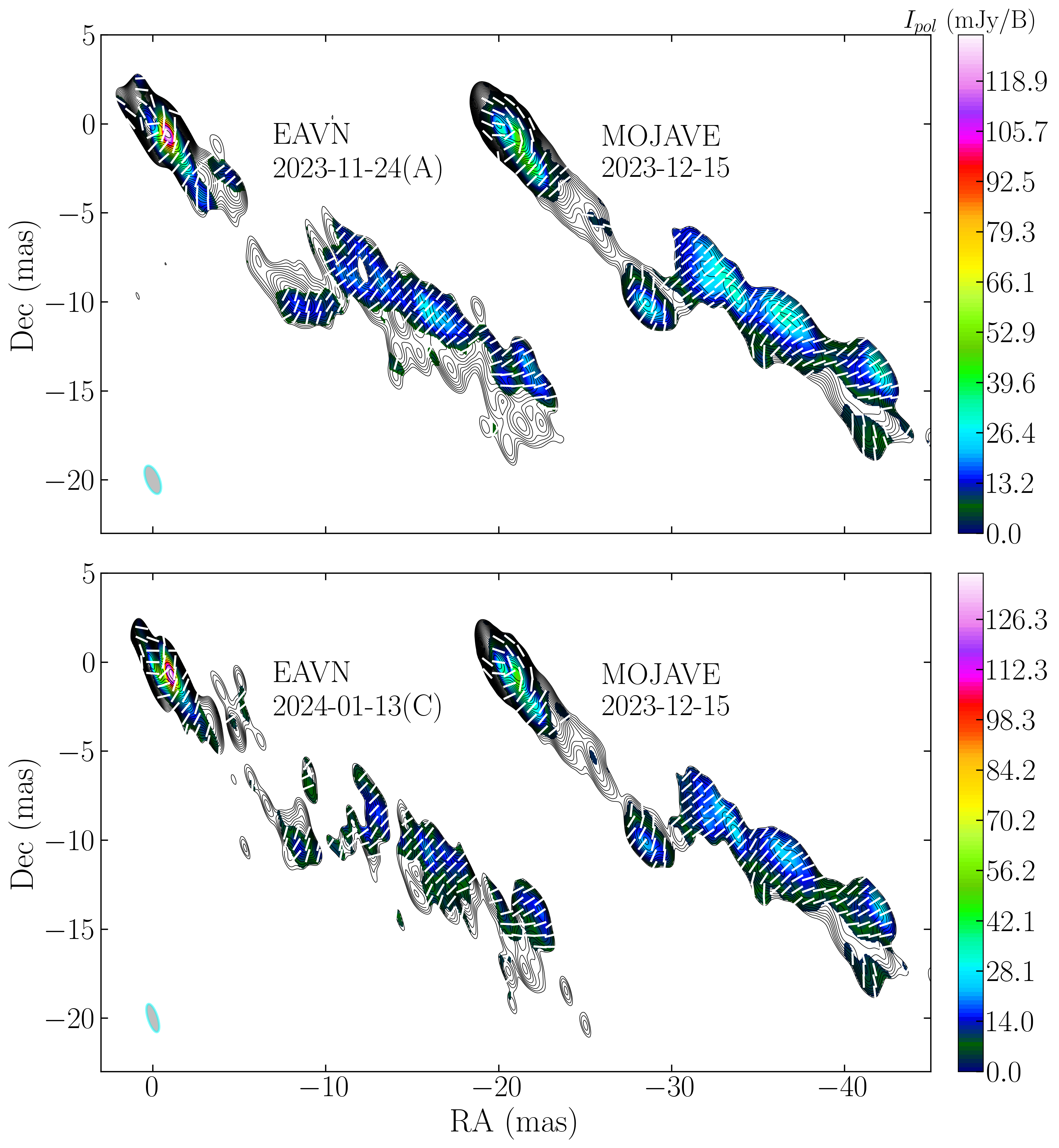}
\caption{Same as Figure~\ref{fig:5} for 3C 273 at 22~GHz. The restoring beam sizes are 1.69~mas $\times$ 0.77~mas with a position angle of 23$^\circ$ for the upper panel, and 1.65~mas $\times$ 0.56~mas with 18$^\circ$ for the lower panel.}
\label{fig:7}
\end{figure*}

\begin{figure}[t]
\centering
\includegraphics[width=\linewidth]{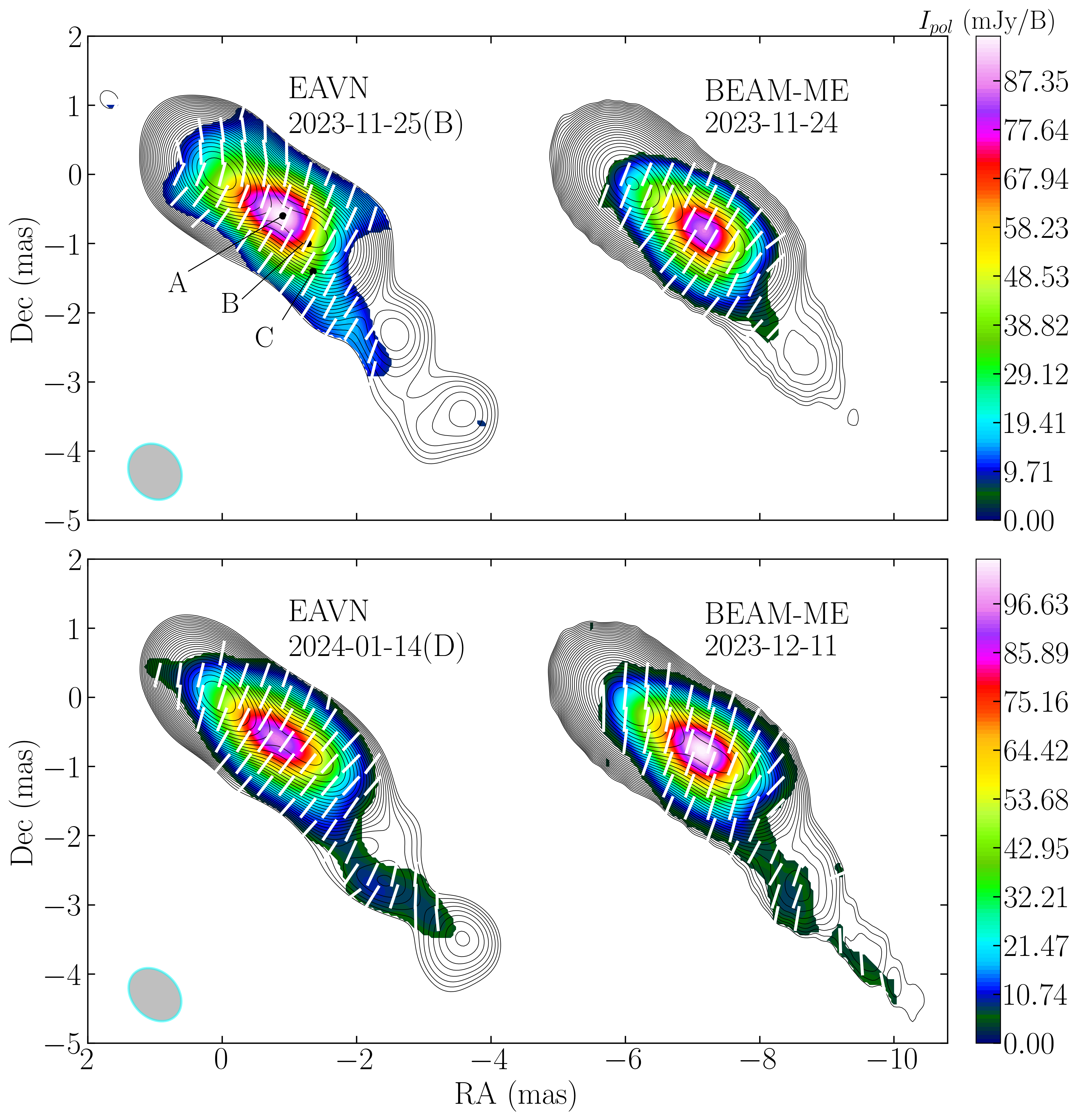}
\caption{Same as Figure~\ref{fig:6} for 3C 273 at 43~GHz. The restoring beam sizes are 0.86~mas $\times$ 0.74~mas with a position angle of 42$^\circ$ for the upper panel, and 0.87~mas $\times$ 0.67~mas with 47$^\circ$ for the lower panel. Points A--C in the upper panel show the locations where polarization properties are compared in Figure~\ref{fig:12}.}
\label{fig:8}
\end{figure}

3C 273 is one of the brightest and most extensively studied quasars, located at a redshift of $z=0.158$. We present the EAVN linear polarization images and compare them with the corresponding VLBA images at 22~GHz (\autoref{fig:7}) and 43~GHz (\autoref{fig:8}). Both the EAVN and VLBA total intensity maps reveal a bent extended jet structure on milliarcsecond (mas) scales, and the polarization structures at both frequencies show overall good agreement between the two arrays.




\subsubsection{OJ 287}\label{sec:3.2.3}
\begin{figure}[t]
\centering
\includegraphics[width=\linewidth]{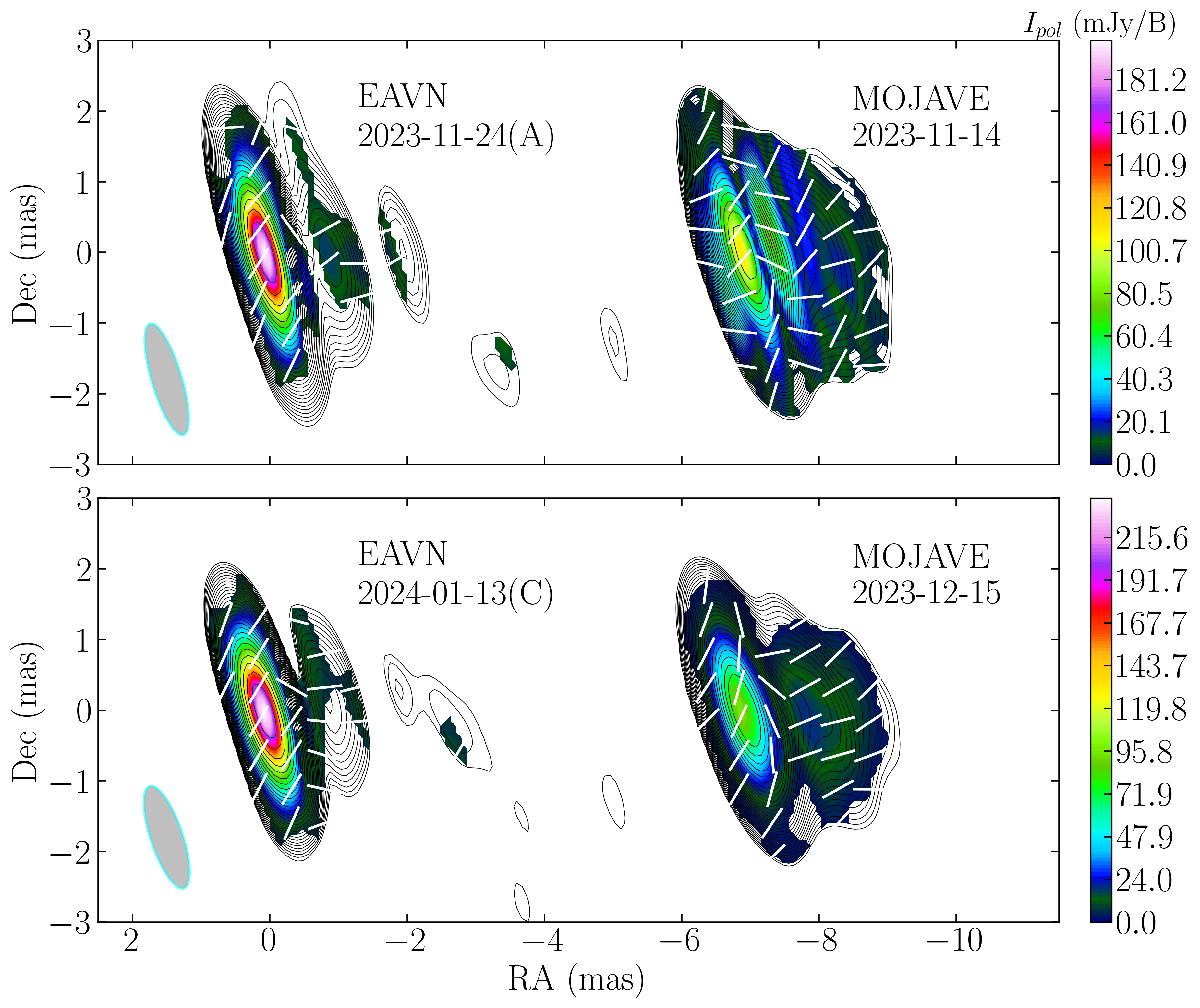}
\caption{Same as Figure~\ref{fig:5} for OJ 287 at 22~GHz. The restoring beam sizes are 1.64~mas $\times$ 0.43~mas with a position angle of 17$^\circ$ for the upper panel, and 1.53~mas $\times$ 0.46~mas with 19$^\circ$ for the lower panel.}
\label{fig:9}
\end{figure}

\begin{figure}[!t]
\centering
\includegraphics[width=\linewidth]{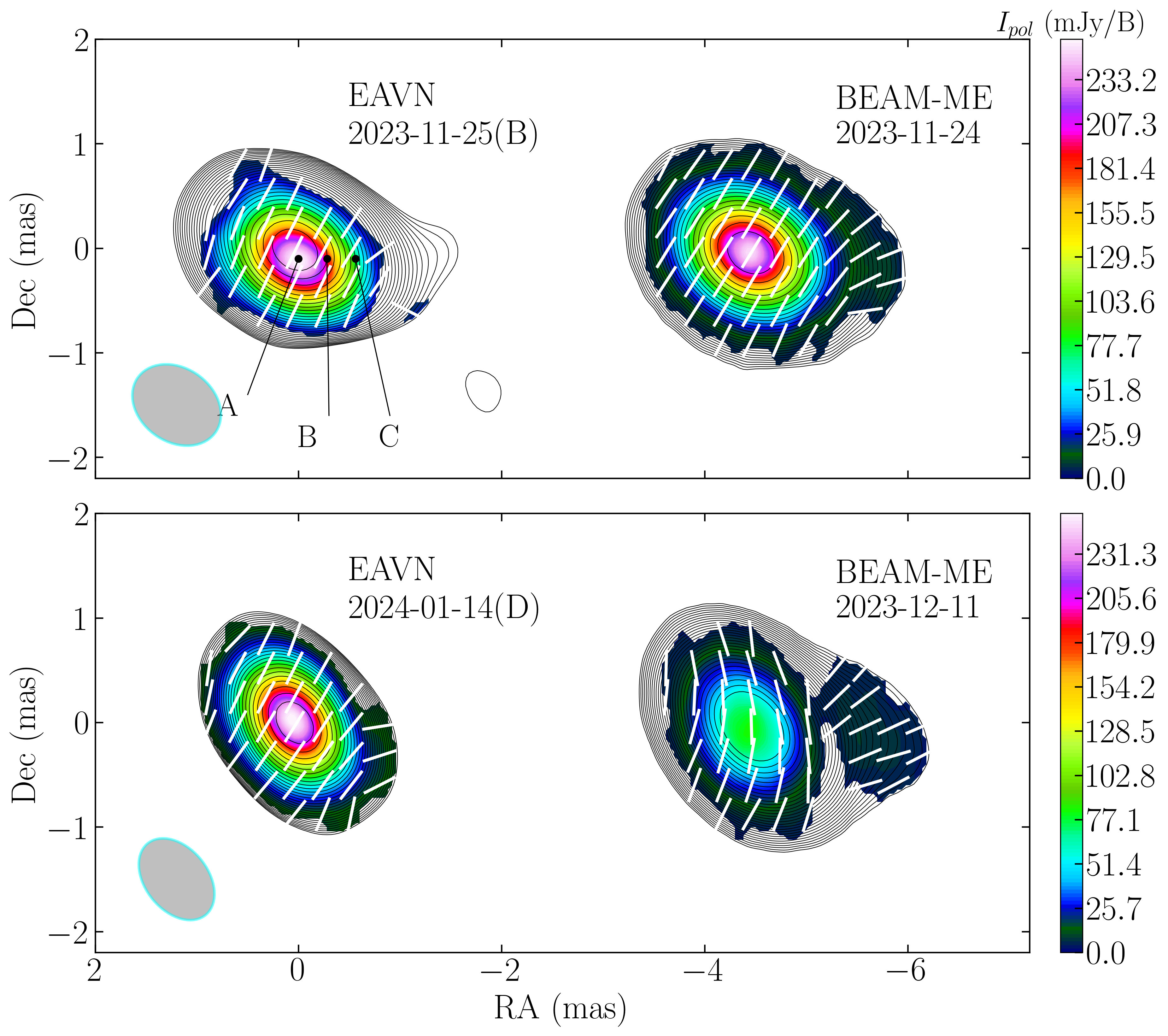}
\caption{Same as Figure~\ref{fig:6} for OJ 287 at 43~GHz. The restoring beam sizes are 0.93~mas $\times$ 0.71~mas with a position angle of 58$^\circ$ for the upper panel, and 0.89~mas $\times$ 0.61~mas with 41$^\circ$ for the lower panel. Points A--C in the upper panel show the locations where polarization properties are compared in Figure~\ref{fig:12}.}
\label{fig:10}
\end{figure}

OJ 287 is a well-known BL Lac object located at a redshift of $z=0.306$, exhibiting quasi-periodic outbursts on a 12-year cycle and rapid short-term variability across multiple wavelengths \citep{sillanpaa1996, valtaoja2000}. It is characterized by fast EVPA rotations of several tens of degrees over just a few days, and substantial changes in radio flux over timescales of weeks to months \citep{holmes1984, pursimo2000, villforth2010, cohen2018}.
We present linear polarization maps of the OJ 287 jet at 22~GHz and 43~GHz in \autoref{fig:9} and \autoref{fig:10}, respectively. In both bands, a core-dominated compact structure is observed, along with a weak outflow toward the west. 
Although OJ 287 is known to exhibit strong variability in polarization intensity over time, a comparison of the two observations conducted just one day apart (upper panel of Figure~\ref{fig:10}) shows good overall consistency in polarization structure.





\section{Discussion}\label{sec:4}

\subsection{D-term Variation in VERA}\label{sec:4.1}
As shown in Figure~\ref{fig:3}-(f) and Figure~\ref{fig:4}-(d,f), several VERA stations—specifically ISG at 22~GHz and IRK/ISG at 43~GHz—exhibit time variability in their D-term phases. We attribute these phase shifts to the field rotator (FR) setup of the VERA stations . The VERA receiver systems are mounted on field rotators (FRs), which rotate to track the apparent motion of objects caused by Earth's rotation. For polarization observations, the FRs are typically deactivated at all stations to ensure that the measured visibilities follow the standard RIME framework, consistent with other EAVN stations.

However, the FRs were not locked to a fixed position for our experiments. Instead, they remained at the position where the previous observation ended. \autoref{tab:3} lists the FR offsets reported between sessions. Notably, some VERA stations are occasionally manually set to $-6^\circ$, $90^\circ$, or $-90^\circ$ between sessions for maintenance purposes, accounting for the occurrence of these specific values. These changes in FR angles between observations can introduce phase rotations in the measured D-terms. In Appendix~\ref{appendix:FR}, we derive how FR offsets affect measured D-term phases within the RIME framework.

\begin{deluxetable}{lccc}
\tablecaption{FR offsets between the sessions in VERA. \label{tab:3}}
\tablehead{
\colhead{} & \colhead{IRK} & \colhead{OGA} & \colhead{ISG}
}
\startdata
B $\rightarrow$ D ($\Delta\,c^{D}$) & -96 & 0 & 84 \\
D $\rightarrow$ E & 0 & 96 & 96 \\
B $\rightarrow$ E ($\Delta\,c^{E}$) & -96 & 96 & 180 \\
\enddata
\tablecomments{Signs denote direction relative to north (positive clockwise). FR offsets were obtained from communication with VERA operators.}
\end{deluxetable}

\begin{figure*}[t!]
\centering
\includegraphics[width=0.75\linewidth]{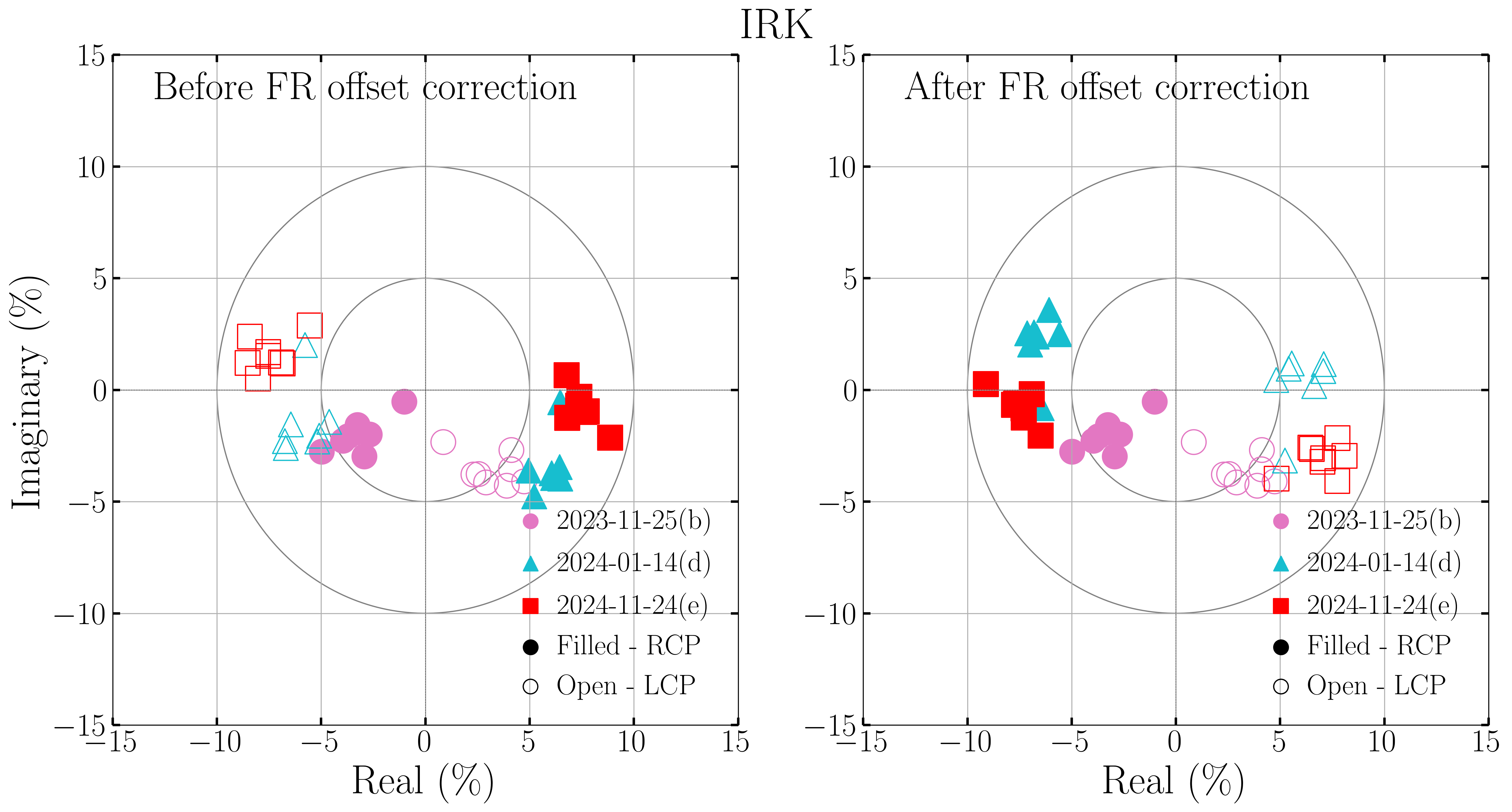}
\includegraphics[width=0.75\linewidth]{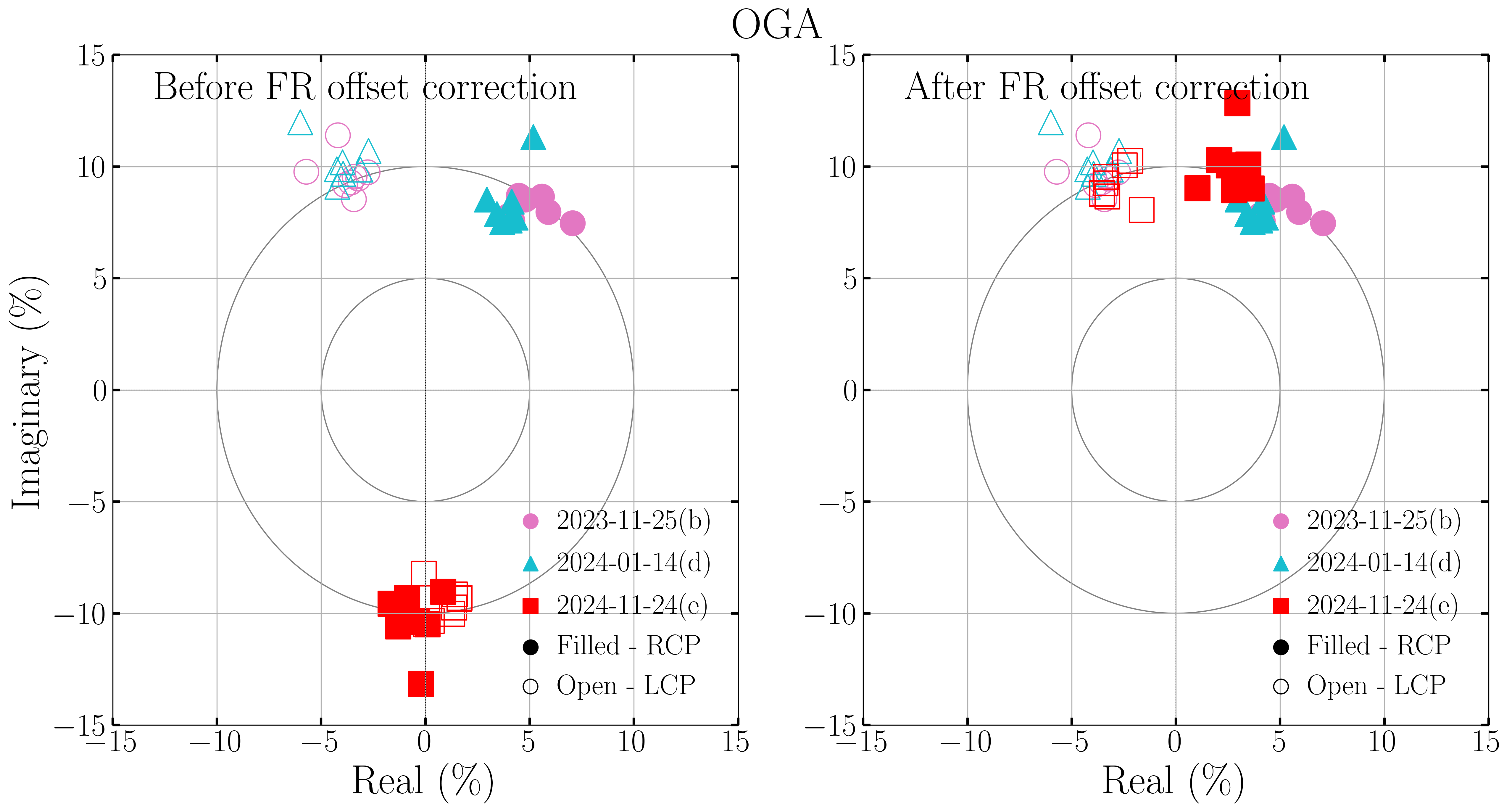}
\includegraphics[width=0.75\linewidth]{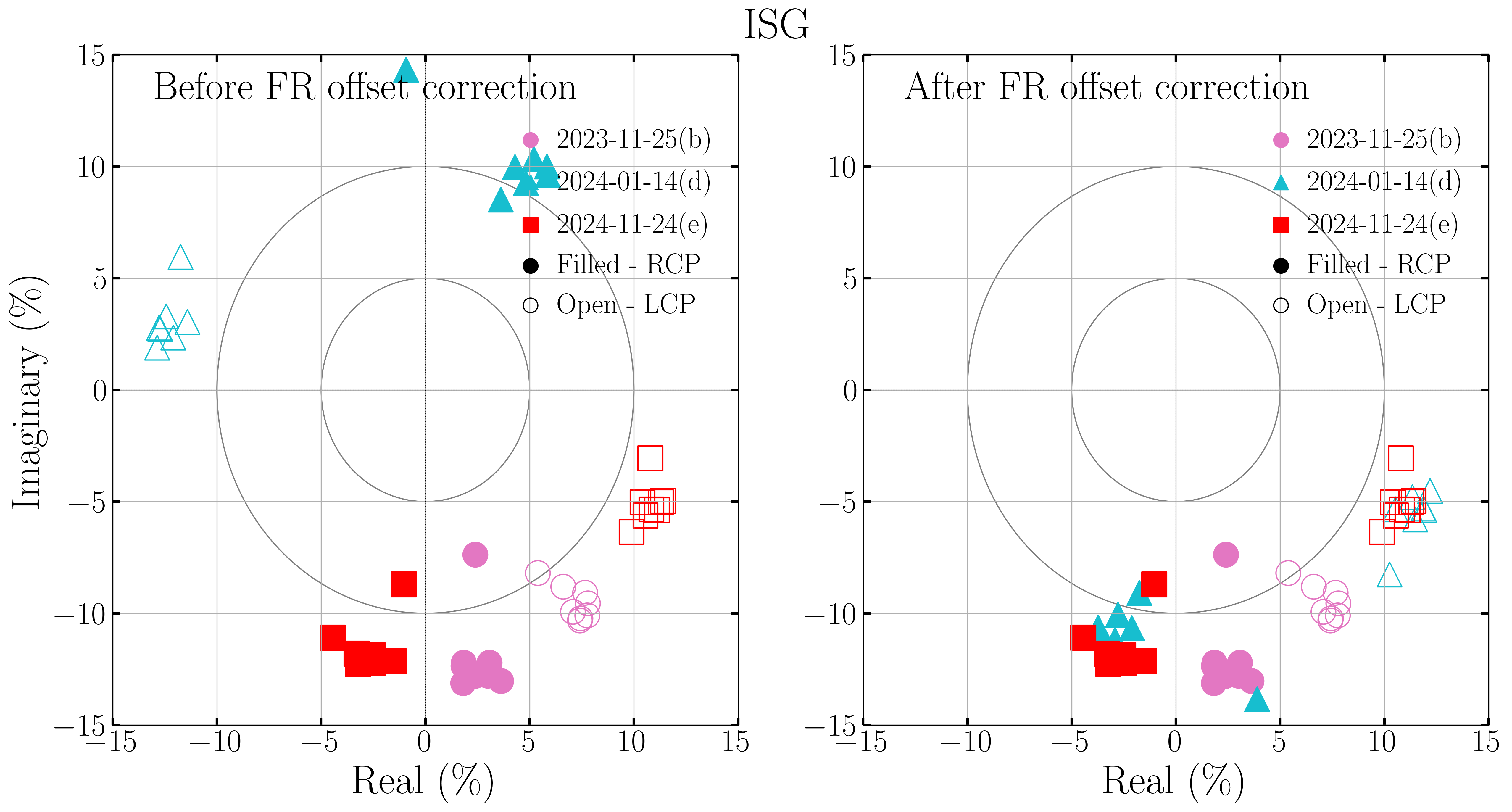}
\caption{Variations of the D-term phase of VERA stations at 43~GHz in the complex plane, including the additional epoch, session E (further details in Kam et al., in preparation). \textit{Left panels}: D-term phases before correcting the FR offsets between sessions. \textit{Right panels}: D-term phases after correcting the FR offsets between sessions.}
\label{fig:11}
\end{figure*}

We corrected the FR offsets between sessions and examined the D-term phase consistency of the VERA stations at 43~GHz across three epochs, including the additional session E. Using the FR offsets $\Delta\,c^{D}$ and $\Delta\,c^{E}$ calculated with session B as the reference (Table~\ref{tab:3}), we applied the corrections; the results are shown in \autoref{fig:11}. Following these corrections, the D-term phases of the VERA stations appear generally stable over time, similar to those of the other stations. These results suggest that the polarimetric leakages of the EAVN stations examined in this study are largely stable over timescales of several months to a year, comparable to other well-established VLBI arrays such as the VLBA. Furthermore, this implies that EVPA calibration can be conducted if D-term phases are known from a nearby epoch, even without external information regarding the source's EVPAs.

Nevertheless, a temporal variation of approximately $10^\circ$ to $30^\circ$ in the D-term phases appears to remain at some stations (including both VERA and specific KVN stations, such as KUS). This residual variation may stem from either genuine systematic instrumental drifts over time or measurement uncertainties in individual sessions. Therefore, further multi-epoch monitoring will be required to robustly constrain whether these temporal variations are intrinsic to the instruments.

\begin{figure*}[t!]
\centering
\includegraphics[width=0.32\textwidth]{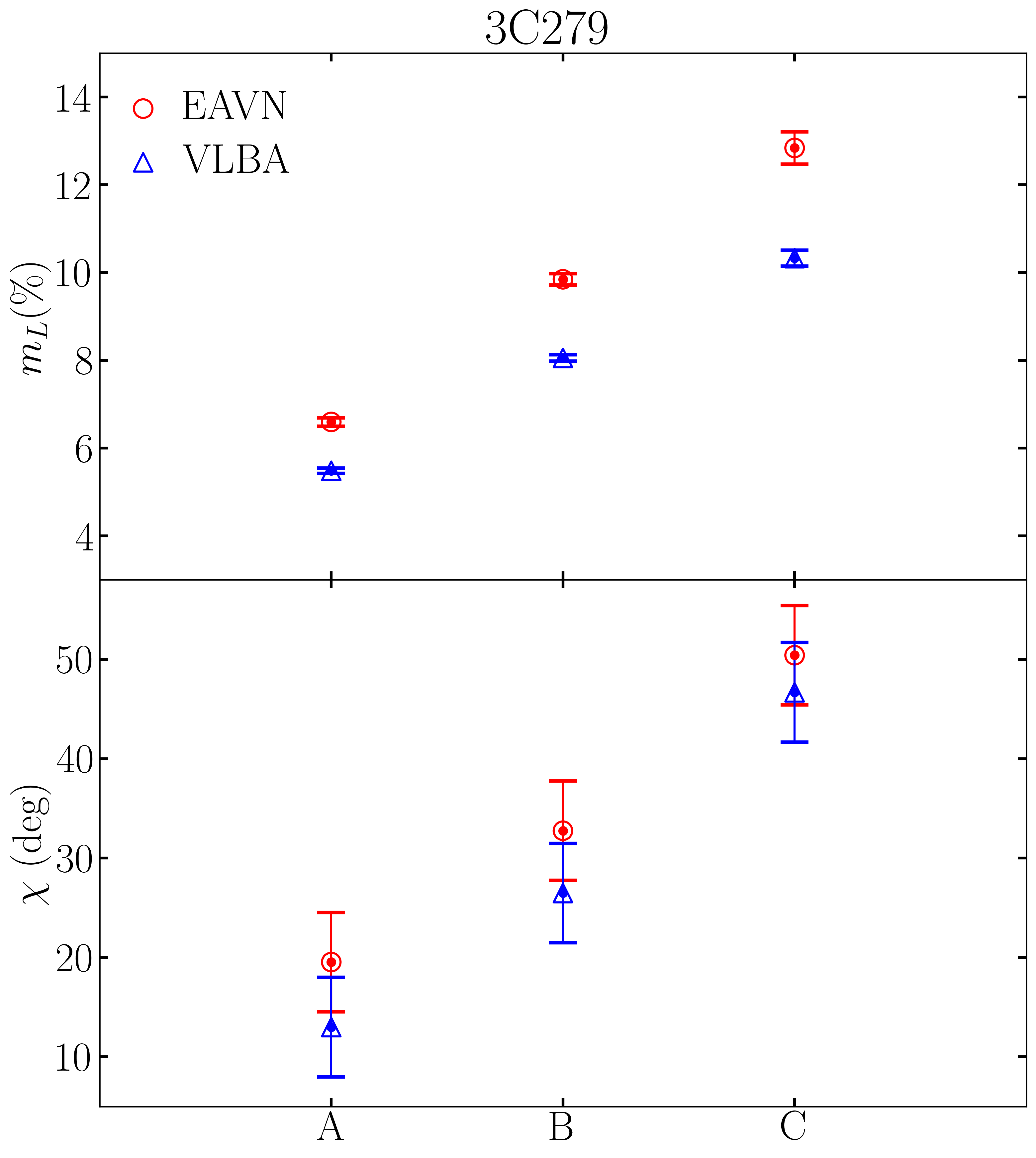}\hfill
\includegraphics[width=0.329\textwidth]{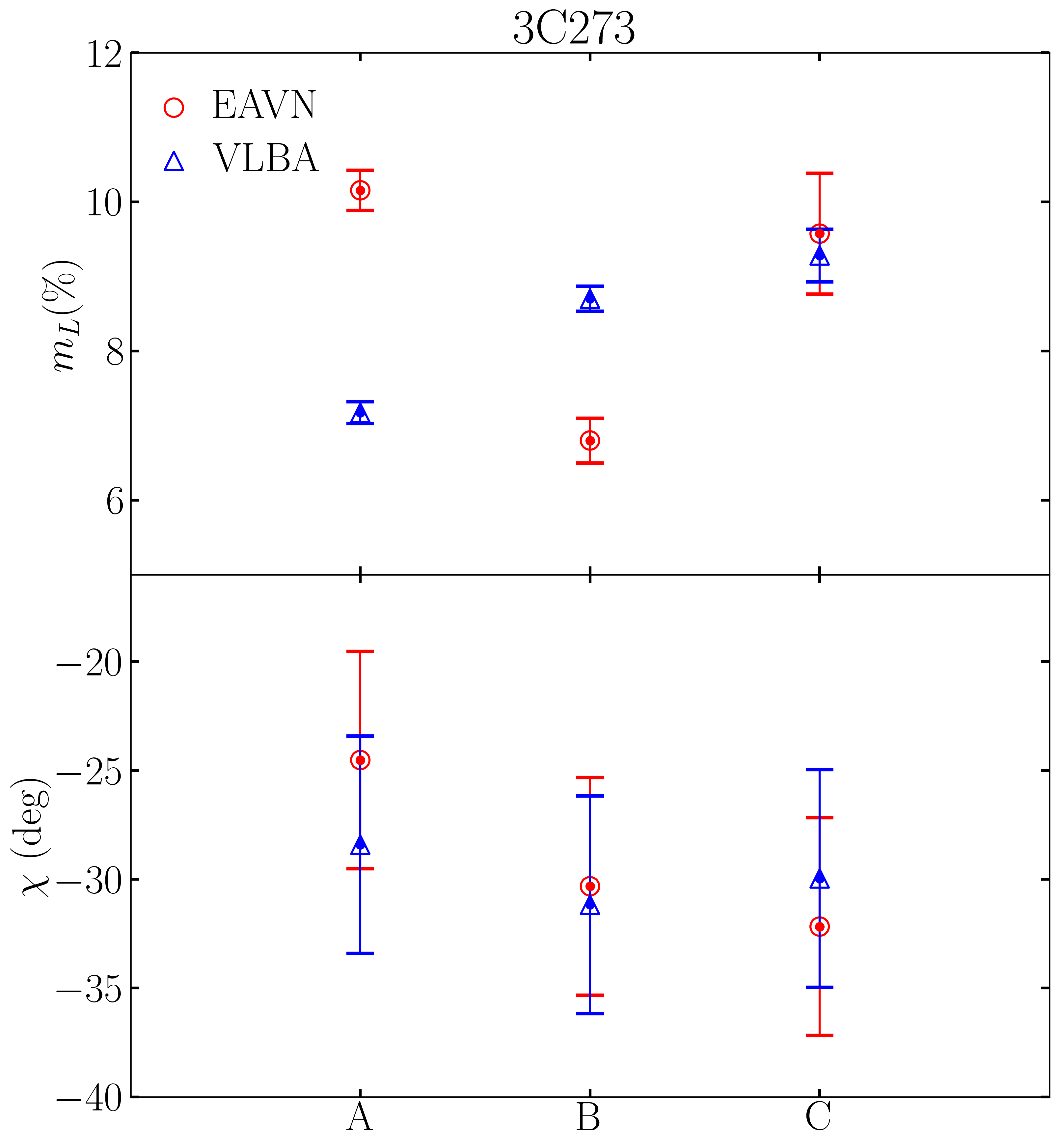}\hfill
\includegraphics[width=0.329\textwidth]{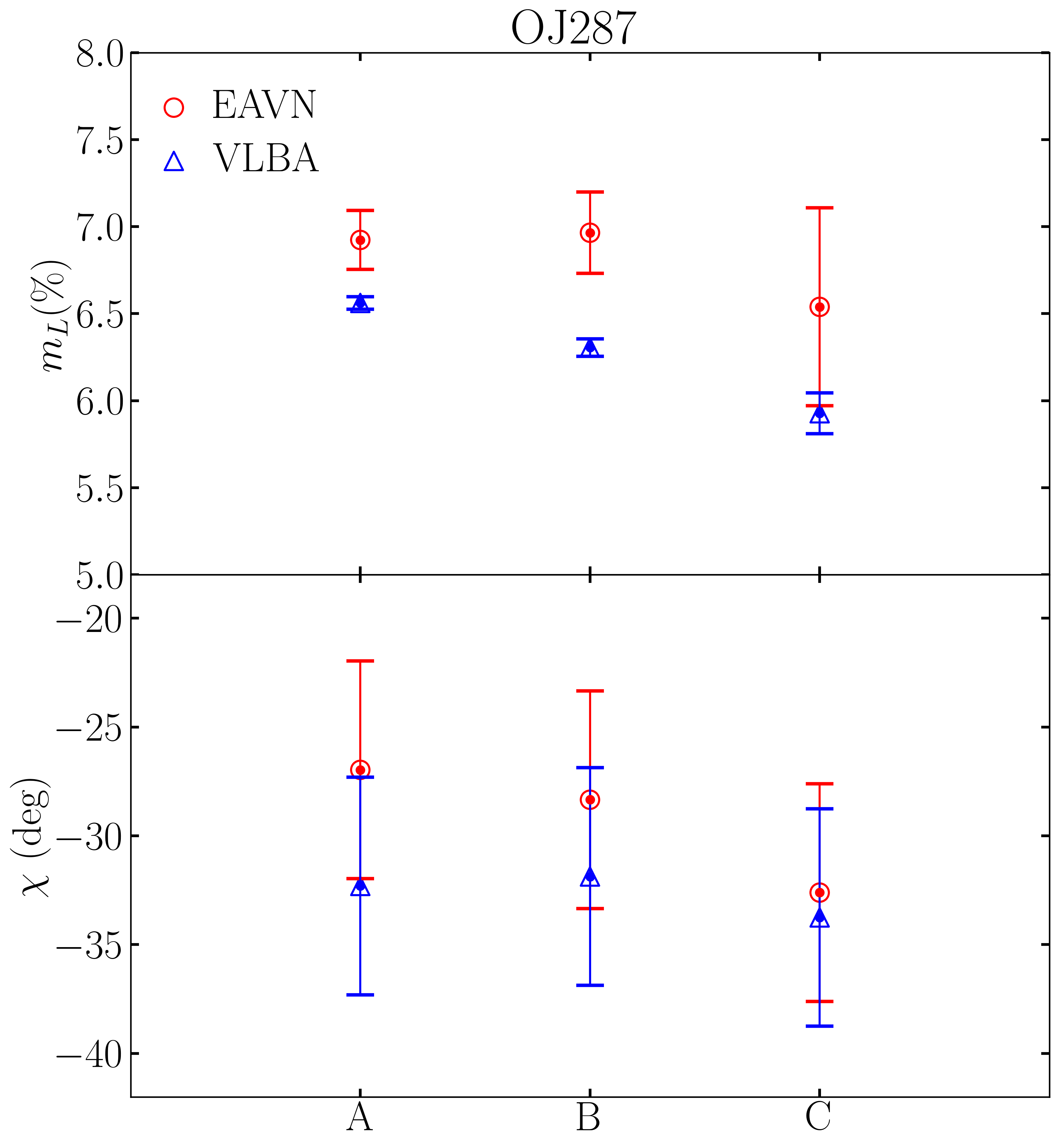}

\caption{Polarization properties of the three sources: (Left) 3C 279, (Middle) 3C 273, and (Right) OJ 287. Fractional polarization (upper panels) and EVPA (lower panels) at points A, B, and C, as indicated in Figures~\ref{fig:6}, \ref{fig:8}, and \ref{fig:10}.}
\label{fig:12}
\end{figure*}

\subsection{Polarization Properties}\label{sec:4.2}
In Section~\ref{sec:3.2}, we qualitatively compared the linear polarization images obtained with the EAVN and VLBA. We note several fundamental limitations in this comparison: the VLBA 15~GHz images and the EAVN 22~GHz images were observed at different frequencies, meaning their polarimetric properties are not necessarily identical; additionally, the EAVN and VLBA observations were separated by weeks to months, making direct comparison difficult due to potential source variability. In contrast, the session B 43~GHz observation and the corresponding VLBA 43~GHz observation were conducted only one day apart at a similar observing frequency, enabling a more robust and quantitative comparison.

\autoref{fig:12} shows a comparison of the fractional polarization and EVPA values calculated at identical jet locations in both the EAVN and VLBA data for the three sources. The uncertainties in fractional polarization and EVPA were estimated following the methods of \citealt{hovatta2012}. For the EVPA, an additional $5^{\circ}$ was quadratically added to the calculated EVPA error to account for systematic uncertainties in the calibration.

Regarding EVPAs, all three sources show good agreement between the EAVN and VLBA measurements, with values consistent within their respective error margins. However, despite being measured at the same locations, the fractional polarization values for 3C 273 differ between the two datasets. This discrepancy likely arises from differences in $(u,v)$-coverage between the two observations, which can result in spatially distributed systematic imaging artifacts in both total intensity and linear polarization images. Nevertheless, the fractional polarization observed in the EAVN images for all three sources tends to be higher than the corresponding VLBA values, especially at locations where the observed linear polarization intensity is close to the peak value. Fractional polarization is known to be sensitive to data calibration accuracy; significant systematic uncertainties, such as residual leakage calibration errors or time-dependent antenna leakages, can locally modulate the fractional polarization over source emission regions. This is because such calibration errors, being antenna-dependent, introduce spurious emissions with alternating signs across the dirty Stokes $Q$ and $U$ images via convolution with the Stokes $I$ structure. Consequently, the generally higher fractional polarization observed in the EAVN images compared to the VLBA images, which appears spatially coherent and non-erratic across multiple sources and regions, implies the absence of critical data or calibration issues. This confirms the EAVN's polarimetric performance for these datasets and indicates that polarimetric data obtained with the EAVN are ready for scientific use.

\section{Summary}\label{sec:con}
This study presents the first systematic verification of the dual-polarization performance of the EAVN at $22$ and $43\,\mathrm{GHz}$, based on observations with the KVN, VERA, and Nanshan stations. The primary target was M87, observed along with the strongly polarized calibrators 3C~279, 3C~273, and OJ~287. Polarization calibration was performed using the novel instrumental-polarization pipeline \textsc{GPCAL}, which successfully fitted the observed cross-hand visibilities across all baselines from short to long. This implies that the EAVN polarimetric data adhere to the standard Radio Interferometer Measurement Equation (RIME) on which \textsc{GPCAL} is based. As a result of the calibration, no significant linear polarization was detected in M87.

We confirm that most antennas in the EAVN exhibited stable D-term amplitudes within the $5$--$10\%$ range over several months, consistent with earlier results for the KVN and VERA individually. While inter-epoch variations in D-term phases were observed at some VERA stations, these are attributed to field-rotator (FR) offsets; after applying the appropriate FR corrections, the D-term phases became stable and mutually consistent across epochs. This demonstrates that the polarimetric leakages of the EAVN stations are stable over timescales of several months to a year, comparable to well-established VLBI arrays such as the VLBA.

The reliability of the resulting linear-polarization maps of 3C~279, 3C~273, and OJ~287 from the EAVN was evaluated by comparing them with near-contemporaneous VLBA observations at similar frequencies. The overall structures of total intensity, polarized intensity, and EVPA show good consistency with the VLBA results. In addition, a quantitative comparison of the polarization properties at $43\,\mathrm{GHz}$, where the two observations were only one day apart, confirms that the EVPA values measured by both networks are consistent within their error ranges. Furthermore, for all sources, the fractional polarization observed with the EAVN tends to be higher than that of the VLBA at locations near the peak polarization intensity. This suggests that there are no critical issues with the data or calibration.

Therefore, we have verified the polarimetric performance of the EAVN for these datasets and conclude that the EAVN is fully equipped with the polarimetric capability to conduct reliable scientific research in the $22$ and $43\,\mathrm{GHz}$ bands. Our work demonstrates such polarimetric capability using KaVA stations together with the Nanshan station, and future observations with increased station participation are expected to further enhance the polarization performance of the EAVN.

\begin{acknowledgments}

The authors appreciate the referee’s constructive comments, which have improved the paper. This work was supported by the National Research Foundation of Korea (NRF) grant funded by the Korea government (MSIT; RS-2024-00449206; RS-2025-02214038). This research has been supported by the POSCO Science Fellowship of POSCO TJ Park Foundation. This research was supported by Global-Learning \& Academic research institution for Master's \textperiodcentered~PhD students, and Postdocs(G-LAMP) Program of the National Research Foundation of Korea(NRF) grant funded by the Ministry of Education(RS-2025-25442355). This research was supported by the Korea Astronomy and Space Science Institute under the R\&D program (Project No. 2025-9-844-00) supervised by the Korea AeroSpace Administration. This work is made use of the East Asia VLBI Network (EAVN), which is operated under cooperative agreement by National Astronomical Observatory of Japan (NAOJ), Korea Astronomy and Space Science Institute (KASI), Shanghai Astronomical Observatory (SHAO), Xinjiang Astronomical Observatory (XAO), Yunnan Astronomical Observatory (YNAO), National Astronomical Research Institute of Thailand (Public Organization) (NARIT), and National Geographic Information Institute (NGII), with the operational support by Ibaraki University (for the operation of Hitachi 32-m and Takahagi 32-m telescopes), Yamaguchi University (for the operation of Yamaguchi 32- m telescope), and Kagoshima University (for the operation of VERA Iriki antenna). We are grateful to the staff of the KVN who helped to operate the array and to correlate the data. The KVN is a facility operated by the KASI (Korea Astronomy and Space Science Institute). The KVN observations and correlations are supported through the high-speed network connections among the KVN sites provided by the KREONET (Korea Research Environment Open NETwork), which is managed and operated by the KISTI (Korea Institute of Science and Technology Information). This work uses data from the MOJAVE database that is maintained by the MOJAVE team \citep{lister2018}. And this work uses VLBA data from the VLBA-BU Blazar Monitoring Program (BEAM-ME and VLBA-BU-BLAZAR; \url{http://www.bu.edu/blazars/BEAM-ME.html}), funded by NASA through the Fermi Guest Investigator Program. The VLBA is an instrument of the National Radio Astronomy Observatory. The National Radio Astronomy Observatory is a facility of the National Science Foundation operated by Associated Universities, Inc.

\end{acknowledgments}

\begin{contribution}

J.P. initiated the project, coordinated the research, and conducted the VLBI observations. Y.L. and J.P. worked on calibration and analysis of the data. K.H. investigated the field rotator issue of the VERA stations. Y.L. wrote the original manuscript. All authors contributed to the discussion of the results presented and commented on the manuscript.

\end{contribution}

\facilities{EAVN, KVN, VLBA}

\software{AIPS \citep{greisen2003}, DIFMAP \citep{Shepherd1997}, GPCAL \citep{park2021,Park2023a, Park2023b}
          }

\appendix

\section{Relationship between FR Offsets and D-term Phases for VERA}\label{appendix:FR}

In this Appendix, we describe the relationship between FR offsets and D-term phases for VERA antennas within the RIME framework \citep{hamaker1996_1, sault1996, hamaker1996_3, smirnov2011}. 
The relationship between the true ($\bar{\mathbf{V}}_{mn}$) and observed visibility ($\mathbf{V}^{\mathrm{obs}}_{mn}$) is given by
\begin{equation}
\label{eq:1}
\mathbf{V}^{\mathrm{obs}}_{mn}=\mathbf{J}_m\,\bar{\mathbf{V}}_{mn}\,\mathbf{J}_n^{H},
\end{equation}
where $H$ denotes the Hermitian operator and $\mathbf{J}$ represents the Jones matrix \citep{jones1941}.

If an FR offset is present on a VERA antenna, the antenna's field rotation angle—equivalent to the parallactic angle for VERA antennas—is rotated by a constant angle ($c_a$, defined as positive clockwise). The Jones matrix can be expressed as
\begin{equation}
\label{eq:2}
\mathbf{J}_a
=
\underbrace{\begin{bmatrix} g_{aR}&0\\0&g_{aL}\end{bmatrix}}_{\mathbf{G}_a}
\underbrace{\begin{bmatrix} 1& d_{aR}\\ d_{aL}&1\end{bmatrix}}_{\mathbf{D}_a}
\underbrace{\begin{bmatrix} e^{-i(\phi_a+c_a)}&0\\0&e^{+i(\phi_a+c_a)}\end{bmatrix}}_{\mathbf{P}_a(\phi_a+c_a)}\!.
\end{equation}
Here, $\mathbf{G}_a$ is the antenna gain matrix, $\mathbf{D}_a$ is the leakage matrix, and $\mathbf{P}_a$ is the field-rotation angle matrix. Note that $\mathbf{P}_a$ becomes a function of $\phi_a+c_a$ due to the FR offset.

We performed parallactic angle correction at an early stage of data reduction without knowledge of the existence of $c_a$; this operation is equivalent to
\begin{equation}
\mathbf{V}'=\mathbf{P}_m^{-1}(\phi_m)\,\mathbf{V}_{\mathrm{obs}}\,[\mathbf{P}_n^{-1}(\phi_n)]^{\mathrm H}.
\end{equation}
Using the fact that both the gain and field-rotation angle matrices contain only diagonal elements and thus commute, one can rewrite this equation as:
\begin{equation}
\label{eq:3}
\begin{aligned}
\mathbf{V}' &= \mathbf{G}_m\;
\big[\mathbf{P}_m^{-1}(\phi_m)\,\mathbf{D}_m\,\mathbf{P}_m(\phi_m)\big]\;
\mathbf{P}_m(c_m)\bar{\mathbf{V}}\;
\mathbf{P}_n^{H}(c_n)\;
\big[\mathbf{P}_n^{-1}(\phi_n)\,\mathbf{D}_n\,\mathbf{P}_n(\phi_n)\big]^{H}\;
\mathbf{G}_n^{H}.
\end{aligned}
\end{equation}
We define the \textit{FR-absorbed} gain matrix as
\begin{equation}
\mathbf{G}_a^{\mathrm{cal}} \;\equiv\; \mathbf{G}_a\,\mathbf{P}_a(c_a)
\;=\;
\begin{bmatrix}
G_{aR}e^{-ic_a} & 0\\
0 & G_{aL}e^{+ic_a}
\end{bmatrix},
\end{equation}
and the \textit{FR-rotated} D-term matrix as
\begin{equation}
\label{eq:5}
\mathbf{D}_a^{\mathrm{rot}} \;\equiv\; \mathbf{P}_a(-c_a)\,\mathbf{D}_a\,\mathbf{P}_a(c_a)
\;=\;
\begin{bmatrix}
1 & d_{aR}e^{+2ic_a}\\
d_{aL}e^{-2ic_a} & 1
\end{bmatrix}.
\end{equation}
Since $\mathbf{P}(c_a)$ is diagonal, it commutes with $\mathbf{P}(\phi_a)$. By inserting $\mathbf{P}(c_m)\mathbf{P}(-c_m)=\mathbf{I}$ to the left of $\mathbf{P}_m(c_m)\bar{\mathbf{V}}$ and $\mathbf{P}(-c_n)\mathbf{P}(c_n)=\mathbf{I}$ to the right of $\bar{\mathbf{V}}\,\mathbf{P}_n^{H}(c_n)$ in Equation (\ref{eq:3}) and regrouping, we obtain:
\begin{equation}
\label{eq:6}
\begin{aligned}
\mathbf{V}'
&= \mathbf{G}_m^{\mathrm{cal}}\;
\big[\mathbf{P}_m^{-1}(\phi_m)\,\mathbf{D}_m^{\mathrm{rot}}\,\mathbf{P}_m(\phi_m)\big]\;
\bar{\mathbf{V}}
\big(\big[\mathbf{P}_n^{-1}(\phi_n)\,\mathbf{D}_n^{\mathrm{rot}}\,\mathbf{P}_n(\phi_n)\big]\big)^{H}\;
\big(\mathbf{G}_n^{\mathrm{cal}}\big)^{H}.
\end{aligned}
\end{equation}
This form is identical to the standard equations adopted by VLBI polarization calibration pipelines (e.g., Equation 7 in \citealt{Park2023a}). The matrices $\mathbf{G}_m^{\mathrm{cal}}$ and $\mathbf{G}_n^{\mathrm{cal}}$ are estimated and removed from the data during the reduction and imaging/self-calibration procedures. Consequently, the new D-terms are phase-rotated with respect to the original D-terms by $\pm2c_a$; specifically, $D_{R,a}^{\mathrm{rot}}=D_{R,a}e^{+2ic_a}$ and $D_{L,a}^{\mathrm{rot}}=D_{L,a}e^{-2ic_a}$. 
Therefore, FR offsets relative to a reference session, denoted by $\Delta\,c^s_a=c_{a}^{s}-c_{a}^{\mathrm{ref}}$, rotate the D-terms by $\mp2\Delta c^s_a$ in phase, where the superscript $s$ indicates the session name. This necessitates the correction of the D-term phase by $\pm2\Delta c^s_a$ between sessions.

\section{Frequency Dependency on D-term Amplitude in EAVN}\label{appendix:amplitude}
\begin{figure*}[h!]
\centering
\includegraphics[height=0.452\textheight]{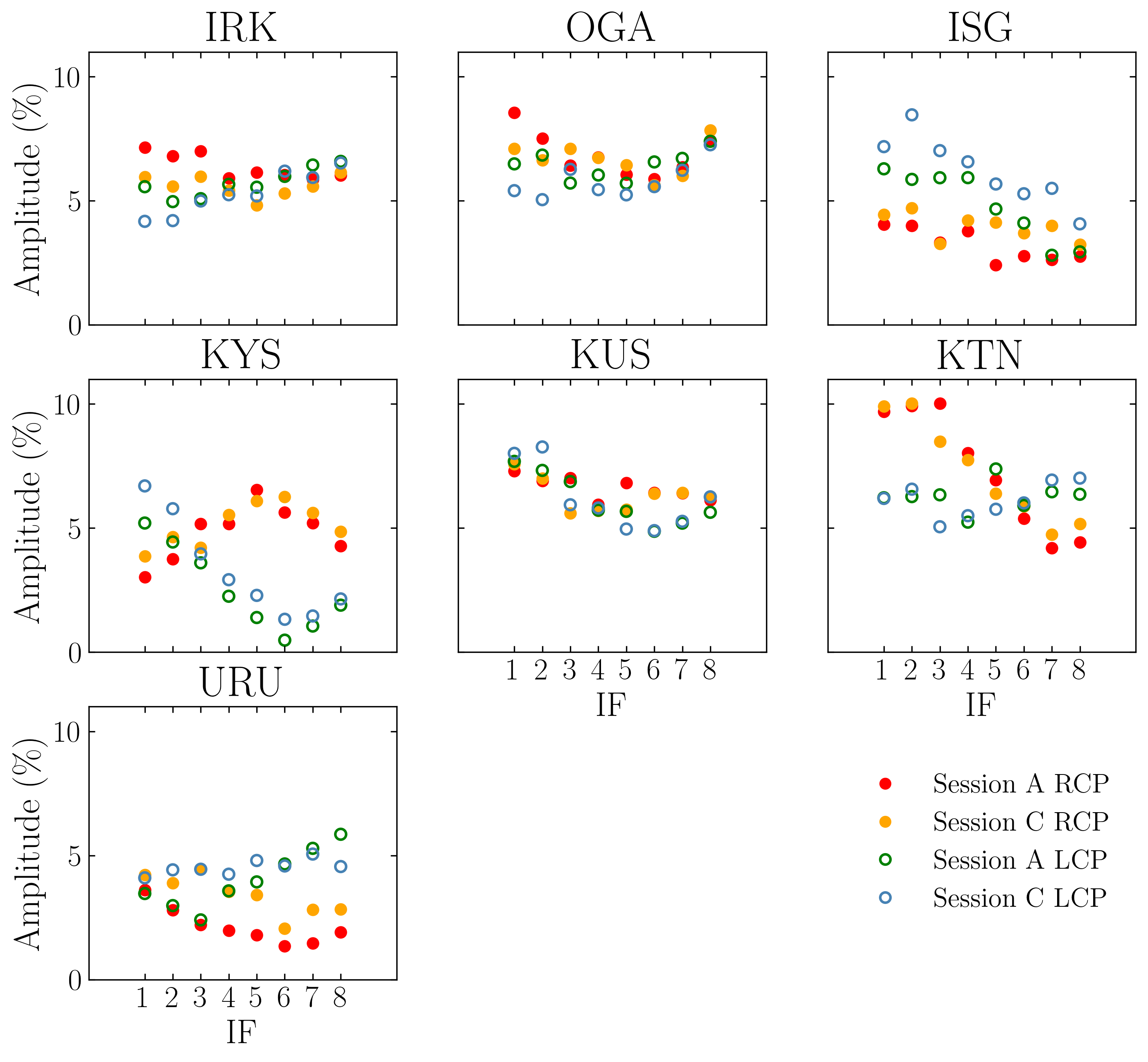}
\caption{D$-$term amplitudes as a function of IF in the K-band. The x$-$axis represents eight IFs ranging from 22.163~GHz to 22.275~GHz at 16~MHz intervals.}
\label{fig:13}
\end{figure*}

\begin{figure*}[h!]
\centering
\includegraphics[height=0.452\textheight]{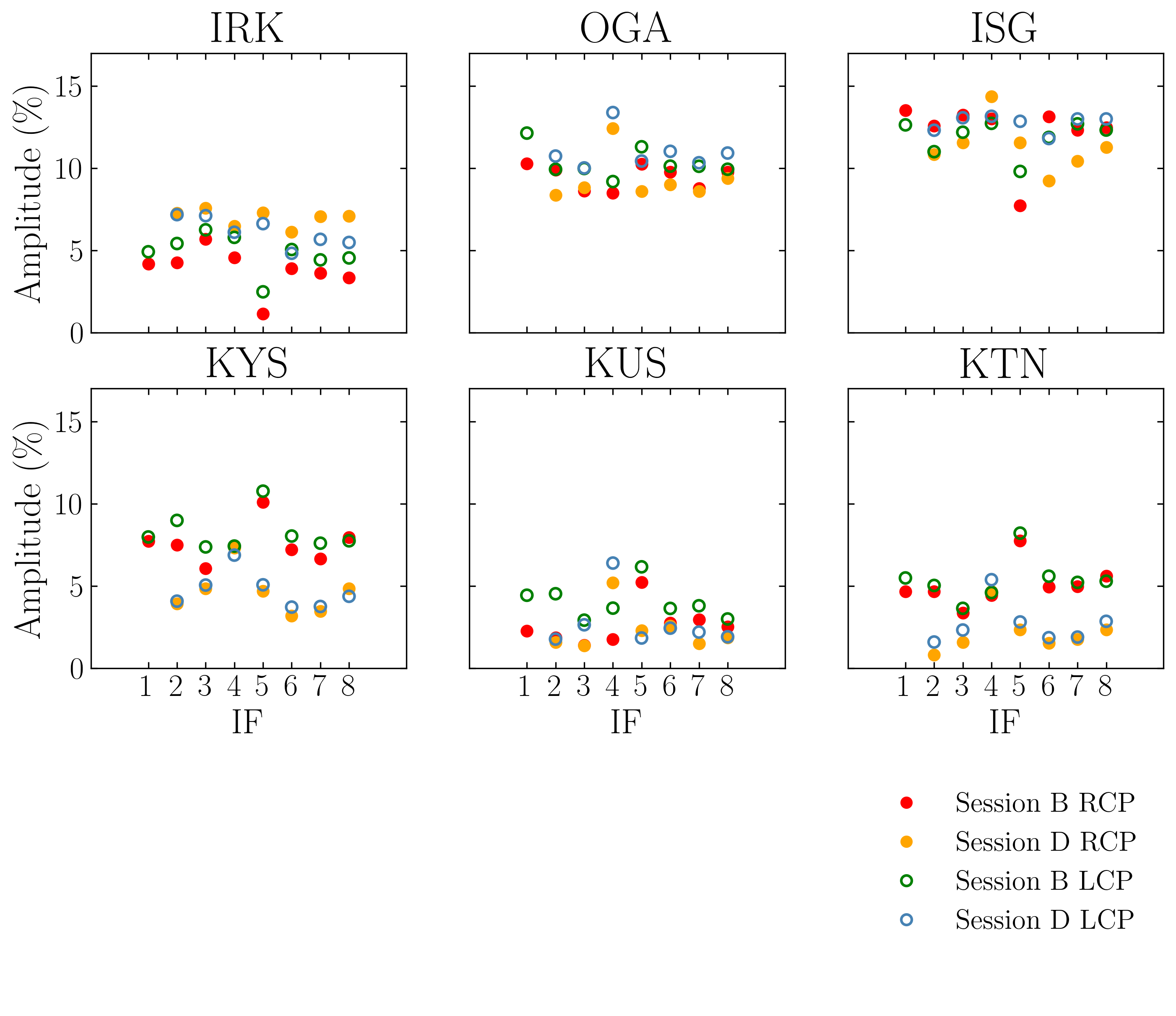}
\caption{D$-$term amplitudes as a function of IF in the Q-band. The x$-$axis represents eight IFs ranging from 43.050~GHz to 43.162~GHz at 16~MHz intervals.}
\label{fig:14}
\end{figure*}

In \autoref{sec:3.1}, although we confirm a moderate frequency dependence of the D-term amplitudes in the K-band, we reported the mean values as representative values across the eight IFs in \autoref{tab:2}. This Appendix provides a detailed view of the D-term amplitudes across all IFs for individual antennas in K- and Q-band, which are presented in \autoref{fig:13} and \autoref{fig:14}, respectively.

\section{Polarization Residual Maps}\label{appendix:residual}
\begin{figure*}[!h]
\centering
\includegraphics[width=0.75\textwidth]{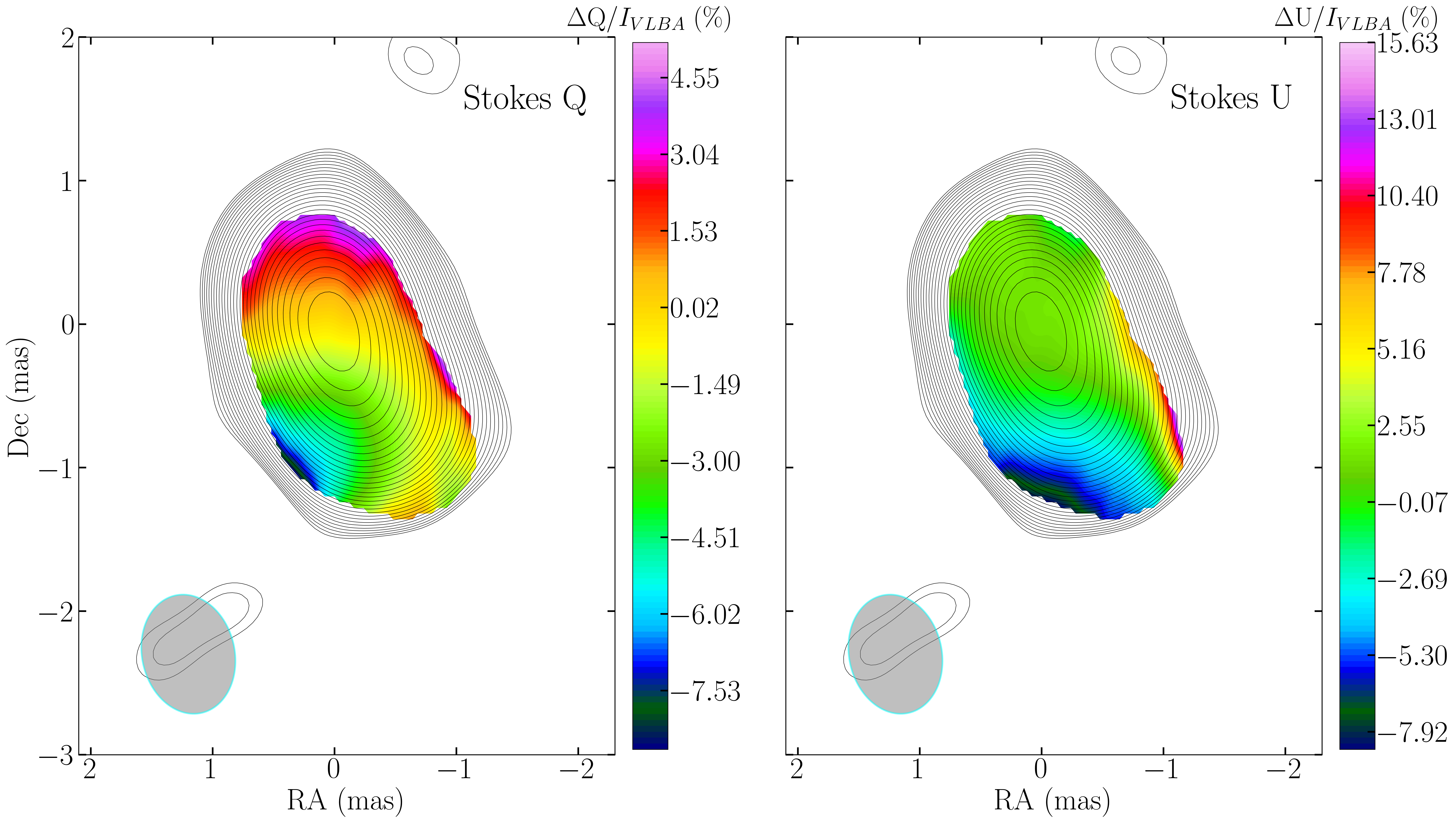}
\includegraphics[width=0.75\textwidth]{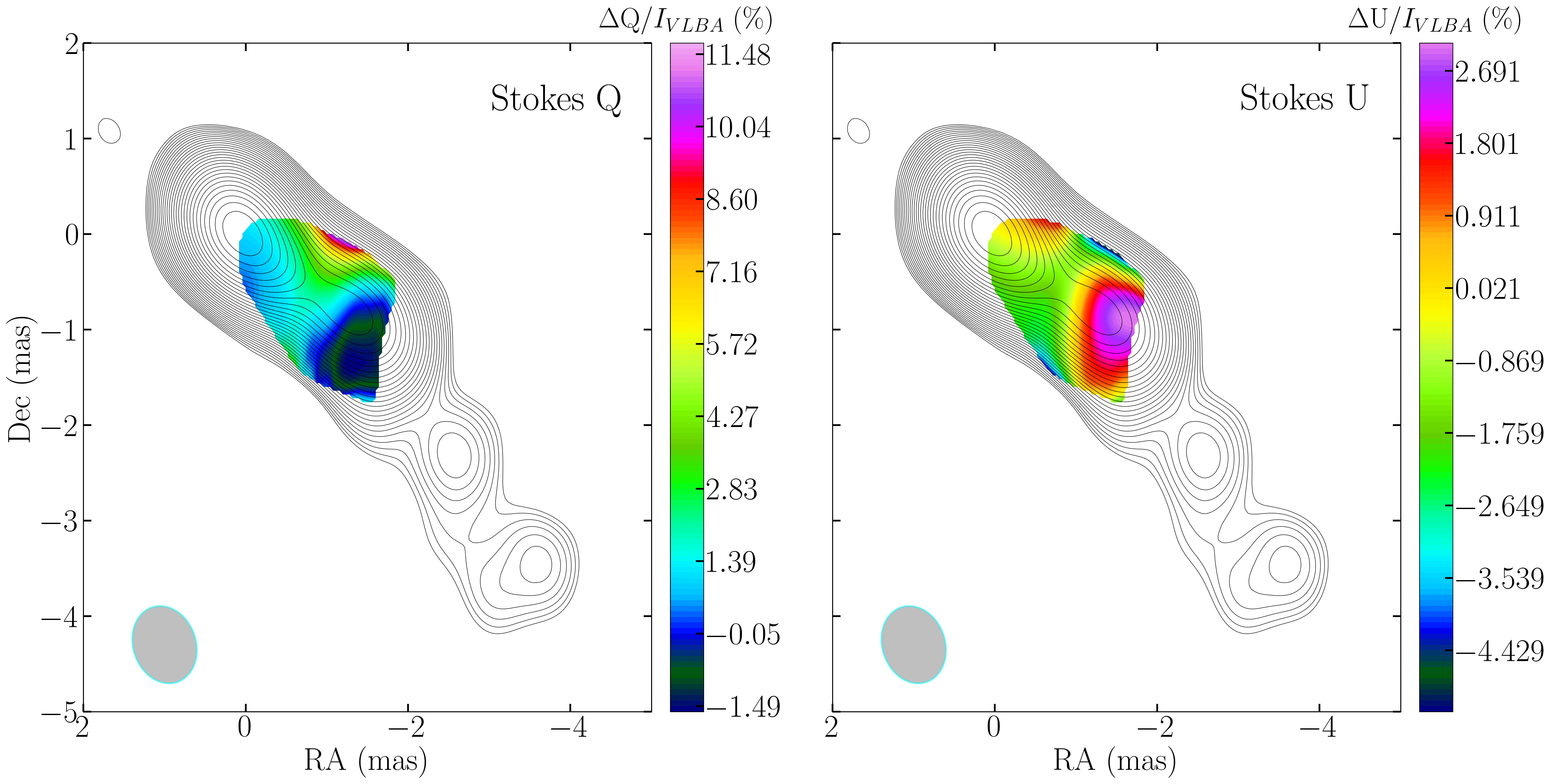}
\hspace*{-0.02\textwidth}
\includegraphics[width=0.75\textwidth]{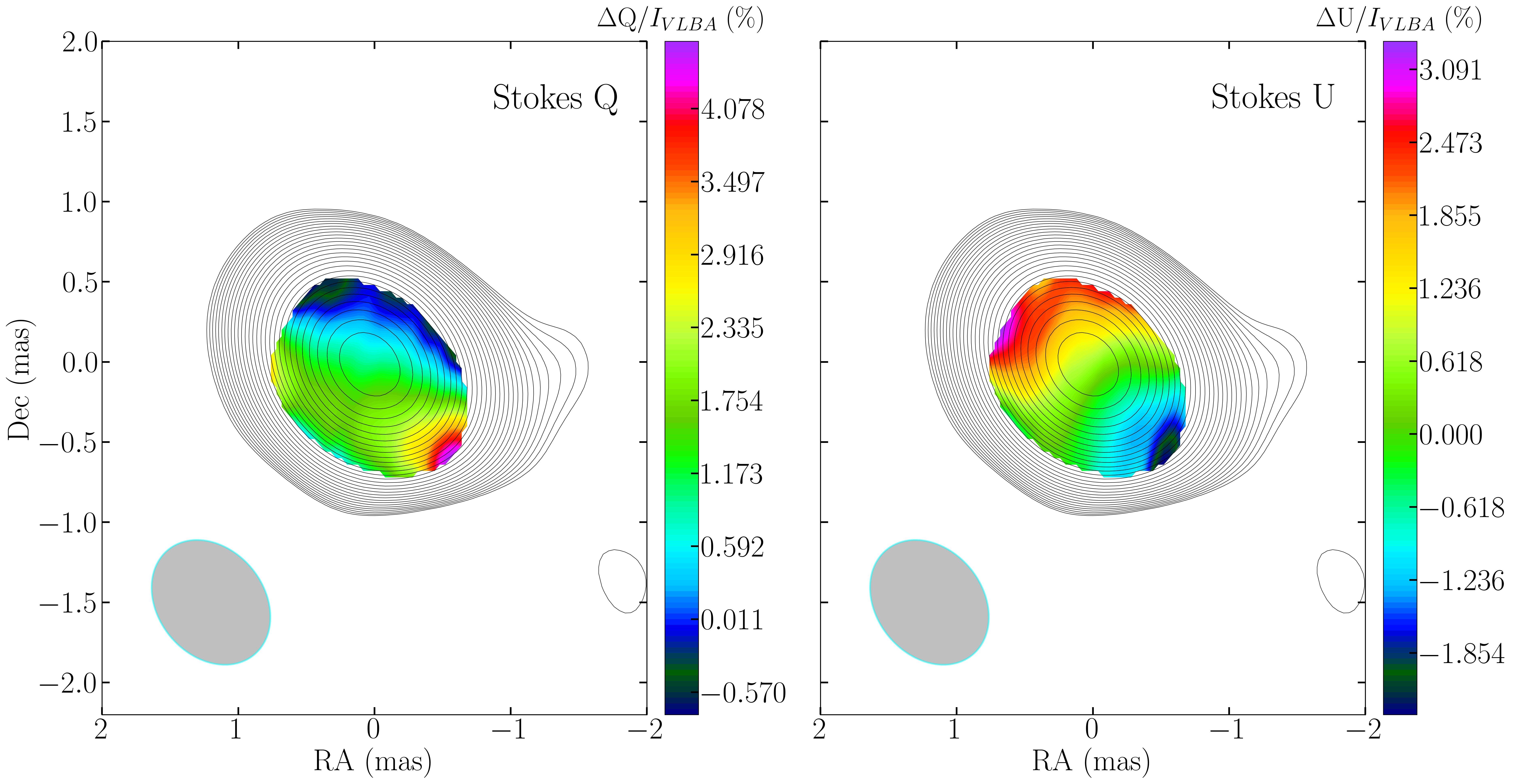}
\caption{Fractional residual maps of Stokes $Q$ and $U$ for the three sources in session B: 3C~279 (top), 3C~273 (middle), and OJ~287 (bottom).}
\label{fig:15}
\end{figure*}

To quantitatively compare the polarization images from the EAVN and VLBA, we presents the fractional residual maps of Stokes $Q$ and $U$ for three sources in \autoref{fig:15}. 
The fractional residuals are defined as $(Q_{\mathrm{EAVN}}-Q_{\mathrm{VLBA}})/I_{\mathrm{VLBA}}$ and $(U_{\mathrm{EAVN}}-U_{\mathrm{VLBA}})/I_{\mathrm{VLBA}}$.
The maps are derived exclusively from session B, where the observations were conducted at a similar observing frequency and only one day apart, ensuring a reliable comparison by minimizing potential source variability.

\bibliography{main}{}

@ARTICLE{Jorstad2017,
       author = {{Jorstad}, Svetlana G. and {Marscher}, Alan P. and {Morozova}, Daria A. and {Troitsky}, Ivan S. and {Agudo}, Iv{\'a}n and {Casadio}, Carolina and {Foord}, Adi and {G{\'o}mez}, Jos{\'e} L. and {MacDonald}, Nicholas R. and {Molina}, Sol N. and {L{\"a}hteenm{\"a}ki}, Anne and {Tammi}, Joni and {Tornikoski}, Merja},
        title = "{Kinematics of Parsec-scale Jets of Gamma-Ray Blazars at 43 GHz within the VLBA-BU-BLAZAR Program}",
      journal = {\apj},
         year = 2017,
        month = sep,
       volume = {846},
       number = {2},
          eid = {98},
        pages = {98},
          doi = {10.3847/1538-4357/aa8407},
archivePrefix = {arXiv},
       eprint = {1711.03983},
 primaryClass = {astro-ph.GA},
       adsurl = {https://ui.adsabs.harvard.edu/abs/2017ApJ...846...98J}
}

@ARTICLE{Niinuma2014,
       author = {{Niinuma}, Kotaro and {Lee}, Sang-Sung and {Kino}, Motoki and {Sohn}, Bong Won and {Akiyama}, Kazunori and {Zhao}, Guang-Yao and {Sawada-Satoh}, Satoko and {Trippe}, Sascha and {Hada}, Kazuhiro and {Jung}, Taehyun and {Hagiwara}, Yoshiaki and {Dodson}, Richard and {Koyama}, Shoko and {Honma}, Mareki and {Nagai}, Hiroshi and {Chung}, Aeree and {Doi}, Akihiro and {Fujisawa}, Kenta and {Han}, Myoung-Hee and {Kim}, Joeng-Sook and {Lee}, Jeewon and {Lee}, Jeong Ae and {Miyazaki}, Atsushi and {Oyama}, Tomoaki and {Sorai}, Kazuo and {Wajima}, Kiyoaki and {Bae}, Jaehan and {Byun}, Do-Young and {Cho}, Se-Hyung and {Choi}, Yoon Kyung and {Chung}, Hyunsoo and {Chung}, Moon-Hee and {Han}, Seog-Tae and {Hirota}, Tomoya and {Hwang}, Jung-Wook and {Je}, Do-Heung and {Jike}, Takaaki and {Jung}, Dong-Kyu and {Jung}, Jin-Seung and {Kang}, Ji-Hyun and {Kang}, Jiman and {Kang}, Yong-Woo and {Kan-ya}, Yukitoshi and {Kanaguchi}, Masahiro and {Kawaguchi}, Noriyuki and {Kim}, Bong Gyu and {Kim}, Hyo Ryoung and {Kim}, Hyun-Goo and {Kim}, Jaeheon and {Kim}, Jongsoo and {Kim}, Kee-Tae and {Kim}, Mikyoung and {Kobayashi}, Hideyuki and {Kono}, Yusuke and {Kurayama}, Tomoharu and {Lee}, Changhoon and {Lee}, Jung-Won and {Lee}, Sang Hyun and {Minh}, Young Chol and {Matsumoto}, Naoko and {Nakagawa}, Akiharu and {Oh}, Chung Sik and {Oh}, Se-Jin and {Park}, Sun-Youp and {Roh}, Duk-Gyoo and {Sasao}, Tetsuo and {Shibata}, Katsunori M. and {Song}, Min-Gyu and {Tamura}, Yoshiaki and {Wi}, Seog-Oh and {Yeom}, Jae-Hwan and {Yun}, Young Joo},
        title = "{VLBI observations of bright AGN jets with the KVN and VERA Array (KaVA): Evaluation of imaging capability}",
      journal = {\pasj},
         year = 2014,
        month = dec,
       volume = {66},
       number = {6},
          eid = {103},
        pages = {103},
          doi = {10.1093/pasj/psu104},
archivePrefix = {arXiv},
       eprint = {1406.4356},
 primaryClass = {astro-ph.IM},
       adsurl = {https://ui.adsabs.harvard.edu/abs/2014PASJ...66..103N}
}

@ARTICLE{Park2026,
       author = {{Park}, Jongho and {Takahashi}, Kazuya and {Toma}, Kenji and {Hada}, Kazuhiro and {Nakamura}, Masanori and {Pu}, Hung-Yi and {Asada}, Keiichi and {Ho}, Paul T.~P. and {Kino}, Motoki and {Kawashima}, Tomohisa and {Kam}, Minchul and {Yi}, Kunwoo and {Cho}, Ilje},
        title = "{Helical Magnetic Field in the Acceleration--Collimation Zone of the M87 Jet}",
      journal = {arXiv e-prints},
         year = 2025,
        month = nov,
          eid = {arXiv:2511.13008},
        pages = {arXiv:2511.13008},
          doi = {10.48550/arXiv.2511.13008},
archivePrefix = {arXiv},
       eprint = {2511.13008},
 primaryClass = {astro-ph.HE},
       adsurl = {https://ui.adsabs.harvard.edu/abs/2025arXiv251113008P}
}

@ARTICLE{Kam2023,
       author = {{Kam}, Minchul and {Trippe}, Sascha and {Byun}, Do-Young and {Park}, Jongho and {Kang}, Sincheol and {Shin}, Naeun and {Lee}, Sang-Sung and {Jung}, Taehyun},
        title = "{Using the Crab Nebula as Polarization Angle Calibrator for the Korean VLBI Network}",
      journal = {Journal of Korean Astronomical Society},
         year = 2023,
        month = jan,
       volume = {56},
        pages = {1-9},
          doi = {10.5303/JKAS.2023.56.1.1},
archivePrefix = {arXiv},
       eprint = {2212.08274},
 primaryClass = {astro-ph.GA},
       adsurl = {https://ui.adsabs.harvard.edu/abs/2023JKAS...56....1K}
}

@ARTICLE{Akiyama2022,
       author = {{Akiyama}, Kazunori and {Algaba}, Juan-Carlos and {An}, Tao and {Asada}, Keiichi and {Asanok}, Kitiyanee and {Byun}, Do-Young and {Chanapote}, Thanapol and {Chen}, Wen and {Chen}, Zhong and {Cheng}, Xiaopeng and {Chibueze}, James O. and {Cho}, Ilje and {Cho}, Se-Hyung and {Chung}, Hyun-Soo and {Cui}, Lang and {Cui}, Yuzhu and {Doi}, Akihiro and {Dong}, Jian and {Fujisawa}, Kenta and {Gou}, Wei and {Guo}, Wen and {Hada}, Kazuhiro and {Hagiwara}, Yoshiaki and {Hirota}, Tomoya and {Hodgson}, Jeffrey A. and {Honma}, Mareki and {Imai}, Hiroshi and {Jaroenjittichai}, Phrudth and {Jiang}, Wu and {Jiang}, Yongbin and {Jiang}, Yongchen and {Jike}, Takaaki and {Jung}, Dong-Kyu and {Jung}, Taehyun and {Kawaguchi}, Noriyuki and {Kim}, Dong-Jin and {Kim}, Hyo-Ryoung and {Kim}, Jaeheon and {Kim}, Jeong-Sook and {Kim}, Kee-Tae and {Kim}, Soon-Wook and {Kino}, Motoki and {Kobayashi}, Hideyuki and {Koyama}, Shoko and {Kramer}, Busaba H. and {Lee}, Jee-Won and {Lee}, Jeong Ae and {Lee}, Sang-Sung and {Lee}, Sang Won and {Li}, Bin and {Li}, Guanghui and {Li}, Xiaofei and {Li}, Zhixuan and {Liu}, Qinghui and {Liu}, Xiang and {Lu}, Ru-Sen and {Motogi}, Kazuhito and {Nakamura}, Masanori and {Niinuma}, Kotaro and {Oh}, Chungsik and {Oh}, Hongjong and {Oh}, Junghwan and {Oh}, Se-Jin and {Oyama}, Tomoaki and {Park}, Jongho and {Poshyachinda}, Saran and {Ro}, Hyunwook and {Roh}, Duk-Gyoo and {Rujopakarn}, Wiphu and {Sakai}, Nobuyuki and {Sawada-Satoh}, Satoko and {Shen}, Zhi-Qiang and {Shibata}, Katsunori M. and {Sohn}, Bong Won and {Soonthornthum}, Boonrucksar and {Sugiyama}, Koichiro and {Sun}, Yunxia and {Takamura}, Mieko and {Tanabe}, Yoshihiro and {Tazaki}, Fumie and {Trippe}, Sascha and {Wajima}, Kiyoaki and {Wang}, Jinqing and {Wang}, Na and {Wang}, Shiqiang and {Wang}, Xuezheng and {Xia}, Bo and {Xu}, Shuangjing and {Yan}, Hao and {Yang}, Wenjun and {Yeom}, Jae-Hwan and {Yi}, Kunwoo and {Yi}, Sang-Oh and {Yonekura}, Yoshinori and {Yoon}, Hasu and {Yu}, Linfeng and {Yuan}, Jianping and {Yun}, Youngjoo and {Zhang}, Bo and {Zhang}, Hua and {Zhang}, Yingkang and {Zhao}, Guang-Yao and {Zhao}, Rongbing and {Zhong}, Weiye and {East Asian VLBI Network Collaboration}},
        title = "{Overview of the Observing System and Initial Scientific Accomplishments of the East Asian VLBI Network (EAVN)}",
      journal = {Galaxies},
         year = 2022,
        month = dec,
       volume = {10},
       number = {6},
          eid = {113},
        pages = {113},
          doi = {10.3390/galaxies10060113},
archivePrefix = {arXiv},
       eprint = {2212.07040},
 primaryClass = {astro-ph.IM},
       adsurl = {https://ui.adsabs.harvard.edu/abs/2022Galax..10..113A}
}

@INPROCEEDINGS{Shepherd1997,
       author = {{Shepherd}, M.~C.},
        title = "{Difmap: an Interactive Program for Synthesis Imaging}",
    booktitle = {Astronomical Data Analysis Software and Systems VI},
         year = 1997,
       editor = {{Hunt}, Gareth and {Payne}, Harry},
       series = {Astronomical Society of the Pacific Conference Series},
       volume = {125},
        month = jan,
        pages = {77},
       adsurl = {https://ui.adsabs.harvard.edu/abs/1997ASPC..125...77S}
}

@ARTICLE{Park2023a,
       author = {{Park}, Jongho and {Asada}, Keiichi and {Byun}, Do-Young},
        title = "{Calibrating VLBI Polarization Data Using GPCAL. I. Frequency-dependent Calibration}",
      journal = {\apj},
         year = 2023,
        month = nov,
       volume = {958},
       number = {1},
          eid = {27},
        pages = {27},
          doi = {10.3847/1538-4357/acfd2f},
archivePrefix = {arXiv},
       eprint = {2310.03242},
 primaryClass = {astro-ph.IM},
       adsurl = {https://ui.adsabs.harvard.edu/abs/2023ApJ...958...27P}
}

@ARTICLE{Park2023b,
       author = {{Park}, Jongho and {Asada}, Keiichi and {Byun}, Do-Young},
        title = "{Calibrating VLBI Polarization Data Using GPCAL. II. Time-dependent Calibration}",
      journal = {\apj},
         year = 2023,
        month = nov,
       volume = {958},
       number = {1},
          eid = {28},
        pages = {28},
          doi = {10.3847/1538-4357/acfd30},
archivePrefix = {arXiv},
       eprint = {2310.03244},
 primaryClass = {astro-ph.IM},
       adsurl = {https://ui.adsabs.harvard.edu/abs/2023ApJ...958...28P}
}

@article{rees1984,
  title={Black hole models for active galactic nuclei},
  author={Rees, Martin J},
  journal={IN: Annual review of astronomy and astrophysics. Volume 22. Palo Alto, CA, Annual Reviews, Inc., 1984, p. 471-506.},
  volume={22},
  pages={471--506},
  year={1984}
}

@ARTICLE{Park2021b,
       author = {{Park}, Jongho and {Hada}, Kazuhiro and {Nakamura}, Masanori and {Asada}, Keiichi and {Zhao}, Guangyao and {Kino}, Motoki},
        title = "{Jet Collimation and Acceleration in the Giant Radio Galaxy NGC 315}",
      journal = {\apj},
         year = 2021,
        month = mar,
       volume = {909},
       number = {1},
          eid = {76},
        pages = {76},
          doi = {10.3847/1538-4357/abd6ee},
archivePrefix = {arXiv},
       eprint = {2012.14154},
 primaryClass = {astro-ph.HE},
       adsurl = {https://ui.adsabs.harvard.edu/abs/2021ApJ...909...76P}
}

@ARTICLE{ZT2002,
       author = {{Zavala}, R.~T. and {Taylor}, G.~B.},
        title = "{Faraday Rotation Measures in the Parsec-Scale Jets of the Radio Galaxies M87, 3C 111, and 3C 120}",
      journal = {\apjl},
         year = 2002,
        month = feb,
       volume = {566},
       number = {1},
        pages = {L9-L12},
          doi = {10.1086/339441},
archivePrefix = {arXiv},
       eprint = {astro-ph/0201458},
 primaryClass = {astro-ph},
       adsurl = {https://ui.adsabs.harvard.edu/abs/2002ApJ...566L...9Z}
}

@ARTICLE{Nikonov2023,
       author = {{Nikonov}, A.~S. and {Kovalev}, Y.~Y. and {Kravchenko}, E.~V. and {Pashchenko}, I.~N. and {Lobanov}, A.~P.},
        title = "{Properties of the jet in M87 revealed by its helical structure imaged with the VLBA at 8 and 15 GHz}",
      journal = {\mnras},
         year = 2023,
        month = dec,
       volume = {526},
       number = {4},
        pages = {5949-5963},
          doi = {10.1093/mnras/stad3061},
archivePrefix = {arXiv},
       eprint = {2307.11660},
 primaryClass = {astro-ph.GA},
       adsurl = {https://ui.adsabs.harvard.edu/abs/2023MNRAS.526.5949N}
}

@ARTICLE{Park2021c,
       author = {{Park}, Jongho and {Asada}, Keiichi and {Nakamura}, Masanori and {Kino}, Motoki and {Pu}, Hung-Yi and {Hada}, Kazuhiro and {Kravchenko}, Evgeniya V. and {Giroletti}, Marcello},
        title = "{A Revised View of the Linear Polarization in the Subparsec Core of M87 at 7 mm}",
      journal = {\apj},
         year = 2021,
        month = dec,
       volume = {922},
       number = {2},
          eid = {180},
        pages = {180},
          doi = {10.3847/1538-4357/ac26bf},
archivePrefix = {arXiv},
       eprint = {2107.13243},
 primaryClass = {astro-ph.HE},
       adsurl = {https://ui.adsabs.harvard.edu/abs/2021ApJ...922..180P}
}

@ARTICLE{Park2024,
       author = {{Park}, Jongho and {Zhao}, Guang-Yao and {Nakamura}, Masanori and {Mizuno}, Yosuke and {Pu}, Hung-Yi and {Asada}, Keiichi and {Takahashi}, Kazuya and {Toma}, Kenji and {Kino}, Motoki and {Cho}, Ilje and {Hada}, Kazuhiro and {Edwards}, Phil G. and {Ro}, Hyunwook and {Kam}, Minchul and {Yi}, Kunwoo and {Lee}, Yunjeong and {Koyama}, Shoko and {Byun}, Do-Young and {Phillips}, Chris and {Reynolds}, Cormac and {Hodgson}, Jeffrey A. and {Lee}, Sang-Sung},
        title = "{Discovery of Limb Brightening in the Parsec-scale Jet of NGC 315 through Global Very Long Baseline Interferometry Observations and Its Implications for Jet Models}",
      journal = {\apjl},
         year = 2024,
        month = oct,
       volume = {973},
       number = {2},
          eid = {L45},
        pages = {L45},
          doi = {10.3847/2041-8213/ad7137},
archivePrefix = {arXiv},
       eprint = {2408.09069},
 primaryClass = {astro-ph.HE},
       adsurl = {https://ui.adsabs.harvard.edu/abs/2024ApJ...973L..45P}
}

@ARTICLE{Weaver2022,
       author = {{Weaver}, Zachary R. and {Jorstad}, Svetlana G. and {Marscher}, Alan P. and {Morozova}, Daria A. and {Troitsky}, Ivan S. and {Agudo}, Iv{\'a}n and {G{\'o}mez}, Jos{\'e} L. and {L{\"a}hteenm{\"a}ki}, Anne and {Tammi}, Joni and {Tornikoski}, Merja},
        title = "{Kinematics of Parsec-scale Jets of Gamma-Ray Blazars at 43 GHz during 10 yr of the VLBA-BU-BLAZAR Program}",
      journal = {\apjs},
         year = 2022,
        month = may,
       volume = {260},
       number = {1},
          eid = {12},
        pages = {12},
          doi = {10.3847/1538-4365/ac589c},
archivePrefix = {arXiv},
       eprint = {2202.12290},
 primaryClass = {astro-ph.HE},
       adsurl = {https://ui.adsabs.harvard.edu/abs/2022ApJS..260...12W}
}

@BOOK{Thompson2017,
       author = {{Thompson}, A. Richard and {Moran}, James M. and {Swenson}, George W., Jr.},
        title = "{Interferometry and Synthesis in Radio Astronomy, 3rd Edition}",
         year = 2017,
          doi = {10.1007/978-3-319-44431-4},
       adsurl = {https://ui.adsabs.harvard.edu/abs/2017isra.book.....T}
}

@article{park2021,
doi = {10.3847/1538-4357/abcc6e},
url = {https://dx.doi.org/10.3847/1538-4357/abcc6e},
year = {2021},
month = {jan},
publisher = {The American Astronomical Society},
volume = {906},
number = {2},
pages = {85},
author = {Park, Jongho and Byun, Do-Young and Asada, Keiichi and Yun, Youngjoo},
title = {GPCAL: A Generalized Calibration Pipeline for Instrumental Polarization in VLBI Data},
journal = {The Astrophysical Journal}
}

@ARTICLE{Blandford2019,
       author = {{Blandford}, Roger and {Meier}, David and {Readhead}, Anthony},
        title = "{Relativistic Jets from Active Galactic Nuclei}",
      journal = {\araa},
         year = 2019,
        month = aug,
       volume = {57},
        pages = {467-509},
          doi = {10.1146/annurev-astro-081817-051948},
archivePrefix = {arXiv},
       eprint = {1812.06025},
 primaryClass = {astro-ph.HE},
       adsurl = {https://ui.adsabs.harvard.edu/abs/2019ARA&A..57..467B}
}

@ARTICLE{Takamura2023,
       author = {{Takamura}, Mieko and {Hada}, Kazuhiro and {Honma}, Mareki and {Oyama}, Tomoaki and {Yamauchi}, Aya and {Suzuki}, Syunsaku and {Hagiwara}, Yoshiaki and {Orienti}, Monica and {D'Ammando}, Filippo and {Park}, Jongho and {Kam}, Minchul and {Doi}, Akihiro},
        title = "{Probing the Heart of Active Narrow-line Seyfert 1 Galaxies with VERA Wideband Polarimetry}",
      journal = {\apj},
         year = 2023,
        month = jul,
       volume = {952},
       number = {1},
          eid = {47},
        pages = {47},
          doi = {10.3847/1538-4357/acd9a8},
archivePrefix = {arXiv},
       eprint = {2306.03139},
 primaryClass = {astro-ph.HE},
       adsurl = {https://ui.adsabs.harvard.edu/abs/2023ApJ...952...47T}
}

@ARTICLE{Park2019,
       author = {{Park}, Jongho and {Hada}, Kazuhiro and {Kino}, Motoki and
         {Nakamura}, Masanori and {Ro}, Hyunwook and {Trippe}, Sascha},
        title = "{Faraday Rotation in the Jet of M87 inside the Bondi Radius: Indication of Winds from Hot Accretion Flows Confining the Relativistic Jet}",
      journal = {\apj},
         year = 2019,
        month = feb,
       volume = {871},
       number = {2},
          eid = {257},
        pages = {257},
          doi = {10.3847/1538-4357/aaf9a9},
archivePrefix = {arXiv},
       eprint = {1812.08386},
 primaryClass = {astro-ph.HE},
       adsurl = {https://ui.adsabs.harvard.edu/abs/2019ApJ...871..257P}
}

@ARTICLE{Cawthorne2013,
       author = {{Cawthorne}, T.~V. and {Jorstad}, S.~G. and {Marscher}, A.~P.},
        title = "{Polarization Structure in the Core of 1803+784: A Signature of Recollimation Shocks?}",
      journal = {\apj},
         year = 2013,
        month = jul,
       volume = {772},
       number = {1},
          eid = {14},
        pages = {14},
          doi = {10.1088/0004-637X/772/1/14},
archivePrefix = {arXiv},
       eprint = {1305.5356},
 primaryClass = {astro-ph.HE},
       adsurl = {https://ui.adsabs.harvard.edu/abs/2013ApJ...772...14C}
}

@ARTICLE{Jorstad2007,
       author = {{Jorstad}, Svetlana G. and {Marscher}, Alan P. and {Stevens}, Jason A. and
         {Smith}, Paul S. and {Forster}, James R. and {Gear}, Walter K. and
         {Cawthorne}, Timothy V. and {Lister}, Matthew L. and
         {Stirling}, Alastair M. and {G{\'o}mez}, Jos{\'e} L. and
         {Greaves}, Jane S. and {Robson}, E. Ian},
        title = "{Multiwaveband Polarimetric Observations of 15 Active Galactic Nuclei at High Frequencies: Correlated Polarization Behavior}",
      journal = {\aj},
         year = 2007,
        month = aug,
       volume = {134},
       number = {2},
        pages = {799-824},
          doi = {10.1086/519996},
archivePrefix = {arXiv},
       eprint = {0705.4273},
 primaryClass = {astro-ph},
       adsurl = {https://ui.adsabs.harvard.edu/abs/2007AJ....134..799J}
}

@ARTICLE{Clausen-Brown2011,
       author = {{Clausen-Brown}, E. and {Lyutikov}, M. and {Kharb}, P.},
        title = "{Signatures of large-scale magnetic fields in active galactic nuclei jets: transverse asymmetries}",
      journal = {\mnras},
         year = 2011,
        month = aug,
       volume = {415},
       number = {3},
        pages = {2081-2092},
          doi = {10.1111/j.1365-2966.2011.18757.x},
archivePrefix = {arXiv},
       eprint = {1101.5149},
 primaryClass = {astro-ph.HE},
       adsurl = {https://ui.adsabs.harvard.edu/abs/2011MNRAS.415.2081C}
}

@ARTICLE{blandford_1977,
       author = {{Blandford}, R.~D. and {Znajek}, R.~L.},
        title = "{Electromagnetic extraction of energy from Kerr black holes.}",
      journal = {\mnras},
         year = "1977",
        month = "May",
       volume = {179},
        pages = {433-456},
          doi = {10.1093/mnras/179.3.433},
       adsurl = {https://ui.adsabs.harvard.edu/\#abs/1977MNRAS.179..433B}
}

@ARTICLE{blandford_1982,
       author = {{Blandford}, R.~D. and {Payne}, D.~G.},
        title = "{Hydromagnetic flows from accretion disks and the production of radio jets.}",
      journal = {\mnras},
         year = "1982",
        month = "Jun",
       volume = {199},
        pages = {883-903},
          doi = {10.1093/mnras/199.4.883},
       adsurl = {https://ui.adsabs.harvard.edu/\#abs/1982MNRAS.199..883B} }

@ARTICLE{Gomez2002,
       author = {{G\'omez}, J.~L. and {Marscher}, A.~P. and {Alberdi}, A. and {Jorstad}, S.~G. and {Agudo}, I.},
        title = "{Polarization calibration of the VLBA using the D-terms.}",
      journal = {VLBA Scientific Memo No. 30},
         year = "2002",
        month = "Mar",
       volume = {},
        pages = {},
          doi = {},
       adsurl = {}}

@ARTICLE{eht2021,
       author = {{Event Horizon Telescope Collaboration} and {Akiyama}, Kazunori and {Algaba}, Juan Carlos and {Alberdi}, Antxon and {Alef}, Walter and {Anantua}, Richard and {Asada}, Keiichi and {Azulay}, Rebecca and {Baczko}, Anne-Kathrin and {Ball}, David and {Balokovi{\'c}}, Mislav and {Barrett}, John and {Benson}, Bradford A. and {Bintley}, Dan and {Blackburn}, Lindy and {Blundell}, Raymond and {Boland}, Wilfred and {Bouman}, Katherine L. and {Bower}, Geoffrey C. and {Boyce}, Hope and {Bremer}, Michael and {Brinkerink}, Christiaan D. and {Brissenden}, Roger and {Britzen}, Silke and {Broderick}, Avery E. and {Broguiere}, Dominique and {Bronzwaer}, Thomas and {Byun}, Do-Young and {Carlstrom}, John E. and {Chael}, Andrew and {Chan}, Chi-kwan and {Chatterjee}, Shami and {Chatterjee}, Koushik and {Chen}, Ming-Tang and {Chen}, Yongjun and {Chesler}, Paul M. and {Cho}, Ilje and {Christian}, Pierre and {Conway}, John E. and {Cordes}, James M. and {Crawford}, Thomas M. and {Crew}, Geoffrey B. and {Cruz-Osorio}, Alejandro and {Cui}, Yuzhu and {Davelaar}, Jordy and {De Laurentis}, Mariafelicia and {Deane}, Roger and {Dempsey}, Jessica and {Desvignes}, Gregory and {Dexter}, Jason and {Doeleman}, Sheperd S. and {Eatough}, Ralph P. and {Falcke}, Heino and {Farah}, Joseph and {Fish}, Vincent L. and {Fomalont}, Ed and {Ford}, H. Alyson and {Fraga-Encinas}, Raquel and {Freeman}, William T. and {Friberg}, Per and {Fromm}, Christian M. and {Fuentes}, Antonio and {Galison}, Peter and {Gammie}, Charles F. and {Garc{\'\i}a}, Roberto and {Gentaz}, Olivier and {Georgiev}, Boris and {Goddi}, Ciriaco and {Gold}, Roman and {G{\'o}mez}, Jos{\'e} L. and {G{\'o}mez-Ruiz}, Arturo I. and {Gu}, Minfeng and {Gurwell}, Mark and {Hada}, Kazuhiro and {Haggard}, Daryl and {Hecht}, Michael H. and {Hesper}, Ronald and {Ho}, Luis C. and {Ho}, Paul and {Honma}, Mareki and {Huang}, Chih-Wei L. and {Huang}, Lei and {Hughes}, David H. and {Ikeda}, Shiro and {Inoue}, Makoto and {Issaoun}, Sara and {James}, David J. and {Jannuzi}, Buell T. and {Janssen}, Michael and {Jeter}, Britton and {Jiang}, Wu and {Jimenez-Rosales}, Alejandra and {Johnson}, Michael D. and {Jorstad}, Svetlana and {Jung}, Taehyun and {Karami}, Mansour and {Karuppusamy}, Ramesh and {Kawashima}, Tomohisa and {Keating}, Garrett K. and {Kettenis}, Mark and {Kim}, Dong-Jin and {Kim}, Jae-Young and {Kim}, Jongsoo and {Kim}, Junhan and {Kino}, Motoki and {Koay}, Jun Yi and {Kofuji}, Yutaro and {Koch}, Patrick M. and {Koyama}, Shoko and {Kramer}, Michael and {Kramer}, Carsten and {Krichbaum}, Thomas P. and {Kuo}, Cheng-Yu and {Lauer}, Tod R. and {Lee}, Sang-Sung and {Levis}, Aviad and {Li}, Yan-Rong and {Li}, Zhiyuan and {Lindqvist}, Michael and {Lico}, Rocco and {Lindahl}, Greg and {Liu}, Jun and {Liu}, Kuo and {Liuzzo}, Elisabetta and {Lo}, Wen-Ping and {Lobanov}, Andrei P. and {Loinard}, Laurent and {Lonsdale}, Colin and {Lu}, Ru-Sen and {MacDonald}, Nicholas R. and {Mao}, Jirong and {Marchili}, Nicola and {Markoff}, Sera and {Marrone}, Daniel P. and {Marscher}, Alan P. and {Mart{\'\i}-Vidal}, Iv{\'a}n and {Matsushita}, Satoki and {Matthews}, Lynn D. and {Medeiros}, Lia and {Menten}, Karl M. and {Mizuno}, Izumi and {Mizuno}, Yosuke and {Moran}, James M. and {Moriyama}, Kotaro and {Moscibrodzka}, Monika and {M{\"u}ller}, Cornelia and {Musoke}, Gibwa and {Mej{\'\i}as}, Alejandro Mus and {Michalik}, Daniel and {Nadolski}, Andrew and {Nagai}, Hiroshi and {Nagar}, Neil M. and {Nakamura}, Masanori and {Narayan}, Ramesh and {Narayanan}, Gopal and {Natarajan}, Iniyan and {Nathanail}, Antonios and {Neilsen}, Joey and {Neri}, Roberto and {Ni}, Chunchong and {Noutsos}, Aristeidis and {Nowak}, Michael A. and {Okino}, Hiroki and {Olivares}, H{\'e}ctor and {Ortiz-Le{\'o}n}, Gisela N. and {Oyama}, Tomoaki and {{\"O}zel}, Feryal and {Palumbo}, Daniel C.~M. and {Park}, Jongho and {Patel}, Nimesh and {Pen}, Ue-Li and {Pesce}, Dominic W. and {Pi{\'e}tu}, Vincent and {Plambeck}, Richard and {PopStefanija}, Aleksandar and {Porth}, Oliver and {P{\"o}tzl}, Felix M. and {Prather}, Ben and {Preciado-L{\'o}pez}, Jorge A. and {Psaltis}, Dimitrios and {Pu}, Hung-Yi and {Ramakrishnan}, Venkatessh and {Rao}, Ramprasad and {Rawlings}, Mark G. and {Raymond}, Alexander W. and {Rezzolla}, Luciano and {Ricarte}, Angelo and {Ripperda}, Bart and {Roelofs}, Freek and {Rogers}, Alan and {Ros}, Eduardo and {Rose}, Mel and {Roshanineshat}, Arash and {Rottmann}, Helge and {Roy}, Alan L. and {Ruszczyk}, Chet and {Rygl}, Kazi L.~J. and {S{\'a}nchez}, Salvador and {S{\'a}nchez-Arguelles}, David and {Sasada}, Mahito},
        title = "{First M87 Event Horizon Telescope Results. VII. Polarization of the Ring}",
      journal = {\apjl},
         year = 2021,
        month = mar,
       volume = {910},
       number = {1},
          eid = {L12},
        pages = {L12},
          doi = {10.3847/2041-8213/abe71d},
archivePrefix = {arXiv},
       eprint = {2105.01169},
 primaryClass = {astro-ph.HE},
       adsurl = {https://ui.adsabs.harvard.edu/abs/2021ApJ...910L..12E}
}

@article{cui2021,
  title={East Asian VLBI Network observations of active galactic nuclei jets: imaging with KaVA+ Tianma+ Nanshan},
  author={Cui, Yu-Zhu and Hada, Kazuhiro and Kino, Motoki and Sohn, Bong-Won and Park, Jongho and Ro, Hyun-Wook and Sawada-Satoh, Satoko and Jiang, Wu and Cui, Lang and Honma, Mareki and others},
  journal={Research in Astronomy and Astrophysics},
  volume={21},
  number={8},
  pages={205},
  year={2021},
  publisher={IOP Publishing}
}

@article{lister2018,
  title={MOJAVE. XV. VLBA 15 GHz total intensity and polarization maps of 437 parsec-scale AGN jets from 1996 to 2017},
  author={Lister, ML and Aller, MF and Aller, HD and Hodge, MA and Homan, DC and Kovalev, YY and Pushkarev, AB and Savolainen, T},
  journal={The Astrophysical Journal Supplement Series},
  volume={234},
  number={1},
  pages={12},
  year={2018},
  publisher={IOP Publishing}
}

@incollection{greisen2003,
  title={AIPS, the VLA, and the VLBA},
  author={Greisen, Eric W},
  booktitle={Information Handling in Astronomy-Historical Vistas},
  pages={109--125},
  year={2003},
  publisher={Springer}
}

@inproceedings{shepherd1994,
  title={DIFMAP: an interactive program for synthesis imaging.},
  author={Shepherd, MC and Pearson, TJ and Taylor, GB},
  booktitle={Bulletin of the Astronomical Society, Vol. 26, No. 2, p. 987-989},
  volume={26},
  pages={987--989},
  year={1994}
}

@ARTICLE{Park2018,
       author = {{Park}, Jongho and {Kam}, Minchul and {Trippe}, Sascha and {Kang}, Sincheol and {Byun}, Do-Young and {Kim}, Dae-Won and {Algaba}, Juan-Carlos and {Lee}, Sang-Sung and {Zhao}, Guang-Yao and {Kino}, Motoki and {Shin}, Naeun and {Hada}, Kazuhiro and {Lee}, Taeseok and {Oh}, Junghwan and {Hodgson}, Jeffrey A. and {Sohn}, Bong Won},
        title = "{Revealing the Nature of Blazar Radio Cores through Multifrequency Polarization Observations with the Korean VLBI Network}",
      journal = {\apj},
         year = 2018,
        month = jun,
       volume = {860},
       number = {2},
          eid = {112},
        pages = {112},
          doi = {10.3847/1538-4357/aac490},
archivePrefix = {arXiv},
       eprint = {1805.04299},
 primaryClass = {astro-ph.HE},
       adsurl = {https://ui.adsabs.harvard.edu/abs/2018ApJ...860..112P}
}

@ARTICLE{kim2015,
       author = {{Kim}, Jae-Young and {Trippe}, Sascha and {Sohn}, Bong Won and {Oh}, Junghwan and {Park}, Jong-Ho and {Lee}, Sang-Sung and {Lee}, Taeseok and {Kim}, Daewon},
        title = "{PAGaN I: Multi-Frequency Polarimetry of AGN Jets with KVN}",
      journal = {Journal of Korean Astronomical Society},
         year = 2015,
        month = oct,
       volume = {48},
       number = {5},
        pages = {285-298},
          doi = {10.5303/JKAS.2015.48.5.285},
archivePrefix = {arXiv},
       eprint = {1510.08150},
 primaryClass = {astro-ph.GA},
       adsurl = {https://ui.adsabs.harvard.edu/abs/2015JKAS...48..285K}
}

@article{hagiwara2022,
  title={Demonstration of ultrawideband polarimetry using VLBI exploration of radio astrometry (VERA)},
  author={Hagiwara, Yoshiaki and Hada, Kazuhiro and Takamura, Mieko and Oyama, Tomoaki and Yamauchi, Aya and Suzuki, Syunsaku},
  journal={Galaxies},
  volume={10},
  number={6},
  pages={114},
  year={2022},
  publisher={MDPI}
}

@article{chatterjee2008,
  title={Correlated multi-wave band variability in the blazar 3C 279 from 1996 to 2007},
  author={Chatterjee, Ritaban and Jorstad, Svetlana G and Marscher, Alan P and Oh, Haruki and McHardy, Ian M and Aller, Margo F and Aller, Hugh D and Balonek, Thomas J and Miller, H Richard and Ryle, Wesley T and others},
  journal={The Astrophysical Journal},
  volume={689},
  number={1},
  pages={79},
  year={2008},
  publisher={IOP Publishing}
}

@article{hayashida2012,
  title={The structure and emission model of the relativistic jet in the quasar 3C 279 inferred from radio to high-energy $\gamma$-ray observations in 2008--2010},
  author={Hayashida, M and Madejski, GM and Nalewajko, K and Sikora, M and Wehrle, AE and Ogle, P and Collmar, W and Larsson, Stefan and Fukazawa, Y and Itoh, R and others},
  journal={The Astrophysical Journal},
  volume={754},
  number={2},
  pages={114},
  year={2012},
  publisher={IOP Publishing}
}

@article{larionov2020,
  title={Multiwavelength behaviour of the blazar 3C 279: decade-long study from $\gamma$-ray to radio},
  author={Larionov, Valeri M and Jorstad, Svetlana G and Marscher, Alan P and Villata, M and Raiteri, CM and Smith, PS and Agudo, Iv{\'a}n and Savchenko, Sergey S and Morozova, DA and Acosta-Pulido, JA and others},
  journal={Monthly Notices of the Royal Astronomical Society},
  volume={492},
  number={3},
  pages={3829--3848},
  year={2020},
  publisher={Oxford University Press}
}

@article{kiehlmann2016,
  title={Polarization angle swings in blazars: The case of 3C 279},
  author={Kiehlmann, S and Savolainen, T and Jorstad, SG and Sokolovsky, KV and Schinzel, FK and Marscher, AP and Larionov, VM and Agudo, I and Akitaya, H and Ben{\'\i}tez, E and others},
  journal={Astronomy \& Astrophysics},
  volume={590},
  pages={A10},
  year={2016},
  publisher={EDP Sciences}
}

@article{strauss1992,
  title={A redshift survey of IRAS galaxies. VII-The infrared and redshift data for the 1.936 Jansky sample},
  author={Strauss, Michael A and Huchra, John P and Davis, Marc and Yahil, Amos and Fisher, Karl B and Tonry, John},
  journal={Astrophysical Journal Supplement Series (ISSN 0067-0049), vol. 83, no. 1, p. 29-63.},
  volume={83},
  pages={29--63},
  year={1992}
}

@article{sillanpaa1996,
  title={Confirmation of the 12-year optical outburst cycle in blazar OJ 287.},
  author={Sillanpaa, A and Takalo, LO and Pursimo, T and Lehto, HJ and Nilsson, K and Teerikorpi, P and Heinaemaeki, P and Kidger, M and De Diego, JA and Gonzalez-Perez, JN and others},
  journal={Astronomy and Astrophysics, v. 305, p. L17},
  volume={305},
  pages={L17},
  year={1996}
}

@article{valtaoja2000,
  title={Radio Monitoring of OJ 287 and Binary Black Hole Models for PeriodicOutbursts},
  author={Valtaoja, E and Ter{\"a}sranta, Harri and Tornikoski, M and Sillanp{\"a}{\"a}, A and Aller, MF and Aller, HD and Hughes, PA},
  journal={The Astrophysical Journal},
  volume={531},
  number={2},
  pages={744},
  year={2000},
  publisher={IOP Publishing}
}

@article{pursimo2000,
  title={Intensive monitoring of OJ 287},
  author={Pursimo, T and Takalo, LO and Sillanp{\"a}{\"a}, A and Kidger, M and Lehto, HJ and Heidt, J and Charles, PA and Aller, H and Aller, M and Beckmann, V and others},
  journal={Astronomy and Astrophysics Supplement Series},
  volume={146},
  number={1},
  pages={141--155},
  year={2000},
  publisher={EDP Sciences}
}

@article{villforth2010,
  title={Variability and stability in blazar jets on time-scales of years: Optical polarization monitoring of OJ 287 in 2005--2009},
  author={Villforth, C and Nilsson, K and Heidt, J and Takalo, LO and Pursimo, T and Berdyugin, A and Lindfors, E and Pasanen, M and Winiarski, M and Drozdz, M and others},
  journal={Monthly Notices of the Royal Astronomical Society},
  volume={402},
  number={3},
  pages={2087--2111},
  year={2010},
  publisher={Blackwell Publishing Ltd Oxford, UK}
}

@article{holmes1984,
  title={A polarization flare in OJ 287},
  author={Holmes, PA and Brand, PWJL and Impey, CD and Williams, PM and Smith, P and Elston, R and Balonek, T and Zeilik, M and Burns, J and Heckert, P and others},
  journal={Monthly Notices of the Royal Astronomical Society},
  volume={211},
  number={3},
  pages={497--506},
  year={1984},
  publisher={Oxford University Press Oxford, UK}
}

@article{cohen2018,
  title={Reversals in the Direction of Polarization Rotation in OJ 287},
  author={Cohen, MH and Aller, HD and Aller, MF and Hovatta, T and Kharb, P and Kovalev, YY and Lister, ML and Meier, DL and Pushkarev, AB and Savolainen, T},
  journal={The Astrophysical Journal},
  volume={862},
  number={1},
  pages={1},
  year={2018},
  publisher={IOP Publishing}
}

@article{asada2002,
  title={A helical magnetic field in the jet of 3C 273},
  author={Asada, Keiichi and Inoue, Makoto and Uchida, Yutaka and Kameno, Seiji and Fujisawa, Kenta and Iguchi, Satoru and Mutoh, Mutsumi},
  journal={Publications of the Astronomical Society of Japan},
  volume={54},
  number={3},
  pages={L39--L43},
  year={2002},
  publisher={Oxford University Press Oxford, UK}
}

@article{hovatta2012,
  title={MOJAVE: monitoring of jets in active galactic nuclei with VLBA experiments. VIII. Faraday rotation in parsec-scale AGN jets},
  author={Hovatta, Talvikki and Lister, Matthew L and Aller, Margo F and Aller, Hugh D and Homan, Daniel C and Kovalev, Yuri Y and Pushkarev, Alexander B and Savolainen, Tuomas},
  journal={The Astronomical Journal},
  volume={144},
  number={4},
  pages={105},
  year={2012},
  publisher={IOP Publishing}
}

@article{hamaker1996_1,
  title={Understanding radio polarimetry. I. Mathematical foundations},
  author={Hamaker, JP and Bregman, JD and Sault, RJ},
  journal={Astronomy and Astrophysics Supplement Series},
  volume={117},
  number={1},
  pages={137--147},
  year={1996},
  publisher={EDP Sciences}
}

@article{sault1996,
  title={Understanding radio polarimetry. II. Instrumental calibration of an interferometer array},
  author={Sault, RJ and Hamaker, JP and Bregman, JD},
  journal={Astronomy and Astrophysics Supplement Series},
  volume={117},
  number={1},
  pages={149--159},
  year={1996},
  publisher={EDP Sciences}
}

@article{hamaker1996_3,
  title={Understanding radio polarimetry. III. Interpreting the IAU/IEEE definitions of the Stokes parameters},
  author={Hamaker, JP and Bregman, JD},
  journal={Astronomy and Astrophysics Supplement Series},
  volume={117},
  number={1},
  pages={161--165},
  year={1996},
  publisher={EDP Sciences}
}

@article{smirnov2011,
  title={Revisiting the radio interferometer measurement equation-II. Calibration and direction-dependent effects},
  author={Smirnov, Oleg M},
  journal={Astronomy \& Astrophysics},
  volume={527},
  pages={A107},
  year={2011},
  publisher={EDP Sciences}
}

@article{jones1941,
  title={A new calculus for the treatment of optical systemsi. description and discussion of the calculus},
  author={Jones, R Clark},
  journal={Journal of the Optical Society of America},
  volume={31},
  number={7},
  pages={488--493},
  year={1941},
  publisher={OSA}
}

@article{lister2005,
  title={MOJAVE: monitoring of jets in active galactic nuclei with VLBA experiments. I. First-epoch 15 GHz linear polarization images},
  author={Lister, Matthew L and Homan, DC},
  journal={The Astronomical Journal},
  volume={130},
  number={4},
  pages={1389},
  year={2005},
  publisher={IOP Publishing}
}

@article{an2018,
  title={Capabilities and prospects of the East Asia very long baseline interferometry network},
  author={An, T and Sohn, BW and Imai, H},
  journal={Nature Astronomy},
  volume={2},
  number={2},
  pages={118--125},
  year={2018},
  publisher={Nature Publishing Group UK London}
}

@article{lee2015,
  title={A New Hardware Correlator in Korea: Performance Evaluation Using KVN Observations},
  author={Lee, Sang-Sung and Oh, Chung Sik and Roh, Duk-Gyoo and Oh, Se-Jin and Kim, Jongsoo and Yeom, Jae-Hwan and Kim, Hyo Ryoung and Jung, Dong-Gyu and Byun, Do-Young and Jung, Taehyun and others},
  journal={arXiv preprint arXiv:1503.07972},
  year={2015}
}
\bibliographystyle{aasjournalv7}

\end{document}